\documentclass[reprint,amsmath,amssymb,aps,]{revtex4-2}
\usepackage[utf8]{inputenc}
\usepackage[resetlabels,labeled]{multibib}
\usepackage{graphicx}
\usepackage[symbol]{footmisc}
\usepackage{dsfont}
\usepackage{pifont}
\usepackage{braket}
\usepackage{bbold}
\usepackage{multirow}
\usepackage{tabularx}
\usepackage{tabu}
\usepackage{bbm}
\usepackage{tikz}
\usepackage{adjustbox}
\usetikzlibrary{quantikz}
\newcolumntype{L}[1]{>{\raggedright\let\newline\\\arraybackslash\hspace{0pt}}m{#1}}
\newcolumntype{C}[1]{>{\centering\let\newline\\\arraybackslash\hspace{0pt}}m{#1}}

\usepackage{hyperref}
\hypersetup{
    colorlinks=true,
    linkcolor=blue,
    filecolor=magenta,      
    urlcolor=blue,
    citecolor=blue,
    pdftitle={Error-detected gates},
    pdfpagemode=FullScreen,
    }

\begin{document}

\title{Error-detectable bosonic entangling gates with a noisy ancilla}

\author{Takahiro Tsunoda}
\thanks{These authors contributed equally to this work. \\ takahiro.tsunoda@yale.edu \\ james.teoh@yale.edu \\}
\affiliation{Departments of Applied Physics and Physics, Yale University, New Haven, CT, USA}
\affiliation{Yale Quantum Institute, Yale University, New Haven, CT, USA}

\author{James D. Teoh}
\thanks{These authors contributed equally to this work. \\ takahiro.tsunoda@yale.edu \\ james.teoh@yale.edu \\}
\affiliation{Departments of Applied Physics and Physics, Yale University, New Haven, CT, USA}
\affiliation{Yale Quantum Institute, Yale University, New Haven, CT, USA}

\author{William D. Kalfus}%
\affiliation{Departments of Applied Physics and Physics, Yale University, New Haven, CT, USA}
\affiliation{Yale Quantum Institute, Yale University, New Haven, CT, USA}

\author{Stijn J. de Graaf}%
\affiliation{Departments of Applied Physics and Physics, Yale University, New Haven, CT, USA}
\affiliation{Yale Quantum Institute, Yale University, New Haven, CT, USA}

\author{Benjamin J. Chapman}%
\affiliation{Departments of Applied Physics and Physics, Yale University, New Haven, CT, USA}
\affiliation{Yale Quantum Institute, Yale University, New Haven, CT, USA}

\author{Jacob C. Curtis}%
\affiliation{Departments of Applied Physics and Physics, Yale University, New Haven, CT, USA}
\affiliation{Yale Quantum Institute, Yale University, New Haven, CT, USA}

\author{Neel Thakur}%
\affiliation{Departments of Applied Physics and Physics, Yale University, New Haven, CT, USA}
\affiliation{Yale Quantum Institute, Yale University, New Haven, CT, USA}

\author{Steven M. Girvin}%
\affiliation{Departments of Applied Physics and Physics, Yale University, New Haven, CT, USA}
\affiliation{Yale Quantum Institute, Yale University, New Haven, CT, USA}

\author{Robert J. Schoelkopf}%
\affiliation{Departments of Applied Physics and Physics, Yale University, New Haven, CT, USA}
\affiliation{Yale Quantum Institute, Yale University, New Haven, CT, USA}

\date{\today}

\begin{abstract}
Bosonic quantum error correction has proven to be a successful approach for extending the coherence of quantum memories, but to execute deep quantum circuits, high-fidelity gates between encoded qubits are needed. To that end, we present a family of error-detectable two-qubit gates for a variety of bosonic encodings. From a new geometric framework based on a ``Bloch sphere'' of bosonic operators, we construct $ZZ_L(\theta)$ and $\text{eSWAP}(\theta)$ gates for the binomial, 4-legged cat, dual-rail and several other bosonic codes. The gate Hamiltonian is simple to engineer, requiring only a programmable beamsplitter between two bosonic qubits and an ancilla dispersively coupled to one qubit. This Hamiltonian can be realized in circuit QED hardware with ancilla transmons and microwave cavities. The proposed theoretical framework was developed for circuit QED but is generalizable to any platform that can effectively generate this Hamiltonian. Crucially, one can also detect first-order errors in the ancilla and the bosonic qubits during the gates. We show that this allows one to reach error-detected gate fidelities at the $10^{-4}$ level with today's hardware, limited only by second-order hardware errors.
\end{abstract}
\maketitle

\section{Introduction}
High-fidelity two-qubit entangling gates are crucial for implementing useful quantum circuits.
For the most part, efforts to enhance two-qubit gate performance have focused on hardware-level improvements that increase qubit coherence lifetimes \cite{koch_charge-insensitive_2007, paik_observation_2011, place_new_2021, ballance_high-fidelity_2016}, $T_{\text{coh}}$, and decrease gate duration $\tau_{\text{gate}}$, while also minimizing unwanted crosstalk \cite{dicarlo_demonstration_2009, chen_qubit_2014, wei_quantum_2021, leung_robust_2018, torrontegui_ultra-fast_2020,OliverGates2021}.
The success in engineering ever lower $\tau_{\text{gate}}/T_{\text{coh}}$ has enabled two-qubit gate fidelities in excess of  99\% in many-qubit processors using trapped ions and superconducting qubits \cite{monroe_large-scale_2014, egan_fault-tolerant_2021, olsacher_scalable_2020, postler_demonstration_2022, arute_quantum_2019, jurcevic_demonstration_2021}.
But beyond these hardware level improvements, are there other resources we can exploit?

In systems with both high-fidelity readout and more than two energy levels \cite{elder_high-fidelity_2020}, we may boost the gate fidelity by engineering a scheme in which extra levels in an ancilla act as flag states for dominant errors \cite{wu_erasure_2022, kang_quantum_2022}.
For gates where no errors are detected, the gate infidelity scales  $\propto(\tau_{\text{gate}}/T_{\text{coh}})^2$ by requiring two hardware errors during the gate to avoid detection. This quadratic scaling---rather than linear scaling---amplifies the benefit of further hardware improvements to circuit performance.

Detecting errors is useful in several important contexts, even when we do not correct them at the gate level. 
For example, in the surface code, a detected error can be treated as an erasure on a particular qubit  \cite{wu_erasure_2022, kang_quantum_2022}. 
For circuits in which erasures dominate over Pauli errors, the fault-tolerance threshold is substantially higher, thereby reducing the hardware requirements.
Meanwhile, in Noisy Intermediate-Scale Quantum (NISQ) algorithms with shallow-depth circuits, simply postselecting shots in which no errors were detected is a viable strategy \cite{preskill_quantum_2018}.

In this paper, we show how a circuit QED (cQED) system \cite{wallraff_strong_2004, reagor_quantum_2016} can implement any excitation-preserving logical two-qubit gate for a wide range of bosonic encodings, including binomial codes \cite{michael_new_2016}, cat codes \cite{Cochrane1999,mirrahimi_dynamically_2014}, rotationally-symmetric codes \cite{grimsmo_quantum_2020} and the dual-rail code \cite{chuang_simple_1995, Teoh2022}, while detecting the dominant ancilla decay and dephasing errors.
Surprisingly, this can be realized using only a programmable beamsplitter interaction between two bosonic modes plus a single three-level ancilla dispersively coupled to one of these modes \cite{gao_entanglement_2019}.
We develop a powerful geometric framework to represent the oscillator dynamics under this Hamiltonian on a Bloch sphere, enabling the straightforward design of two-qubit gates on bosonic modes, while also providing a natural means for incorporating ancilla error detection.
Building on previous work on fault-tolerant single-qubit gates for bosonic codes~\cite{ma_path-independent_2020, reinhold_error-corrected_2020}, we make use of the $\ket{e}$ and $\ket{f}$ levels of the three-level ancilla as flag states for all first-order ancilla errors during two-qubit gates, such as a single phase flip errors or decay errors.

To demonstrate the power of this protocol, we simulate the performance of logical two-qubit gates on binomial and dual-rail encoded qubits.  With typical hardware parameters and error rates, we find an expected gate infidelity below $10^{-4}$ and which scales as $(\tau_\text{gate}/T_\text{coh})^2$. Gate failure, which results from detecting an error, occurs with probabilities below $10^{-2}$ and scale as $(\tau_\text{gate}/T_\text{coh})$

\section{Overview of bosonic two-qubit gate design}

\begin{figure*}[t]
    \begin{tabular}{c c}
    \includegraphics[width=0.8\linewidth]{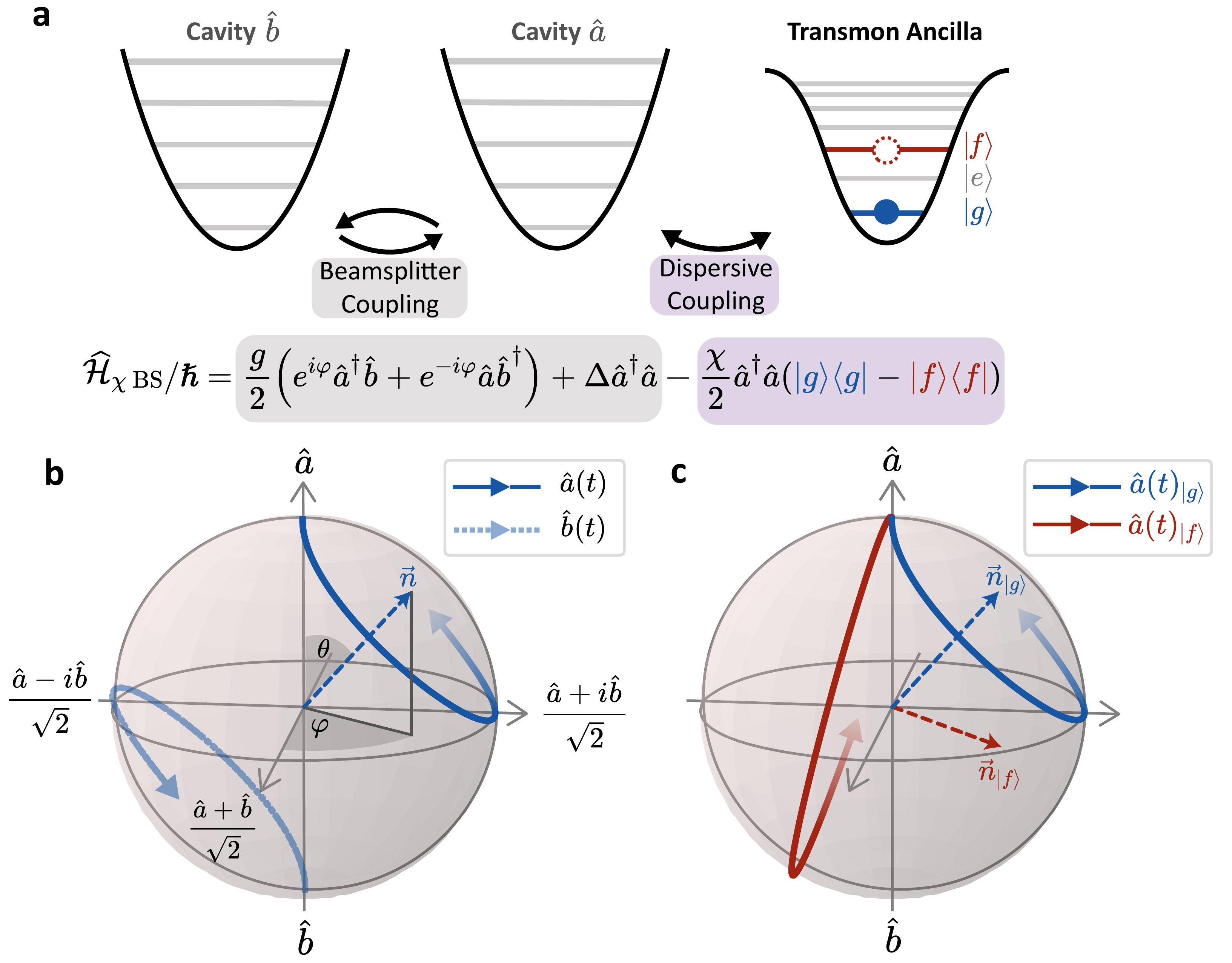}
    \end{tabular}

\caption{
\textbf{Operator Bloch sphere framework for designing entangling gates for bosonic qubits} \textbf{(a)} Physical system and Hamiltonian $\widehat{\mathcal{H}}_{\chi \text{BS}}$ used to generate the gates. Two bosonic modes are coupled by a programmable beamsplitter interaction (grey). Simultaneously, one of the modes is dispersively coupled (purple) to an ancilla, which may be a transmon. We operate the transmon in the $gf$ manifold, reserving the $\ket{e}$ state for error detection. \textbf{(b)} Applying the beamsplitter interaction causes the bosonic operators to evolve in time in the Heisenberg picture. The new modes are linear combinations of the original mode operators $\hat{a}$ and $\hat{b}$ and this time evolution can be represented geometrically by the trajectories shown in blue. Trajectories orbit the precession vector $\vec{n}$, which is fully determined by Hamiltonian parameters: $g$ and $\Delta$ set the polar angle, $\theta$ whereas $\varphi$ sets the azimuthal angle.  \textbf{(c)} When an ancilla is dispersively coupled to one of the bosonic modes, the precession vector $\vec{n}$ becomes dependent on the ancilla state, yielding distinct vectors $\vec{n}_g$ and $\vec{n}_f$. Furthermore, if the ancilla is in a superposition of $\ket{g}$ and $\ket{f}$, the bosonic modes will be in a superposition of the two trajectories shown in the figure. The $\hat{b}(t)$ trajectories are antipodal to the $\hat{a}(t)$ trajectories and are not shown on this Bloch sphere.
}

\label{fig:bosonic_bloch_sphere}
\end{figure*}

All proposed entangling gates are based on a simple Hamiltonian, from which we can design an entire family of logical entangling gates for a variety of bosonic encodings. This Hamiltonian combines a beamsplitter interaction between bosonic modes with a dispersive interaction between an ancilla and one of the modes as shown in  Fig. \ref{fig:bosonic_bloch_sphere}a. It is written as:
\begin{align}
    \widehat{\mathcal{H}}_{\chi \text{BS}} &= \widehat{\mathcal{H}}_{\chi}+\widehat{\mathcal{H}}_{\text{BS}},
\end{align}
where
\begin{align}
    \widehat{\mathcal{H}}_{\chi}/\hbar&= -\frac{\chi}{2}\hat{\sigma}_z\hat{a}^\dagger\hat{a}, \\
    \widehat{\mathcal{H}}_{\text{BS}}/\hbar &=\frac{g(t)}{2}\left(e^{i\varphi(t)}\hat{a}^\dagger\hat{b} + e^{-i\varphi(t)}\hat{a}\hat{b}^\dagger\right) + \Delta(t) \hat{a}^\dagger\hat{a},
\label{eq:detuned_bs}
\end{align}
and $\hat{\sigma}_z \equiv \ket{g}\bra{g} - \ket{f}\bra{f}$ is the Pauli $Z$ operator in the two-level subspace defined by the $\ket{g}$ and $\ket{f}$ levels of the ancilla. A three-level ancilla is used solely because it allows us to reserve the $\ket{e}$ level for detecting a single ancilla decay event.  The annihilation operators $\hat{a}$ and $\hat{b}$ act on the two bosonic modes, $g$ is the strength of the beamsplitter interaction, $\Delta$ is an effective detuning between two modes and $\chi$ is the strength of the dispersive interaction between the ancilla (in the $gf$-manifold) and mode $\hat{a}$. 
We have written this Hamiltonian in a frame where the dispersive interaction is symmetric, shifting the frequency of $\hat{a}$ by $\pm\chi/2$ dependent on the ancilla state.

Such a Hamiltonian $\widehat{\mathcal{H}}_{\chi}$ is routinely engineered in cQED systems \cite{wallraff_strong_2004, kirchmair_observation_2013, blais_circuit_2021}, where superconducting cavities serve as bosonic storage modes and a transmon is used as an ancilla. 
In a superconducting circuit, the beamsplitter interaction between two cavities can be generated by a driven non-linear coupling element, such as a transmon \cite{zhang_engineering_2019, gao_entanglement_2019}, SNAIL \cite{StijnBS2022, zhou_modular_2022} or SQUID \cite{YaoBS2022, sirois_coherent-state_2015}.
Since this parametric interaction may be actuated by one or more microwave drive signals, many of the parameters in the Hamiltonian such as the coupling strength $g$, its phase $\varphi$, and detuning $\Delta$ can all be rapidly and easily varied via standard microwave techniques.
Thus we can exploit the full time-dependent control of $g$, $\varphi$, and $\Delta$ to engineer new operations, even though the strength of the dispersive interaction $\chi$ is usually fixed. We assume that actuating the beamsplitter Hamiltonian does not introduce any new significant sources of error to our system~\cite{StijnBS2022,YaoBS2022}.

Although an ancilla couples to only one of the bosonic modes, in the presence of a beamsplitter interaction both modes interact with the ancilla, thereby enabling various non-trivial two-mode operations.
The size of the joint Hilbert space, however, can make the dynamics of this Hamiltonian difficult to interpret. 
In general, this Hamiltonian can drive logical states out of the codespace.  A simple example is the Hong-Ou-Mandel effect, which can take two qubits encoded in the ``Fock 0-1'' encoding ($\ket{0_L}=\ket{0},\ket{1_L}=\ket{1}$) to superpositions of $\ket{0}$ and $
\ket{2}$ when starting in the state $\ket{1_L1_L}$. 
This is different from typical implementations for two-level qubits \cite{rigetti_fully_2010, dicarlo_demonstration_2009, chen_qubit_2014} where logical states remain in the qubit subspace throughout the gate. 

We show that this complexity can be handled by mapping the dynamics to an ``operator Bloch sphere'' in Sec.~\ref{section:OBS}. 
In Sec.~\ref{section:AC logical unitaries} we identify useful unitaries generated by $\widehat{\mathcal{H}}_{\chi \text{BS}}$, namely the ancilla-controlled unitaries $\textbf{cZZ}_L$ and $\textbf{cSWAP}$. 
In Sec.~\ref{sec:family_of_gates}, we interleave the ancilla-controlled unitaries with ancilla rotations to turn these into parametrized $ZZ_L(\theta)$ and $\text{eSWAP}(\theta)$ logical gates. 
Combining these with existing $Z_L(\theta)$ gates allows us to realize any excitation-preserving two-qubit logical gate (see Appendix~\ref{app:family of logical gates}). 
Then, in Sec.~\ref{sec:EDgates} we describe how these gate constructions also allow us to detect errors in the ancilla and preserve the error-detection properties of the bosonic encodings.
Finally, in Sec.~\ref{sec:simulations} we show in simulations that we can achieve error-detected two-qubit gate fidelities exceeding $\sim99.99\%$ with reasonable hardware coherence times and gate success probabilities exceeding $\sim99\%$.

\section{Operator Bloch sphere framework for beamsplitter interactions}
\label{section:OBS}
We introduce the ``operator Bloch sphere''  to visualize the dynamics generated by $\widehat{\mathcal{H}}_{\chi \text{BS}}$ and port existing intuition for single-qubit control on the Bloch sphere to the design of two-qubit gates for bosonic qubits.
Working in the Heisenberg picture allows us to avoid tracking the evolution of each full two-mode state.
To begin, we consider evolution under $\widehat{\mathcal{H}}_{\text{BS}}$, then generalize to the full $\widehat{\mathcal{H}}_{\chi \text{BS}}$ Hamiltonian.

Inspired by Schwinger's angular momentum formalism of bosonic operators \cite{schwinger_angular_1952}, we rewrite $\widehat{\mathcal{H}}_{\text{BS}}$ with the angular momentum operators $\widehat{L}_I = \frac{1}{2} (\hat{a}^{\dag}\hat{a} + \hat{b}^{\dag}\hat{b})$, $\widehat{L}_X = \frac{1}{2} (\hat{a}^{\dag}\hat{b} + \hat{a}\hat{b}^{\dag})$, $\widehat{L}_Y = \frac{1}{2i} (\hat{a}^{\dag}\hat{b} - \hat{a}\hat{b}^{\dag})$ and $\widehat{L}_Z = \frac{1}{2} (\hat{a}^{\dag}\hat{a} - \hat{b}^{\dag}\hat{b})$, which allows us to rewrite $\widehat{\mathcal{H}}_{\text{BS}}$ as
\begin{eqnarray}
\widehat{\mathcal{H}}_{\text{BS}}(g,\varphi, \Delta)/\hbar &=& g\cos{\varphi} \widehat{L}_X - g\sin{\varphi} \widehat{L}_Y + \Delta \widehat{L}_Z + \Delta \widehat{L}_I. \nonumber
\label{eq:detuned_bs_angular_momentum}
\end{eqnarray}
For the case where the parameters $g$, $\varphi$, $\Delta$ are constant, the Heisenberg representation of the mode operators can be obtained by transforming the mode operators via the unitary operator $\widehat{U}=\exp{(-i\widehat{\mathcal{H}}_{\text{BS}}t/\hbar)}$,

\begin{eqnarray}
\begin{pmatrix}
\hat{a}(t) \\
\hat{b}(t) \\
\end{pmatrix}
=
\begin{pmatrix}
\widehat{U}\hat{a}\widehat{U}^{\dag} \\
\widehat{U}\hat{b}\widehat{U}^{\dag} \\
\end{pmatrix}
&=& e^{-i\frac{\Delta}{2} t}
R_{\vec{n}}(\Omega t) 
\begin{pmatrix} \hat{a} \\ \hat{b} \\ \end{pmatrix}, \label{eq:2}
\label{bs_abcd}
\end{eqnarray}
where $R_{\vec{n}}(\Omega t)=(\cos{\frac{\Omega t}{2}} I - i\sin{\frac{\Omega t}{2}} \vec{n}\cdot\vec{\sigma})$ is a matrix in SU(2), which can be interpreted as a rotation around a precession vector $\vec{n}=[\sin{\theta}\cos{\varphi}, -\sin{\theta}\sin{\varphi}, \cos{\theta}]$ at rate $\Omega=\sqrt{g^2 + \Delta^2}$. The polar angle of the precession vector is determined by the ratio of the coupling strength $g$ and the detuning $\Delta$ such that $\cos{\theta}=\Delta/\sqrt{g^2+\Delta^2}$ and $\sin{\theta}=g/\sqrt{g^2+\Delta^2}$.
Analogous to state evolution on a qubit Bloch sphere, we plot the mode transformations at each point in time to form trajectories on the operator Bloch sphere as shown in Fig.~\ref{fig:bosonic_bloch_sphere}b.
Here, the north pole represents the initial mode operator $\hat{a}$ and the solid arrow represents the trajectory of the transformed mode operator $\hat{a}(t)$. Similarly, the south pole represents the initial mode operator $\hat{b}$ and the dashed arrow represents the trajectory of the transformed mode operator $\hat{b}(t)$.
The trajectory can be fully controlled by modulating the complex amplitude of the beamsplitter interaction, which is routinely done in the cQED systems \cite{gao_entanglement_2019, StijnBS2022, YaoBS2022}.
The trajectories from the north and south pole are antipodal to one another and therefore we will only show the transformation of $\hat{a}$ going forward.
The end points of the trajectories indicate the final mode transformations of the original $\hat{a},\hat{b}$ operators.

The effect of the ancilla's interaction, $\widehat{\mathcal{H}}_{\chi}$, appears
as an ancilla-state-dependent detuning $\Delta=\Delta^\prime\pm\frac{\chi}{2}$ where $\Delta^\prime$ now represents the detuning of the beamsplitter drives from resonance. We can now write the dispersive beamsplitter Hamiltonian as
\begin{align}
\widehat{\mathcal{H}}_{\chi\text{BS}}&=\widehat{\mathcal{H}}_{\text{BS}}(g,\varphi, \Delta'-\frac{\chi}{2})\otimes\ket{g}\bra{g}\\
&+ \widehat{\mathcal{H}}_{\text{BS}}(g,\varphi, \Delta'+\frac{\chi}{2})\otimes\ket{f}\bra{f}.\nonumber
\end{align}
Since the total detuning of the beamsplitter becomes dependent on the ancilla state, there now exist two different ``conditional'' precession vectors with different $z$-components, allowing one to construct ancilla-controlled mode trajectories. This is illustrated in Fig.~\ref{fig:bosonic_bloch_sphere}c.

\begin{figure}[h]
\centering
    \begin{tabular}{c c}
    \includegraphics[width=1.0\linewidth]{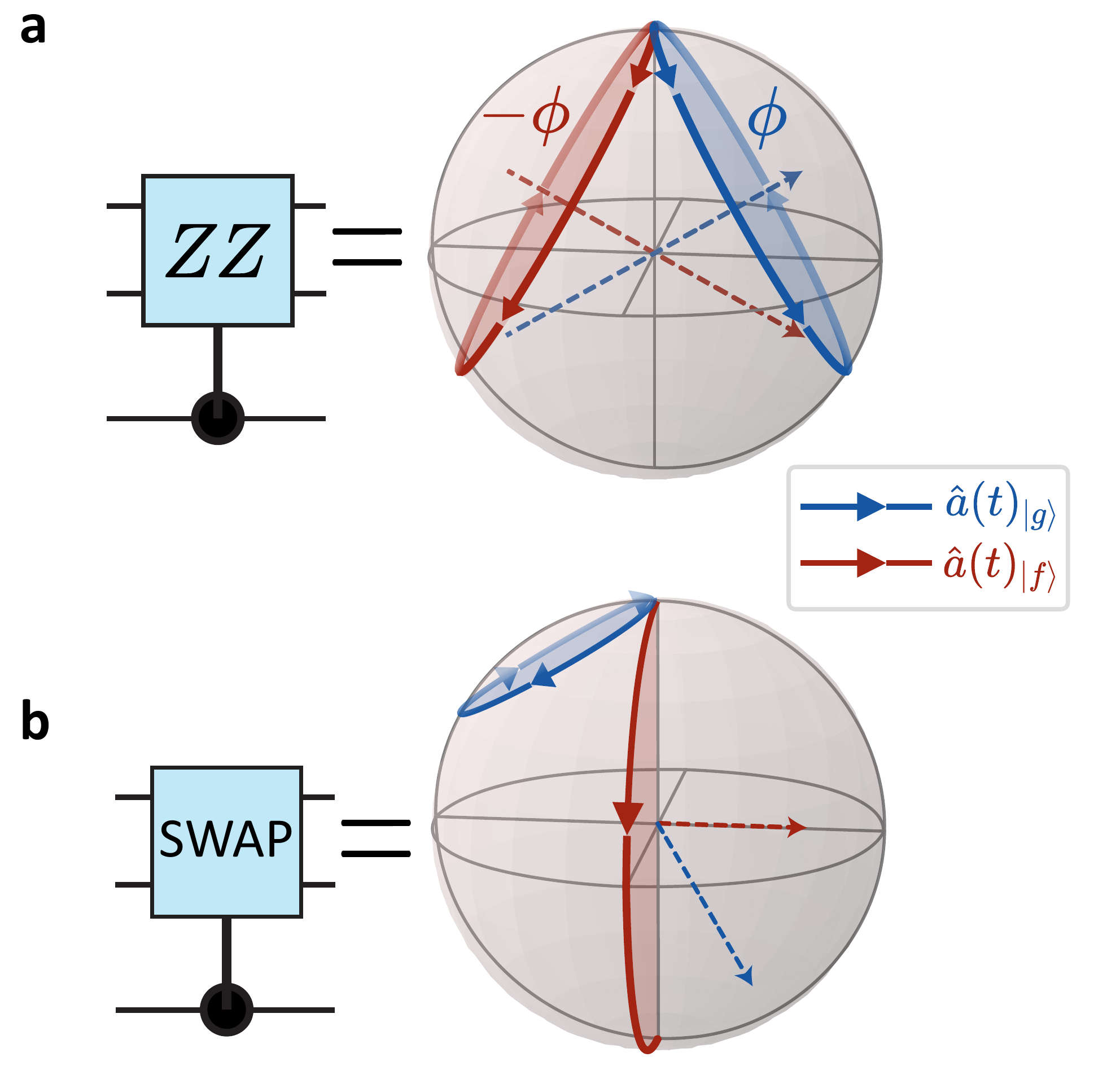}
    \end{tabular}

\caption{
\textbf{cZZ and cSWAP on the operator Bloch sphere realized with the Hamiltonian parameters and durations listed in Table~\ref{tab:pump_conditions}}. 
\textbf{(a)} Trajectories for $\textbf{cZZ}_L$. 
After an amount of time $T$, the trajectories complete a full orbit at the end of the gate and enclose area $\pm\phi$. By choosing Hamiltonian parameters such that $\phi=\pi/2$ , these ancilla-dependent mode transformations are equivalent to performing a \text{cZZ} unitary for the Fock 01 and dual-rail codes. When $\phi=\pi/4$ this is a \text{cZZ} unitary for the binomial and 4-legged cat code (see Appendix \ref{app:family of primitive gates}). \textbf{(b)} Trajectories for \textbf{cSWAP}. The $\hat{a}(t)_{\ket{g}}$ trajectory returns to the north pole, indicating that we perform identity on the two bosonic modes up to a geometric phase. The $\hat{a}(t)_{\ket{f}}$ trajectory ends at the south Pole, which corresponds to a complete swapping of the two modes up to a geometric phase. 
}
\label{fig:JP and CSWAP}
\end{figure}

All possible dynamics generated by the detuned beamsplitter Hamiltonian can be represented on the operator Bloch sphere.
The three degrees of freedom in the dispersive beamsplitter Hamiltonian $g$, $\varphi$, and $\Delta$ determine the axis and rate of precession. This holds true even when these parameters have time dependence, which leads to time-varying precession axes and precession rates.
We emphasize that the operator Bloch sphere picture is necessary to visualize the time dynamics generated by a continuous beamsplitter interaction, over which we have fine control of the Hamiltonian parameters. This differs from the more discrete beamsplitter treatments used in linear optics \cite{reck_experimental_1994}.

\section{Ancilla-controlled logical unitaries}
\label{section:AC logical unitaries}
The operator Bloch sphere picture is a powerful tool for finding new and interesting ancilla-controlled unitaries generated by $\widehat{\mathcal{H}}_{\chi\text{BS}}$. 
Specifically, we show a way to realize both $\textbf{cZZ}_L$ and $\textbf{cSWAP}$ in our simple proposed hardware layout.

We define ancilla-controlled unitaries as unitaries where we perform identity on the bosonic modes if the ancilla is in $\ket{g}$, and a (non-entangling) two-qubit gate if the ancilla is in $\ket{f}$. 
At the end of the unitary, the bosonic states must return to the logical codespace. 
This constraint restricts us to trajectories which start and end at the poles of the operator Bloch sphere, corresponding to either SWAP or identity operations. 
However, there is a crucial feature: the solid angle enclosed by these trajectories determines the geometric phase imparted to the bosonic modes, and is used as a resource to enact logical operations (see Appendix \ref{app:interpretation_phase}). 
We later use this effect to engineer the ancilla-controlled $ZZ_L$, $\textbf{cZZ}_L$, and ancilla-controlled SWAP, $\textbf{cSWAP}$, unitaries as shown in Fig.~\ref{fig:JP and CSWAP}. Moreover, by combining these unitaries with arbitrary ancilla rotations, we can construct a family of excitation-preserving gates such as $ZZ_L(\theta)$, $\text{iSWAP}(\theta)$ and $\text{fSim}(\theta_1, \theta_2)$ on the logical subspace (see Appendix~\ref{app:family of logical gates}).

Designing trajectories that enclose a specific geometric phase can be used to build useful unitaries. The geometric phase is set by the term $e^{-i\frac{\Delta}{2} t}$ in  Eq.~\ref{bs_abcd}. Completely enclosing a solid angle $\phi$ corresponds to performing the unitary $\hat{R}_\phi = e^{i\phi(\hat{a}^\dagger \hat{a} + \hat{b}^\dagger \hat{b})}$ on the bosonic modes. For many bosonic encodings, $e^{i\frac{\pi}{n} \hat{a}^\dagger \hat{a}}=Z_L$ for $n \in \mathbb{Z}$ and hence $ZZ_L = e^{i\frac{\pi}{n}(\hat{a}^\dagger \hat{a} + \hat{b}^\dagger \hat{b})}$. Therefore, by varying the relative strengths of the microwave-controlled Hamiltonian parameters, we can choose the enclosed geometric phase to match the $ZZ_L$ operator for a particular bosonic code. 

\begin{table}[t]
    {\renewcommand{\arraystretch}{2}
    \begin{tabular}{C{1.8cm}C{1.8cm}C{1.8cm}C{1.8cm}}
        \toprule
        Hamiltonian parameters & $\textbf{cZZ}_L$ for Fock 01 or dual-rail & $\textbf{cZZ}_L$ for binomial or 4-cat & $\textbf{cSWAP}$\\
        \hline
        $g$ & $\frac{\sqrt{3}}{2}\chi$ & $\frac{\sqrt{15}}{2}\chi$ & $\frac{\chi}{\sqrt{3}}$\\
        $\Delta$ & 0 & 0 & $\frac{\chi}{2}$\\
        $T$ & $\frac{2\pi}{\chi}$ & $\frac{\pi}{\chi}$ & $\frac{\sqrt{3}\pi}{\chi}$\\
        \toprule
    \end{tabular}
    }
    \caption{\textbf{Pump conditions for primitive operations.} The beamsplitter rate $g$, pump detuning $\Delta$ and pulse duration required to perform key primitive operations (up to ancilla-state dependent local rotations).}
    \label{tab:pump_conditions}
\end{table}

Trajectories that depend on the two ancilla states generate three types of ancilla-controlled unitaries that return to the codespace. (a) Both trajectories may return to the starting pole, (b) one trajectory returns to the starting pole whilst the other returns to the opposite pole, (c) both trajectories may return to the opposite pole. Although these trajectories are a small subset of all the possible trajectories we could engineer, 
each case represents a different, useful ancilla-controlled logical operation.

First, we consider evolution where the two trajectories conditioned on the ancilla state return to their starting poles (Fig.~\ref{fig:JP and CSWAP}a). 
The geometric phase accumulation means we perform the ancilla-controlled unitary 
\begin{equation}
e^{+i\phi(\hat{a}^\dagger \hat{a} + \hat{b}^\dagger \hat{b})}\otimes\ket{g}\bra{g}+e^{-i\phi(\hat{a}^\dagger \hat{a} + \hat{b}^\dagger \hat{b})}\otimes\ket{f}\bra{f}.
\label{eq:control-joint-parity}
\end{equation}
We can use this geometric phase accumulation to perform logical operations on the bosonic modes. For many bosonic encodings, $Z_L$ takes the form $e^{i\frac{\pi}{n} a^\dagger a}$ for a code with $n$-fold rotational symmetry \cite{grimsmo_quantum_2020} and hence $ZZ_L = e^{i\frac{\pi}{n}(\hat{a}^\dagger \hat{a} + \hat{b}^\dagger \hat{b})}$.

When $\phi = \frac{\pi}{2n}$ or $\phi = \pi(1-\frac{1}{2n})$ this is equivalent to the $\textbf{cZZ}_L$ unitary 
\begin{equation}
\widehat{\mathds{1}}\otimes\ket{g}\bra{g}+ ZZ_L\otimes\ket{f}\bra{f},
\end{equation}
up to the rotation operator $e^{-i\frac{\pi}{2n}(\hat{a}^\dagger \hat{a} + \hat{b}^\dagger \hat{b})}$ which is easily tracked in software.
The required Hamiltonian parameters are found from the general formula for the solid angle, $\phi$. For orbits about a fixed precession vector, this is given by $\phi=4\pi(1-\cos(\theta))=4\pi(1-\frac{\chi}{2\Omega})$. The parameters for the $\textbf{cZZ}_L$ gate are shown in Table~\ref{tab:pump_conditions} for bosonic codes where $Z_L = e^{i\pi\hat{a}^\dagger\hat{a}}$ or $Z_L = e^{i\frac{\pi}{2}\hat{a}^\dagger\hat{a}}$. Note that since the interaction strengths $g/2\pi$ and $\chi/2\pi$ may both be several MHz \cite{StijnBS2022, YaoBS2022}, all of these gates on multiphoton encoded qubits may be performed in times of $0.1-1\ \mu$s, 3 to 4 orders of magnitude faster than typical microwave cavity decay rates \cite{reagor_quantum_2016}.

With another set of Hamiltonian parameters, we can create the $\textbf{cSWAP}$ operation (Fig.~\ref{fig:JP and CSWAP}b) defined as
\begin{equation}
\widehat{\mathds{1}}\otimes\ket{g}\bra{g} + {\rm SWAP}\otimes\ket{f}\bra{f}.
\end{equation}
In this case, we need the trajectory conditioned on $\ket{f}$ to end at the opposite pole whilst the trajectory conditioned on $\ket{g}$ completes an orbit about a different precession vector to return to the initial pole. For the parameters presented in Table~\ref{tab:pump_conditions}, this implements the unitary
\begin{equation}
e^{i\pi\left(1-\frac{\sqrt{3}}{2}\right)(\hat{a}^\dagger \hat{a} + \hat{b}^\dagger \hat{b})}\otimes\ket{g}\bra{g} + e^{-i\frac{\pi}{2}(\hat{a}^\dagger \hat{a} + \hat{b}^\dagger \hat{b})} {\rm SWAP}\otimes\ket{f}\bra{f}.
\end{equation}
By adding appropriate delays in the gate sequence (see Appendix~\ref{app:family of primitive gates}), one can null unwanted geometric phase accumulations to realize the $\textbf{cSWAP}$ unitary.
This operation was experimentally realized in \cite{StijnBS2022}.

Finally, when both trajectories end at the opposite pole, we perform a SWAP between the bosonic modes that is independent of the ancilla state (up to geometric phase accumulation), which we call an ``unconditional SWAP''. Without using our framework, this operation is hard to realize when the ancilla is in a superposition of states, due to the static nature of the dispersive interaction. Unconditional SWAPs allow us to extend our ancilla-controlled unitaries that act on more than two bosonic modes (see Appendix~\ref{app:family of primitive gates}) .

\section{A family of logical two-qubit gates}
\label{sec:family_of_gates}
 Importantly, with just the two primitives $\textbf{cZZ}_L$ and \textbf{cSWAP}, we can use arbitrary rotations on the ancilla to now perform a continuous family of entangling gates on the bosonic logical subspace.\\
 
\begin{figure}[t]
\centering
    \begin{tabular}{c c}
    \includegraphics[width=1\linewidth]{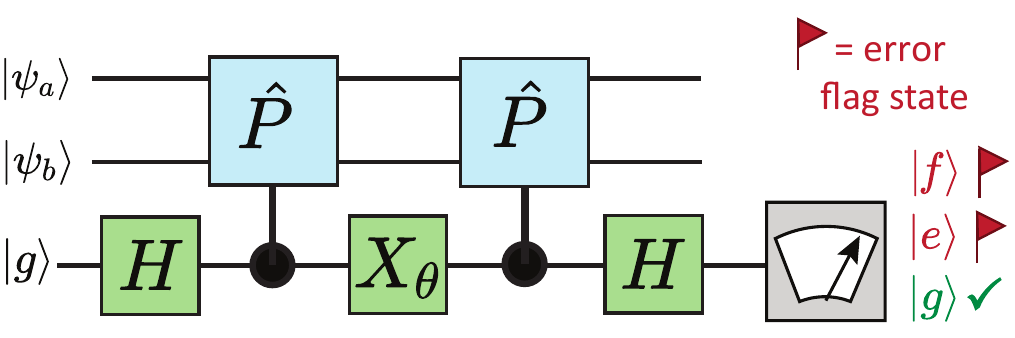}
    \end{tabular}
\caption{\textbf{Error-detected circuits for bosonic entangling gates.} We interleave ancilla-controlled unitaries (in blue) with ancilla rotations in the $g-f$ manifold (in green) to construct entangling gates. Provided $\widehat{P}^2=\hat{\mathds{1}}$, we implement the general gate $\widehat{P}(\theta)= \exp{(-i\frac{\theta}{2} \hat{P})}$.
By using the $\textbf{cZZ}_L$ operation introduced earlier, we execute this circuit with $\widehat{P}=ZZ_L$ to perform the entangling gate $ZZ_L(\theta)$ on our bosonic qubits. Similarly, with \textbf{cSWAP} we can perform an eSWAP($\theta$) entangling gate. These gates are easily parameterized via the middle $X_\theta$ ancilla rotation. Crucially, we can detect ancilla errors during the gate by measuring the state of the ancilla at the end of the circuit. Measuring the ancilla in $\ket{e}$ or $\ket{f}$ flags that an error has occurred. We accept the gate when we measure ancilla in $\ket{g}$.
}
\label{fig:exponential_gadget}
\end{figure}

Inspired by gate teleportation techniques \cite{shor_fault-tolerant_1997, gottesman_demonstrating_1999, zhou_methodology_2000, bravyi_universal_2005}, ancilla-controlled unitary gates on the logical subspace can be ‘exponentiated’ by a construction shown in Fig.~\ref{fig:exponential_gadget} \cite{nielsen_chuang_2010} to create gates that only act on the bosonic modes with the ancilla starting and ending in $\ket{g}$. A key advantage of this approach is that by checking that the ancilla returns to $\ket{g}$, we can detect whether ancilla errors have occurred during the gate.

The exponentiation circuit we present is very general and allows one to realize the unitary
\begin{eqnarray}
\widehat{P}(\theta) = \exp{\left(-i\frac{\theta}{2} \widehat{P}\right)} = \cos{\frac{\theta}{2}} \widehat{I} - i \sin{\frac{\theta}{2}} \widehat{P},
\end{eqnarray}
from the ancilla controlled unitary \textbf{c}$\mathbf{\widehat{P}}$ where $\widehat{P}$ is any ``Pauli-like'' operator acting on the two-qubit logical subspace that satisfies $\widehat{P}^2=\widehat{\mathds{1}}$ (in other words $\widehat{P}$ is Hermitian and unitary). The full unitary implemented on the qubit-ancilla system by this circuit is 
\begin{equation}
    \widehat{P}(\theta)\otimes\ket{g}\bra{g}+\widehat{P}(-\theta)\otimes\ket{f}\bra{f}.   
\end{equation}
The circuit construction in Fig.~\ref{fig:exponential_gadget} inherently detects a single ancilla dephasing error (see Appendix~\ref{app:error detected gate construction}) but not ancilla decay events, which necessitates the use of a three-level ancilla.

Setting $\widehat{P}=\mathrm{SWAP}$ with the $\textbf{cSWAP}$ unitary yields a new construction for the exponential-${\rm SWAP}$ ($\rm eSWAP$) gate \cite{gao_entanglement_2019, lau_universal_2016}. Similarly, with the $\textbf{cZZ}_L$ unitary we construct the $ZZ_L(\theta)$ gate which has not yet been realized for bosonic qubits. An advantage of these constructions is the ease of varying the angle $\theta$, which is controlled simply by varying the angle of the intermediate ancilla rotation in Fig.~\ref{fig:exponential_gadget}.  
Both these gates are maximally entangling for $\theta = \pi/2$, 
and $ZZ(\frac{\pi}{2})$ is equivalent to a CNOT gate up to single qubit gates.
By combining eSWAP($\theta$) and $ZZ_L(\theta)$ with single qubit $Z_L(\theta)$ gates, we can implement any excitation-preserving logical two-qubit gate on the two bosonic qubits (see Appendix~\ref{app:family of logical gates}).  

The $Z_L(\theta)$ gate can be implemented either by using the same construction with ancilla-controlled rotations of a single bosonic mode (which naturally arises from the dispersive interaction) or by using a fault-tolerant Selective Number-dependent Arbitary Phase (SNAP) gate \cite{heeres_cavity_2015, reinhold_error-corrected_2020}.

Another powerful application of the ancilla-controlled logical gates is to perform a QND logical measurement of the operator $\widehat{P}$. This is carried out by preparing the ancilla in $\ket{+}_{gf}\equiv (\ket{g} + \ket{f})/\sqrt{2}$, applying $\mathbf{c\widehat{P}}$ and then measuring the ancilla in the $\ket{\pm}_{gf}$ basis. For $\textbf{cSWAP}$ this amounts to a SWAP test~\cite{StijnBS2022}. Similarly $\textbf{cZZ}_L$ can be turned into a QND logical measurement of the $ZZ_L$ operator. This operation finds use in measurement-based alternatives to entangling gates \cite{briegel_measurement-based_2009} and can form a component of a Bell measurement. Unlike the gate construction, in principle these measurements can correct single ancilla decay errors and all orders of ancilla dephasing. This is explored further in Appendix \ref{app:error-corrected measurement}.

\section{Hardware efficient error-detected gates}
\label{sec:EDgates}
Perhaps the most exciting aspect of these gates is the ability to detect hardware errors at any time during (or before) the gates in both the ancilla and in the bosonic modes.
Crucially, a successful gate should always return the ancilla to $\ket{g}$. 
If we measure the ancilla to be in any state other than $\ket{g}$ at the end of the gate, we flag the gate as having experienced an error. Bosonic errors such as photon loss remain detectable after the gate (e.g, via photon number parity measurements). In the usual case where the ancilla has worse coherence than the bosonic modes, it is essential to prevent ancilla errors from propagating onto the bosonic modes. Otherwise, we lose the advantages of using a bosonic code in the first place. 

Here we describe the critical error detection properties of the protocol for ancilla decay, ancilla dephasing and photon loss in the bosonic modes. An in-depth discussion can be found in Appendix \ref{app:error_closure}. 
First, we analyze errors that occur during the ancilla-controlled unitaries, which have the longest duration in our gate sequences. The gate constructions are naturally robust to ancilla dephasing but not ancilla decay.
As previously mentioned, we circumvent this problem by operating the ancilla in the manifold spanned by $\ket{g}$ and $\ket{f}$, reserving the $\ket{e}$ level to detect if a single ancilla decay has occurred. 
This is roughly equivalent to realizing a simple noise-biased ancilla and is generalizable to ancillae in other platforms.
Alternatively, two-levels of a true noise-biased ancilla \cite{puri_stabilized_2019} also suffice for implementing these circuits, provided the ancilla rotations also preserve the noise bias.  

Ancilla dephasing can be described by the jump operator $\hat{\sigma}_z$. Since this operator commutes with  $\widehat{\mathcal{H}}_{\chi \text{BS}}$, the ancilla-controlled unitaries are error-transparent \cite{rosenblum_fault-tolerant_2018} to dephasing. In other words, dephasing during the unitary evolution is the same as performing the ancilla-controlled unitary correctly, and then applying the operator $\hat{\sigma}_z$ afterwards.

For the logical gate construction we use two ancilla-controlled unitaries, and whilst each one of these is error transparent to dephasing, the overall circuit is not due to the ancilla rotation by angle $\theta$ in the middle. Regardless of where in the circuit the dephasing jump occurs, we always observe $\ket{f}$ at the end. With no dephasing jumps we always readout $\ket{g}$. If the jump occurs during the last ancilla controlled-unitary, the correct $ZZ_L(\theta)$ gate is performed on the bosonic modes. However, if the jump occurs during the first control-unitary, we perform $ZZ_L(-\theta)$ instead. Since we cannot distinguish between the two cases, the correction unitary is unknown, and so we can only detect these errors rather than correct them.

We are also robust to dephasing and decay errors that happen during the much shorter ancilla rotations. One can show that the correct gate is still performed when we measure the ancilla to be in $\ket{g}$ (see Appendix~\ref{app:error_closure}). This means the error-detected gate fidelity could in principle exceed coherence-limited single qubit gate fidelities in transmon qubits.  

$\widehat{\mathcal{H}}_{\chi \text{BS}}$ preserves total photon number, and therefore photon loss during the gate remains detectable after the gate when using an appropriate bosonic code. Photon loss during the ancilla-controlled unitaries also dephases the ancilla and hence this error is only detectable, just as for ordinary ancilla dephasing. 

\begin{figure*}[t]
\centering
    \includegraphics[width=1.0\linewidth]{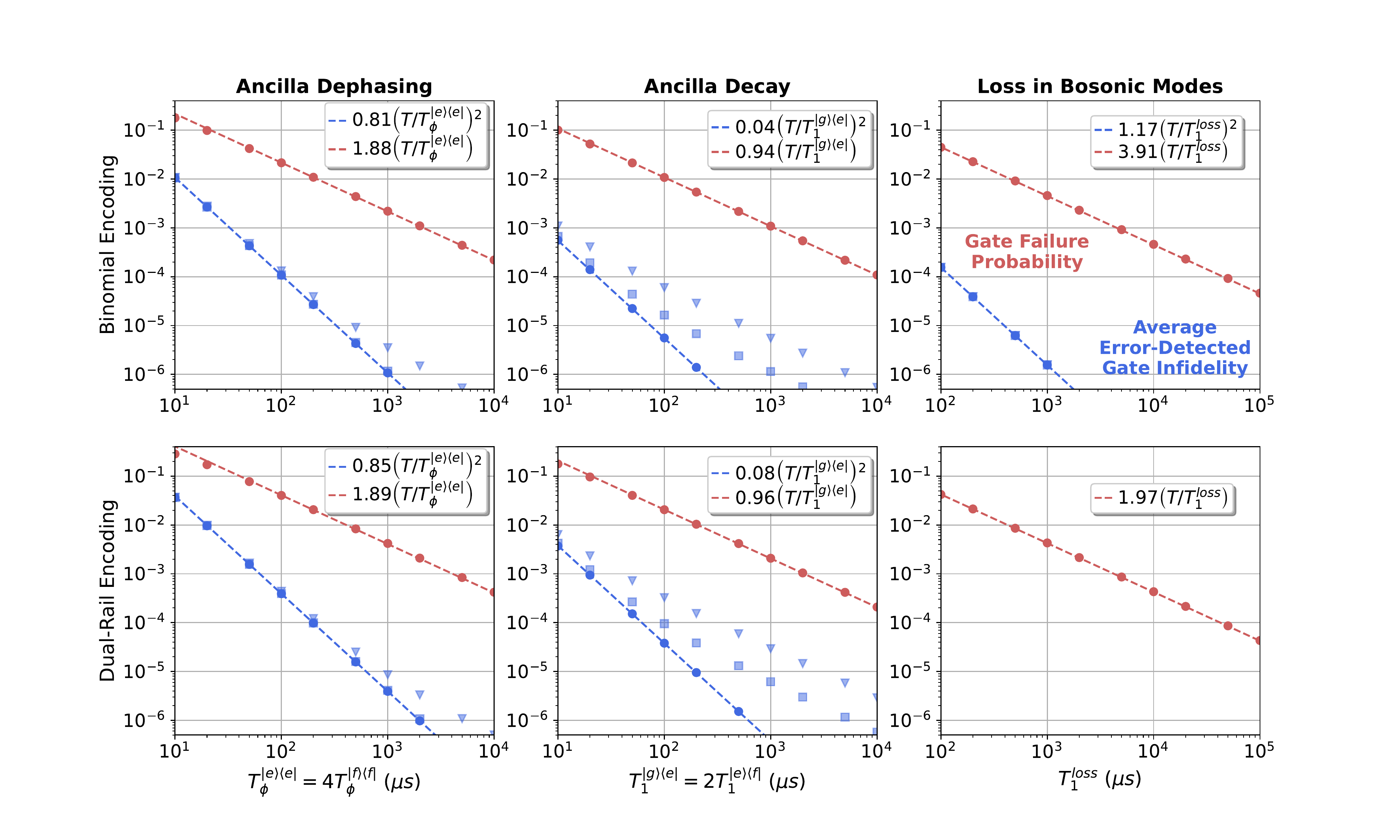}
    \label{fig_sims}
\caption{
\textbf{Quadratic scaling of $ZZ(\frac{\pi}{2})$ gate infidelity with hardware coherence times, reaching error-detected infidelities below $\mathbf{10^{-4}}$.}
Numerical simulation of error-detected $ZZ(\frac{\pi}{2})$ gate on two logical qubits in both the binomial encoding (top row) and the dual-rail encoding (bottom row) under three separate error channels---ancilla decay, ancilla dephasing, and photon loss in the bosonic modes. ``Gate failure probability'' (red circles) refers to the fraction of gate attempts where we measure the ancilla to be in $\ket{e}$ or $\ket{f}$ at the end (left and middle columns), or when we detect photon loss in the bosonic modes via idealized syndrome measurements (right column). ``Average error-detected gate infidelity''(blue circles) is the state transfer fidelity of the evolution of the 36 cardinal two-qubit logical states (which can be shown to be equivalent to the average fidelity over the joint two-qubit space \cite{nielsen_simple_2002}) after postselecting on $\ket{g}$, assuming no measurement error. We additionally calculate this infidelity in the presence of 1\% (blue squares) and 5\%(blue triangle) ancilla readout errors (see Appendix \ref{app:simulation}). Gate failure results from a single jump error (a first-order error) happening during the gate, and therefore this probability scales linearly with coherence time. Since we can detect these errors, the error-detected gate infidelity is only limited by double jump errors (second-order errors) and hence scales quadratically, allowing us to reach extremely low error-detected gate infidelities with typical coherence times. Numerical fits to illustrate these scalings are shown by the blue and red dashed lines (see Appendix \ref{app:sim_scalings}).}
\label{fig:simulation}
\end{figure*}
\section{Numerical simulations of $\mathbf{ZZ_L(\pi/2)}$ gates}
\label{sec:simulations}
We solve the Lindblad master equation in QuTiP \cite{johansson_qutip_2013} to calculate several performance metrics for the $ZZ(\theta = \pi/2)$ gate in two bosonic codes: the lowest order binomial code \cite{michael_new_2016} and the dual-rail encoding \cite{chuang_simple_1995}. 
During the simulation, the system experiences either ancilla decay, ancilla dephasing, or photon loss in the bosonic modes.
From these simulations, we obtain the gate success probability and the error-detected gate infidelity, $\bar{\varepsilon}_{\text{ED}}$
To highlight how these metrics scale with hardware errors, we enable one error channel at a time as shown in Fig.~\ref{fig:simulation}. 
Our proposed gates have the ability to detect all first-order jump errors associated with each error channel, and therefore only second-order jump errors limit the error-detected gate infidelity.

The numerical simulations verify this property, as shown by the quadratic scaling of $\bar{\varepsilon}_{\text{ED}}$ with coherence time. 
Values of $\bar{\varepsilon}_{\text{ED}}$ well below $10^{-4}$ are achieved with feasible coherence times, which would outperform any previously-demonstrated two-qubit gates. The price one pays is that the gate now has a small failure probability of $\sim 1\%$, which scales linearly with hardware coherence rates.

For these simulations we choose $\chi_{gf}/2\pi=-1\text{ MHz}$, and therefore the required beamsplitter rates needed for the $ZZ_L(\pi/2)$ gates are approximately $1\text{ MHz}$. 
Recently, hardware with required system parameters has been implemented \cite{gao_programmable_2018, gao_entanglement_2019} and high-fidelity beamsplitters with rates of up to $5\text{ MHz}$ have been recently demonstrated \cite{YaoBS2022}. 

We simulate the $ZZ_L(\pi/2)$ gate using the circuit in Fig.~\ref{fig:exponential_gadget}, with Hamiltonian parameters listed for $\textbf{cZZ}_L$ in Table~\ref{tab:pump_conditions} for the binomial and dual-rail encodings. We choose constant pulse profiles for the beamsplitter amplitude $g$ and for the ancilla rotation pulses, allowing us to focus on the incoherent error and neglect any coherent error associated with imperfect pulse shaping. This also highlights the gates' robustness to errors during the ancilla rotations themselves (duration $50 \text{ ns}$) meaning $\bar{\varepsilon}_{\text{ED}}$ can surpass typical transmon $\pi$-pulse fidelities (see Appendix \ref{app:error detected gate construction}).

More realistic simulations can include higher-order nonlinearities in the Hamiltonian, which are sources of coherent errors in the gates. Nonetheless, these errors are expected to scale quadratically with gate duration. We consider the effects of cavity self-Kerr, cross-Kerr between the cavities and the $\chi’$ correction to the dispersive coupling in Appendix~\ref{app:simulation}, and show that they introduce infidelity at the $\sim10^{-4}$ level for realistic parameters. 

\section{Conclusion}
We have introduced error-detectable two-qubit entangling gates for a wide range of bosonic qubits, including the dual-rail, binomial, and 4-legged cat encodings. Our gates are based on a tunable beamsplitter interaction between two bosonic modes and an ancilla dispersively coupled to just one of the modes. The time evolution during the gates can be visualized in a new geometric framework which we call the ``operator Bloch Sphere''. In this picture, we can easily derive the Hamiltonian parameters needed to perform ancilla-controlled unitaries $\textbf{cZZ}_L$ and $\textbf{cSWAP}$. With these building blocks, we have shown how to construct new, fast gate implementations for the $ZZ_L(\theta)$ and eSWAP$(\theta)$ gate that could be readily realized on current hardware. By using a three-level transmon as our ancilla, we are able to flag the dominant hardware errors, transmon decay and dephasing, and prevent them from significantly impacting the gate fidelities. 
Because of the nature of the first-order error protection, we expect quadratic improvement of the postselected gate infidelity as we improve the hardware lifetime and the gate speed.
With today’s cQED hardware coherences, we expect to reach error-detected gate fidelities below $10^{-4}$ with $99\%$ gate success probability. 
We have verified the quadratic scaling of these fidelities with hardware coherence times in simulation. This implies that an increase in coherence time by a factor of $10$ yields a gate infidelity that decreases by a factor of $100$. 

Error-detected gates are immediately useful for near-term short-depth circuits and hence NISQ-era applications \cite{preskill_quantum_2018}, where one can postselect for error-free circuit runs. Furthermore, error-detected gates at the hardware level can form the bedrock of a fully error-corrected computation. Knowing exactly which gates had errors allows one to erase and reset the affected qubits. Stabilizer codes, including the surface code, are highly resilient to erasure errors \cite{wu_erasure_2022, kang_quantum_2022}.
The constructions we present are also transferrable to other qubit platforms, such as phonon modes in trapped ions where one has access to conditional beamsplitter interactions and where hyperfine states play the role of the ancilla \cite{hou_coherently_2022}.  
By co-designing the logical encoding and the physical gate construction, one can be insensitive to first-order errors and boost the two-qubit gate fidelity.
We believe such hardware-level error detection offers a new direction in improving gates other than brute-force hardware improvement and broadens design considerations in developing novel quantum gates.
\\

\section{Acknowledgement}
We thank Shruti Puri, Patrick Winkel, Kevin C. Smith, Daniel K. Weiss, Sophia Xue, Harshvardhan K. Babla for helpful discussions.
We also appreciate Luigi Frunzio, Jahan Claes and Alec Eickbusch for critical feedback on the manuscript. 
This research was supported by the U.S. Army Research Office (ARO) under grant W911NF-18-1-0212, and by the U.S. Department of Energy, Office of Science, National Quantum Information Science Research Centers, Co-design Center for Quantum Advantage (C2QA) under contract number DE-SC0012704. The views and conclusions contained in this document are those of the authors and should not be interpreted as representing official policies, either expressed or implied, of the ARO or the U.S. Government. The U.S. Government. is authorized to reproduce and distribute reprints for Government purpose notwithstanding any copyright notation herein. 
We thank the Yale Center for Research Computing for technical support and high performance computing resources.
R.J.S. is a founder and shareholder of Quantum Circuits, Inc.

\bibliography{refs2add.bib}

\begin{thebibliography}{64}%
\makeatletter
\providecommand \@ifxundefined [1]{%
 \@ifx{#1\undefined}
}%
\providecommand \@ifnum [1]{%
 \ifnum #1\expandafter \@firstoftwo
 \else \expandafter \@secondoftwo
 \fi
}%
\providecommand \@ifx [1]{%
 \ifx #1\expandafter \@firstoftwo
 \else \expandafter \@secondoftwo
 \fi
}%
\providecommand \natexlab [1]{#1}%
\providecommand \enquote  [1]{``#1''}%
\providecommand \bibnamefont  [1]{#1}%
\providecommand \bibfnamefont [1]{#1}%
\providecommand \citenamefont [1]{#1}%
\providecommand \href@noop [0]{\@secondoftwo}%
\providecommand \href [0]{\begingroup \@sanitize@url \@href}%
\providecommand \@href[1]{\@@startlink{#1}\@@href}%
\providecommand \@@href[1]{\endgroup#1\@@endlink}%
\providecommand \@sanitize@url [0]{\catcode `\\12\catcode `\$12\catcode
  `\&12\catcode `\#12\catcode `\^12\catcode `\_12\catcode `\%12\relax}%
\providecommand \@@startlink[1]{}%
\providecommand \@@endlink[0]{}%
\providecommand \url  [0]{\begingroup\@sanitize@url \@url }%
\providecommand \@url [1]{\endgroup\@href {#1}{\urlprefix }}%
\providecommand \urlprefix  [0]{URL }%
\providecommand \Eprint [0]{\href }%
\providecommand \doibase [0]{https://doi.org/}%
\providecommand \selectlanguage [0]{\@gobble}%
\providecommand \bibinfo  [0]{\@secondoftwo}%
\providecommand \bibfield  [0]{\@secondoftwo}%
\providecommand \translation [1]{[#1]}%
\providecommand \BibitemOpen [0]{}%
\providecommand \bibitemStop [0]{}%
\providecommand \bibitemNoStop [0]{.\EOS\space}%
\providecommand \EOS [0]{\spacefactor3000\relax}%
\providecommand \BibitemShut  [1]{\csname bibitem#1\endcsname}%
\let\auto@bib@innerbib\@empty
\bibitem [{\citenamefont {Koch}\ \emph {et~al.}(2007)\citenamefont {Koch},
  \citenamefont {Yu}, \citenamefont {Gambetta}, \citenamefont {Houck},
  \citenamefont {Schuster}, \citenamefont {Majer}, \citenamefont {Blais},
  \citenamefont {Devoret}, \citenamefont {Girvin},\ and\ \citenamefont
  {Schoelkopf}}]{koch_charge-insensitive_2007}%
  \BibitemOpen
  \bibfield  {author} {\bibinfo {author} {\bibfnamefont {J.}~\bibnamefont
  {Koch}}, \bibinfo {author} {\bibfnamefont {T.~M.}\ \bibnamefont {Yu}},
  \bibinfo {author} {\bibfnamefont {J.}~\bibnamefont {Gambetta}}, \bibinfo
  {author} {\bibfnamefont {A.~A.}\ \bibnamefont {Houck}}, \bibinfo {author}
  {\bibfnamefont {D.~I.}\ \bibnamefont {Schuster}}, \bibinfo {author}
  {\bibfnamefont {J.}~\bibnamefont {Majer}}, \bibinfo {author} {\bibfnamefont
  {A.}~\bibnamefont {Blais}}, \bibinfo {author} {\bibfnamefont {M.~H.}\
  \bibnamefont {Devoret}}, \bibinfo {author} {\bibfnamefont {S.~M.}\
  \bibnamefont {Girvin}},\ and\ \bibinfo {author} {\bibfnamefont {R.~J.}\
  \bibnamefont {Schoelkopf}},\ }\bibfield  {title} {\bibinfo {title}
  {Charge-insensitive qubit design derived from the {Cooper} pair box},\ }\href
  {https://doi.org/10.1103/PhysRevA.76.042319} {\bibfield  {journal} {\bibinfo
  {journal} {Physical Review A}\ }\textbf {\bibinfo {volume} {76}},\ \bibinfo
  {pages} {042319} (\bibinfo {year} {2007})}\BibitemShut {NoStop}%
\bibitem [{\citenamefont {Paik}\ \emph {et~al.}(2011)\citenamefont {Paik},
  \citenamefont {Schuster}, \citenamefont {Bishop}, \citenamefont {Kirchmair},
  \citenamefont {Catelani}, \citenamefont {Sears}, \citenamefont {Johnson},
  \citenamefont {Reagor}, \citenamefont {Frunzio}, \citenamefont {Glazman},
  \citenamefont {Girvin}, \citenamefont {Devoret},\ and\ \citenamefont
  {Schoelkopf}}]{paik_observation_2011}%
  \BibitemOpen
  \bibfield  {author} {\bibinfo {author} {\bibfnamefont {H.}~\bibnamefont
  {Paik}}, \bibinfo {author} {\bibfnamefont {D.~I.}\ \bibnamefont {Schuster}},
  \bibinfo {author} {\bibfnamefont {L.~S.}\ \bibnamefont {Bishop}}, \bibinfo
  {author} {\bibfnamefont {G.}~\bibnamefont {Kirchmair}}, \bibinfo {author}
  {\bibfnamefont {G.}~\bibnamefont {Catelani}}, \bibinfo {author}
  {\bibfnamefont {A.~P.}\ \bibnamefont {Sears}}, \bibinfo {author}
  {\bibfnamefont {B.~R.}\ \bibnamefont {Johnson}}, \bibinfo {author}
  {\bibfnamefont {M.~J.}\ \bibnamefont {Reagor}}, \bibinfo {author}
  {\bibfnamefont {L.}~\bibnamefont {Frunzio}}, \bibinfo {author} {\bibfnamefont
  {L.~I.}\ \bibnamefont {Glazman}}, \bibinfo {author} {\bibfnamefont {S.~M.}\
  \bibnamefont {Girvin}}, \bibinfo {author} {\bibfnamefont {M.~H.}\
  \bibnamefont {Devoret}},\ and\ \bibinfo {author} {\bibfnamefont {R.~J.}\
  \bibnamefont {Schoelkopf}},\ }\bibfield  {title} {\bibinfo {title}
  {Observation of {High} {Coherence} in {Josephson} {Junction} {Qubits}
  {Measured} in a {Three}-{Dimensional} {Circuit} {QED} {Architecture}},\
  }\href {https://doi.org/10.1103/PhysRevLett.107.240501} {\bibfield  {journal}
  {\bibinfo  {journal} {Physical Review Letters}\ }\textbf {\bibinfo {volume}
  {107}},\ \bibinfo {pages} {240501} (\bibinfo {year} {2011})}\BibitemShut
  {NoStop}%
\bibitem [{\citenamefont {Place}\ \emph {et~al.}(2021)\citenamefont {Place},
  \citenamefont {Rodgers}, \citenamefont {Mundada}, \citenamefont {Smitham},
  \citenamefont {Fitzpatrick}, \citenamefont {Leng}, \citenamefont {Premkumar},
  \citenamefont {Bryon}, \citenamefont {Vrajitoarea}, \citenamefont {Sussman},
  \citenamefont {Cheng}, \citenamefont {Madhavan}, \citenamefont {Babla},
  \citenamefont {Le}, \citenamefont {Gang}, \citenamefont {Jäck},
  \citenamefont {Gyenis}, \citenamefont {Yao}, \citenamefont {Cava},
  \citenamefont {de~Leon},\ and\ \citenamefont {Houck}}]{place_new_2021}%
  \BibitemOpen
  \bibfield  {author} {\bibinfo {author} {\bibfnamefont {A.~P.~M.}\
  \bibnamefont {Place}}, \bibinfo {author} {\bibfnamefont {L.~V.~H.}\
  \bibnamefont {Rodgers}}, \bibinfo {author} {\bibfnamefont {P.}~\bibnamefont
  {Mundada}}, \bibinfo {author} {\bibfnamefont {B.~M.}\ \bibnamefont
  {Smitham}}, \bibinfo {author} {\bibfnamefont {M.}~\bibnamefont
  {Fitzpatrick}}, \bibinfo {author} {\bibfnamefont {Z.}~\bibnamefont {Leng}},
  \bibinfo {author} {\bibfnamefont {A.}~\bibnamefont {Premkumar}}, \bibinfo
  {author} {\bibfnamefont {J.}~\bibnamefont {Bryon}}, \bibinfo {author}
  {\bibfnamefont {A.}~\bibnamefont {Vrajitoarea}}, \bibinfo {author}
  {\bibfnamefont {S.}~\bibnamefont {Sussman}}, \bibinfo {author} {\bibfnamefont
  {G.}~\bibnamefont {Cheng}}, \bibinfo {author} {\bibfnamefont
  {T.}~\bibnamefont {Madhavan}}, \bibinfo {author} {\bibfnamefont {H.~K.}\
  \bibnamefont {Babla}}, \bibinfo {author} {\bibfnamefont {X.~H.}\ \bibnamefont
  {Le}}, \bibinfo {author} {\bibfnamefont {Y.}~\bibnamefont {Gang}}, \bibinfo
  {author} {\bibfnamefont {B.}~\bibnamefont {Jäck}}, \bibinfo {author}
  {\bibfnamefont {A.}~\bibnamefont {Gyenis}}, \bibinfo {author} {\bibfnamefont
  {N.}~\bibnamefont {Yao}}, \bibinfo {author} {\bibfnamefont {R.~J.}\
  \bibnamefont {Cava}}, \bibinfo {author} {\bibfnamefont {N.~P.}\ \bibnamefont
  {de~Leon}},\ and\ \bibinfo {author} {\bibfnamefont {A.~A.}\ \bibnamefont
  {Houck}},\ }\bibfield  {title} {\bibinfo {title} {New material platform for
  superconducting transmon qubits with coherence times exceeding 0.3
  milliseconds},\ }\href {https://doi.org/10.1038/s41467-021-22030-5}
  {\bibfield  {journal} {\bibinfo  {journal} {Nature Communications}\ }\textbf
  {\bibinfo {volume} {12}},\ \bibinfo {pages} {1779} (\bibinfo {year}
  {2021})}\BibitemShut {NoStop}%
\bibitem [{\citenamefont {Ballance}\ \emph {et~al.}(2016)\citenamefont
  {Ballance}, \citenamefont {Harty}, \citenamefont {Linke}, \citenamefont
  {Sepiol},\ and\ \citenamefont {Lucas}}]{ballance_high-fidelity_2016}%
  \BibitemOpen
  \bibfield  {author} {\bibinfo {author} {\bibfnamefont {C.}~\bibnamefont
  {Ballance}}, \bibinfo {author} {\bibfnamefont {T.}~\bibnamefont {Harty}},
  \bibinfo {author} {\bibfnamefont {N.}~\bibnamefont {Linke}}, \bibinfo
  {author} {\bibfnamefont {M.}~\bibnamefont {Sepiol}},\ and\ \bibinfo {author}
  {\bibfnamefont {D.}~\bibnamefont {Lucas}},\ }\bibfield  {title} {\bibinfo
  {title} {High-{Fidelity} {Quantum} {Logic} {Gates} {Using} {Trapped}-{Ion}
  {Hyperfine} {Qubits}},\ }\href
  {https://doi.org/10.1103/PhysRevLett.117.060504} {\bibfield  {journal}
  {\bibinfo  {journal} {Physical Review Letters}\ }\textbf {\bibinfo {volume}
  {117}},\ \bibinfo {pages} {060504} (\bibinfo {year} {2016})}\BibitemShut
  {NoStop}%
\bibitem [{\citenamefont {DiCarlo}\ \emph {et~al.}(2009)\citenamefont
  {DiCarlo}, \citenamefont {Chow}, \citenamefont {Gambetta}, \citenamefont
  {Bishop}, \citenamefont {Johnson}, \citenamefont {Schuster}, \citenamefont
  {Majer}, \citenamefont {Blais}, \citenamefont {Frunzio}, \citenamefont
  {Girvin},\ and\ \citenamefont {Schoelkopf}}]{dicarlo_demonstration_2009}%
  \BibitemOpen
  \bibfield  {author} {\bibinfo {author} {\bibfnamefont {L.}~\bibnamefont
  {DiCarlo}}, \bibinfo {author} {\bibfnamefont {J.~M.}\ \bibnamefont {Chow}},
  \bibinfo {author} {\bibfnamefont {J.~M.}\ \bibnamefont {Gambetta}}, \bibinfo
  {author} {\bibfnamefont {L.~S.}\ \bibnamefont {Bishop}}, \bibinfo {author}
  {\bibfnamefont {B.~R.}\ \bibnamefont {Johnson}}, \bibinfo {author}
  {\bibfnamefont {D.~I.}\ \bibnamefont {Schuster}}, \bibinfo {author}
  {\bibfnamefont {J.}~\bibnamefont {Majer}}, \bibinfo {author} {\bibfnamefont
  {A.}~\bibnamefont {Blais}}, \bibinfo {author} {\bibfnamefont
  {L.}~\bibnamefont {Frunzio}}, \bibinfo {author} {\bibfnamefont {S.~M.}\
  \bibnamefont {Girvin}},\ and\ \bibinfo {author} {\bibfnamefont {R.~J.}\
  \bibnamefont {Schoelkopf}},\ }\bibfield  {title} {\bibinfo {title}
  {Demonstration of two-qubit algorithms with a superconducting quantum
  processor},\ }\href {https://doi.org/10.1038/nature08121} {\bibfield
  {journal} {\bibinfo  {journal} {Nature}\ }\textbf {\bibinfo {volume} {460}},\
  \bibinfo {pages} {240} (\bibinfo {year} {2009})}\BibitemShut {NoStop}%
\bibitem [{\citenamefont {Chen}\ \emph {et~al.}(2014)\citenamefont {Chen},
  \citenamefont {Neill}, \citenamefont {Roushan}, \citenamefont {Leung},
  \citenamefont {Fang}, \citenamefont {Barends}, \citenamefont {Kelly},
  \citenamefont {Campbell}, \citenamefont {Chen}, \citenamefont {Chiaro},
  \citenamefont {Dunsworth}, \citenamefont {Jeffrey}, \citenamefont {Megrant},
  \citenamefont {Mutus}, \citenamefont {O’Malley}, \citenamefont {Quintana},
  \citenamefont {Sank}, \citenamefont {Vainsencher}, \citenamefont {Wenner},
  \citenamefont {White}, \citenamefont {Geller}, \citenamefont {Cleland},\ and\
  \citenamefont {Martinis}}]{chen_qubit_2014}%
  \BibitemOpen
  \bibfield  {author} {\bibinfo {author} {\bibfnamefont {Y.}~\bibnamefont
  {Chen}}, \bibinfo {author} {\bibfnamefont {C.}~\bibnamefont {Neill}},
  \bibinfo {author} {\bibfnamefont {P.}~\bibnamefont {Roushan}}, \bibinfo
  {author} {\bibfnamefont {N.}~\bibnamefont {Leung}}, \bibinfo {author}
  {\bibfnamefont {M.}~\bibnamefont {Fang}}, \bibinfo {author} {\bibfnamefont
  {R.}~\bibnamefont {Barends}}, \bibinfo {author} {\bibfnamefont
  {J.}~\bibnamefont {Kelly}}, \bibinfo {author} {\bibfnamefont
  {B.}~\bibnamefont {Campbell}}, \bibinfo {author} {\bibfnamefont
  {Z.}~\bibnamefont {Chen}}, \bibinfo {author} {\bibfnamefont {B.}~\bibnamefont
  {Chiaro}}, \bibinfo {author} {\bibfnamefont {A.}~\bibnamefont {Dunsworth}},
  \bibinfo {author} {\bibfnamefont {E.}~\bibnamefont {Jeffrey}}, \bibinfo
  {author} {\bibfnamefont {A.}~\bibnamefont {Megrant}}, \bibinfo {author}
  {\bibfnamefont {J.}~\bibnamefont {Mutus}}, \bibinfo {author} {\bibfnamefont
  {P.}~\bibnamefont {O’Malley}}, \bibinfo {author} {\bibfnamefont
  {C.}~\bibnamefont {Quintana}}, \bibinfo {author} {\bibfnamefont
  {D.}~\bibnamefont {Sank}}, \bibinfo {author} {\bibfnamefont {A.}~\bibnamefont
  {Vainsencher}}, \bibinfo {author} {\bibfnamefont {J.}~\bibnamefont {Wenner}},
  \bibinfo {author} {\bibfnamefont {T.}~\bibnamefont {White}}, \bibinfo
  {author} {\bibfnamefont {M.~R.}\ \bibnamefont {Geller}}, \bibinfo {author}
  {\bibfnamefont {A.}~\bibnamefont {Cleland}},\ and\ \bibinfo {author}
  {\bibfnamefont {J.~M.}\ \bibnamefont {Martinis}},\ }\bibfield  {title}
  {\bibinfo {title} {Qubit {Architecture} with {High} {Coherence} and {Fast}
  {Tunable} {Coupling}},\ }\href
  {https://doi.org/10.1103/PhysRevLett.113.220502} {\bibfield  {journal}
  {\bibinfo  {journal} {Physical Review Letters}\ }\textbf {\bibinfo {volume}
  {113}},\ \bibinfo {pages} {220502} (\bibinfo {year} {2014})}\BibitemShut
  {NoStop}%
\bibitem [{\citenamefont {Wei}\ \emph {et~al.}(2021)\citenamefont {Wei},
  \citenamefont {Magesan}, \citenamefont {Lauer}, \citenamefont {Srinivasan},
  \citenamefont {Bogorin}, \citenamefont {Carnevale}, \citenamefont {Keefe},
  \citenamefont {Kim}, \citenamefont {Klaus}, \citenamefont {Landers},
  \citenamefont {Sundaresan}, \citenamefont {Wang}, \citenamefont {Zhang},
  \citenamefont {Steffen}, \citenamefont {Dial}, \citenamefont {McKay},\ and\
  \citenamefont {Kandala}}]{wei_quantum_2021}%
  \BibitemOpen
  \bibfield  {author} {\bibinfo {author} {\bibfnamefont {K.~X.}\ \bibnamefont
  {Wei}}, \bibinfo {author} {\bibfnamefont {E.}~\bibnamefont {Magesan}},
  \bibinfo {author} {\bibfnamefont {I.}~\bibnamefont {Lauer}}, \bibinfo
  {author} {\bibfnamefont {S.}~\bibnamefont {Srinivasan}}, \bibinfo {author}
  {\bibfnamefont {D.~F.}\ \bibnamefont {Bogorin}}, \bibinfo {author}
  {\bibfnamefont {S.}~\bibnamefont {Carnevale}}, \bibinfo {author}
  {\bibfnamefont {G.~A.}\ \bibnamefont {Keefe}}, \bibinfo {author}
  {\bibfnamefont {Y.}~\bibnamefont {Kim}}, \bibinfo {author} {\bibfnamefont
  {D.}~\bibnamefont {Klaus}}, \bibinfo {author} {\bibfnamefont
  {W.}~\bibnamefont {Landers}}, \bibinfo {author} {\bibfnamefont
  {N.}~\bibnamefont {Sundaresan}}, \bibinfo {author} {\bibfnamefont
  {C.}~\bibnamefont {Wang}}, \bibinfo {author} {\bibfnamefont {E.~J.}\
  \bibnamefont {Zhang}}, \bibinfo {author} {\bibfnamefont {M.}~\bibnamefont
  {Steffen}}, \bibinfo {author} {\bibfnamefont {O.~E.}\ \bibnamefont {Dial}},
  \bibinfo {author} {\bibfnamefont {D.~C.}\ \bibnamefont {McKay}},\ and\
  \bibinfo {author} {\bibfnamefont {A.}~\bibnamefont {Kandala}},\ }\bibfield
  {title} {\bibinfo {title} {Quantum crosstalk cancellation for fast entangling
  gates and improved multi-qubit performance},\ }\href
  {http://arxiv.org/abs/2106.00675} {\bibfield  {journal} {\bibinfo  {journal}
  {arXiv:2106.00675 [quant-ph]}\ } (\bibinfo {year} {2021})}\BibitemShut
  {NoStop}%
\bibitem [{\citenamefont {Leung}\ \emph {et~al.}(2018)\citenamefont {Leung},
  \citenamefont {Landsman}, \citenamefont {Figgatt}, \citenamefont {Linke},
  \citenamefont {Monroe},\ and\ \citenamefont {Brown}}]{leung_robust_2018}%
  \BibitemOpen
  \bibfield  {author} {\bibinfo {author} {\bibfnamefont {P.~H.}\ \bibnamefont
  {Leung}}, \bibinfo {author} {\bibfnamefont {K.~A.}\ \bibnamefont {Landsman}},
  \bibinfo {author} {\bibfnamefont {C.}~\bibnamefont {Figgatt}}, \bibinfo
  {author} {\bibfnamefont {N.~M.}\ \bibnamefont {Linke}}, \bibinfo {author}
  {\bibfnamefont {C.}~\bibnamefont {Monroe}},\ and\ \bibinfo {author}
  {\bibfnamefont {K.~R.}\ \bibnamefont {Brown}},\ }\bibfield  {title} {\bibinfo
  {title} {Robust 2-{Qubit} {Gates} in a {Linear} {Ion} {Crystal} {Using} a
  {Frequency}-{Modulated} {Driving} {Force}},\ }\href
  {https://doi.org/10.1103/PhysRevLett.120.020501} {\bibfield  {journal}
  {\bibinfo  {journal} {Physical Review Letters}\ }\textbf {\bibinfo {volume}
  {120}},\ \bibinfo {pages} {020501} (\bibinfo {year} {2018})}\BibitemShut
  {NoStop}%
\bibitem [{\citenamefont {Torrontegui}\ \emph {et~al.}(2020)\citenamefont
  {Torrontegui}, \citenamefont {Heinrich}, \citenamefont {Hussain},
  \citenamefont {Blatt},\ and\ \citenamefont
  {García-Ripoll}}]{torrontegui_ultra-fast_2020}%
  \BibitemOpen
  \bibfield  {author} {\bibinfo {author} {\bibfnamefont {E.}~\bibnamefont
  {Torrontegui}}, \bibinfo {author} {\bibfnamefont {D.}~\bibnamefont
  {Heinrich}}, \bibinfo {author} {\bibfnamefont {M.~I.}\ \bibnamefont
  {Hussain}}, \bibinfo {author} {\bibfnamefont {R.}~\bibnamefont {Blatt}},\
  and\ \bibinfo {author} {\bibfnamefont {J.~J.}\ \bibnamefont
  {García-Ripoll}},\ }\bibfield  {title} {\bibinfo {title} {Ultra-fast
  two-qubit ion gate using sequences of resonant pulses},\ }\href
  {https://doi.org/10.1088/1367-2630/abbab6} {\bibfield  {journal} {\bibinfo
  {journal} {New Journal of Physics}\ }\textbf {\bibinfo {volume} {22}},\
  \bibinfo {pages} {103024} (\bibinfo {year} {2020})}\BibitemShut {NoStop}%
\bibitem [{\citenamefont {Sung}\ \emph {et~al.}(2021)\citenamefont {Sung},
  \citenamefont {Ding}, \citenamefont {Braum\"uller}, \citenamefont
  {Veps\"al\"ainen}, \citenamefont {Kannan}, \citenamefont {Kjaergaard},
  \citenamefont {Greene}, \citenamefont {Samach}, \citenamefont {McNally},
  \citenamefont {Kim}, \citenamefont {Melville}, \citenamefont {Niedzielski},
  \citenamefont {Schwartz}, \citenamefont {Yoder}, \citenamefont {Orlando},
  \citenamefont {Gustavsson},\ and\ \citenamefont {Oliver}}]{OliverGates2021}%
  \BibitemOpen
  \bibfield  {author} {\bibinfo {author} {\bibfnamefont {Y.}~\bibnamefont
  {Sung}}, \bibinfo {author} {\bibfnamefont {L.}~\bibnamefont {Ding}}, \bibinfo
  {author} {\bibfnamefont {J.}~\bibnamefont {Braum\"uller}}, \bibinfo {author}
  {\bibfnamefont {A.}~\bibnamefont {Veps\"al\"ainen}}, \bibinfo {author}
  {\bibfnamefont {B.}~\bibnamefont {Kannan}}, \bibinfo {author} {\bibfnamefont
  {M.}~\bibnamefont {Kjaergaard}}, \bibinfo {author} {\bibfnamefont
  {A.}~\bibnamefont {Greene}}, \bibinfo {author} {\bibfnamefont {G.~O.}\
  \bibnamefont {Samach}}, \bibinfo {author} {\bibfnamefont {C.}~\bibnamefont
  {McNally}}, \bibinfo {author} {\bibfnamefont {D.}~\bibnamefont {Kim}},
  \bibinfo {author} {\bibfnamefont {A.}~\bibnamefont {Melville}}, \bibinfo
  {author} {\bibfnamefont {B.~M.}\ \bibnamefont {Niedzielski}}, \bibinfo
  {author} {\bibfnamefont {M.~E.}\ \bibnamefont {Schwartz}}, \bibinfo {author}
  {\bibfnamefont {J.~L.}\ \bibnamefont {Yoder}}, \bibinfo {author}
  {\bibfnamefont {T.~P.}\ \bibnamefont {Orlando}}, \bibinfo {author}
  {\bibfnamefont {S.}~\bibnamefont {Gustavsson}},\ and\ \bibinfo {author}
  {\bibfnamefont {W.~D.}\ \bibnamefont {Oliver}},\ }\bibfield  {title}
  {\bibinfo {title} {Realization of high-fidelity cz and $zz$-free iswap gates
  with a tunable coupler},\ }\href {https://doi.org/10.1103/PhysRevX.11.021058}
  {\bibfield  {journal} {\bibinfo  {journal} {Phys. Rev. X}\ }\textbf {\bibinfo
  {volume} {11}},\ \bibinfo {pages} {021058} (\bibinfo {year}
  {2021})}\BibitemShut {NoStop}%
\bibitem [{\citenamefont {Monroe}\ \emph {et~al.}(2014)\citenamefont {Monroe},
  \citenamefont {Raussendorf}, \citenamefont {Ruthven}, \citenamefont {Brown},
  \citenamefont {Maunz}, \citenamefont {Duan},\ and\ \citenamefont
  {Kim}}]{monroe_large-scale_2014}%
  \BibitemOpen
  \bibfield  {author} {\bibinfo {author} {\bibfnamefont {C.}~\bibnamefont
  {Monroe}}, \bibinfo {author} {\bibfnamefont {R.}~\bibnamefont {Raussendorf}},
  \bibinfo {author} {\bibfnamefont {A.}~\bibnamefont {Ruthven}}, \bibinfo
  {author} {\bibfnamefont {K.~R.}\ \bibnamefont {Brown}}, \bibinfo {author}
  {\bibfnamefont {P.}~\bibnamefont {Maunz}}, \bibinfo {author} {\bibfnamefont
  {L.-M.}\ \bibnamefont {Duan}},\ and\ \bibinfo {author} {\bibfnamefont
  {J.}~\bibnamefont {Kim}},\ }\bibfield  {title} {\bibinfo {title} {Large-scale
  modular quantum-computer architecture with atomic memory and photonic
  interconnects},\ }\href {https://doi.org/10.1103/PhysRevA.89.022317}
  {\bibfield  {journal} {\bibinfo  {journal} {Physical Review A}\ }\textbf
  {\bibinfo {volume} {89}},\ \bibinfo {pages} {022317} (\bibinfo {year}
  {2014})}\BibitemShut {NoStop}%
\bibitem [{\citenamefont {Egan}\ \emph {et~al.}(2021)\citenamefont {Egan},
  \citenamefont {Debroy}, \citenamefont {Noel}, \citenamefont {Risinger},
  \citenamefont {Zhu}, \citenamefont {Biswas}, \citenamefont {Newman},
  \citenamefont {Li}, \citenamefont {Brown}, \citenamefont {Cetina},\ and\
  \citenamefont {Monroe}}]{egan_fault-tolerant_2021}%
  \BibitemOpen
  \bibfield  {author} {\bibinfo {author} {\bibfnamefont {L.}~\bibnamefont
  {Egan}}, \bibinfo {author} {\bibfnamefont {D.~M.}\ \bibnamefont {Debroy}},
  \bibinfo {author} {\bibfnamefont {C.}~\bibnamefont {Noel}}, \bibinfo {author}
  {\bibfnamefont {A.}~\bibnamefont {Risinger}}, \bibinfo {author}
  {\bibfnamefont {D.}~\bibnamefont {Zhu}}, \bibinfo {author} {\bibfnamefont
  {D.}~\bibnamefont {Biswas}}, \bibinfo {author} {\bibfnamefont
  {M.}~\bibnamefont {Newman}}, \bibinfo {author} {\bibfnamefont
  {M.}~\bibnamefont {Li}}, \bibinfo {author} {\bibfnamefont {K.~R.}\
  \bibnamefont {Brown}}, \bibinfo {author} {\bibfnamefont {M.}~\bibnamefont
  {Cetina}},\ and\ \bibinfo {author} {\bibfnamefont {C.}~\bibnamefont
  {Monroe}},\ }\bibfield  {title} {\bibinfo {title} {Fault-tolerant control of
  an error-corrected qubit},\ }\href
  {https://doi.org/10.1038/s41586-021-03928-y} {\bibfield  {journal} {\bibinfo
  {journal} {Nature}\ }\textbf {\bibinfo {volume} {598}},\ \bibinfo {pages}
  {281} (\bibinfo {year} {2021})}\BibitemShut {NoStop}%
\bibitem [{\citenamefont {Olsacher}\ \emph {et~al.}(2020)\citenamefont
  {Olsacher}, \citenamefont {Postler}, \citenamefont {Schindler}, \citenamefont
  {Monz}, \citenamefont {Zoller},\ and\ \citenamefont
  {Sieberer}}]{olsacher_scalable_2020}%
  \BibitemOpen
  \bibfield  {author} {\bibinfo {author} {\bibfnamefont {T.}~\bibnamefont
  {Olsacher}}, \bibinfo {author} {\bibfnamefont {L.}~\bibnamefont {Postler}},
  \bibinfo {author} {\bibfnamefont {P.}~\bibnamefont {Schindler}}, \bibinfo
  {author} {\bibfnamefont {T.}~\bibnamefont {Monz}}, \bibinfo {author}
  {\bibfnamefont {P.}~\bibnamefont {Zoller}},\ and\ \bibinfo {author}
  {\bibfnamefont {L.~M.}\ \bibnamefont {Sieberer}},\ }\bibfield  {title}
  {\bibinfo {title} {Scalable and {Parallel} {Tweezer} {Gates} for {Quantum}
  {Computing} with {Long} {Ion} {Strings}},\ }\href
  {https://doi.org/10.1103/PRXQuantum.1.020316} {\bibfield  {journal} {\bibinfo
   {journal} {PRX Quantum}\ }\textbf {\bibinfo {volume} {1}},\ \bibinfo {pages}
  {020316} (\bibinfo {year} {2020})}\BibitemShut {NoStop}%
\bibitem [{\citenamefont {Postler}\ \emph {et~al.}(2022)\citenamefont
  {Postler}, \citenamefont {Heu{\ss}en}, \citenamefont {Pogorelov},
  \citenamefont {Rispler}, \citenamefont {Feldker}, \citenamefont {Meth},
  \citenamefont {Marciniak}, \citenamefont {Stricker}, \citenamefont
  {Ringbauer}, \citenamefont {Blatt}, \citenamefont {Schindler}, \citenamefont
  {Müller},\ and\ \citenamefont {Monz}}]{postler_demonstration_2022}%
  \BibitemOpen
  \bibfield  {author} {\bibinfo {author} {\bibfnamefont {L.}~\bibnamefont
  {Postler}}, \bibinfo {author} {\bibfnamefont {S.}~\bibnamefont {Heu{\ss}en}},
  \bibinfo {author} {\bibfnamefont {I.}~\bibnamefont {Pogorelov}}, \bibinfo
  {author} {\bibfnamefont {M.}~\bibnamefont {Rispler}}, \bibinfo {author}
  {\bibfnamefont {T.}~\bibnamefont {Feldker}}, \bibinfo {author} {\bibfnamefont
  {M.}~\bibnamefont {Meth}}, \bibinfo {author} {\bibfnamefont {C.~D.}\
  \bibnamefont {Marciniak}}, \bibinfo {author} {\bibfnamefont {R.}~\bibnamefont
  {Stricker}}, \bibinfo {author} {\bibfnamefont {M.}~\bibnamefont {Ringbauer}},
  \bibinfo {author} {\bibfnamefont {R.}~\bibnamefont {Blatt}}, \bibinfo
  {author} {\bibfnamefont {P.}~\bibnamefont {Schindler}}, \bibinfo {author}
  {\bibfnamefont {M.}~\bibnamefont {Müller}},\ and\ \bibinfo {author}
  {\bibfnamefont {T.}~\bibnamefont {Monz}},\ }\bibfield  {title} {\bibinfo
  {title} {Demonstration of fault-tolerant universal quantum gate operations},\
  }\href {https://doi.org/10.1038/s41586-022-04721-1} {\bibfield  {journal}
  {\bibinfo  {journal} {Nature}\ }\textbf {\bibinfo {volume} {605}},\ \bibinfo
  {pages} {675} (\bibinfo {year} {2022})}\BibitemShut {NoStop}%
\bibitem [{\citenamefont {Arute}\ \emph {et~al.}(2019)\citenamefont {Arute},
  \citenamefont {Arya}, \citenamefont {Babbush}, \citenamefont {Bacon},
  \citenamefont {Bardin}, \citenamefont {Barends}, \citenamefont {Biswas},
  \citenamefont {Boixo}, \citenamefont {Brandao}, \citenamefont {Buell},
  \citenamefont {Burkett}, \citenamefont {Chen}, \citenamefont {Chen},
  \citenamefont {Chiaro}, \citenamefont {Collins}, \citenamefont {Courtney},
  \citenamefont {Dunsworth}, \citenamefont {Farhi}, \citenamefont {Foxen},
  \citenamefont {Fowler}, \citenamefont {Gidney}, \citenamefont {Giustina},
  \citenamefont {Graff}, \citenamefont {Guerin}, \citenamefont {Habegger},
  \citenamefont {Harrigan}, \citenamefont {Hartmann}, \citenamefont {Ho},
  \citenamefont {Hoffmann}, \citenamefont {Huang}, \citenamefont {Humble},
  \citenamefont {Isakov}, \citenamefont {Jeffrey}, \citenamefont {Jiang},
  \citenamefont {Kafri}, \citenamefont {Kechedzhi}, \citenamefont {Kelly},
  \citenamefont {Klimov}, \citenamefont {Knysh}, \citenamefont {Korotkov},
  \citenamefont {Kostritsa}, \citenamefont {Landhuis}, \citenamefont
  {Lindmark}, \citenamefont {Lucero}, \citenamefont {Lyakh}, \citenamefont
  {Mandrà}, \citenamefont {McClean}, \citenamefont {McEwen}, \citenamefont
  {Megrant}, \citenamefont {Mi}, \citenamefont {Michielsen}, \citenamefont
  {Mohseni}, \citenamefont {Mutus}, \citenamefont {Naaman}, \citenamefont
  {Neeley}, \citenamefont {Neill}, \citenamefont {Niu}, \citenamefont {Ostby},
  \citenamefont {Petukhov}, \citenamefont {Platt}, \citenamefont {Quintana},
  \citenamefont {Rieffel}, \citenamefont {Roushan}, \citenamefont {Rubin},
  \citenamefont {Sank}, \citenamefont {Satzinger}, \citenamefont {Smelyanskiy},
  \citenamefont {Sung}, \citenamefont {Trevithick}, \citenamefont
  {Vainsencher}, \citenamefont {Villalonga}, \citenamefont {White},
  \citenamefont {Yao}, \citenamefont {Yeh}, \citenamefont {Zalcman},
  \citenamefont {Neven},\ and\ \citenamefont {Martinis}}]{arute_quantum_2019}%
  \BibitemOpen
  \bibfield  {author} {\bibinfo {author} {\bibfnamefont {F.}~\bibnamefont
  {Arute}}, \bibinfo {author} {\bibfnamefont {K.}~\bibnamefont {Arya}},
  \bibinfo {author} {\bibfnamefont {R.}~\bibnamefont {Babbush}}, \bibinfo
  {author} {\bibfnamefont {D.}~\bibnamefont {Bacon}}, \bibinfo {author}
  {\bibfnamefont {J.~C.}\ \bibnamefont {Bardin}}, \bibinfo {author}
  {\bibfnamefont {R.}~\bibnamefont {Barends}}, \bibinfo {author} {\bibfnamefont
  {R.}~\bibnamefont {Biswas}}, \bibinfo {author} {\bibfnamefont
  {S.}~\bibnamefont {Boixo}}, \bibinfo {author} {\bibfnamefont {F.~G. S.~L.}\
  \bibnamefont {Brandao}}, \bibinfo {author} {\bibfnamefont {D.~A.}\
  \bibnamefont {Buell}}, \bibinfo {author} {\bibfnamefont {B.}~\bibnamefont
  {Burkett}}, \bibinfo {author} {\bibfnamefont {Y.}~\bibnamefont {Chen}},
  \bibinfo {author} {\bibfnamefont {Z.}~\bibnamefont {Chen}}, \bibinfo {author}
  {\bibfnamefont {B.}~\bibnamefont {Chiaro}}, \bibinfo {author} {\bibfnamefont
  {R.}~\bibnamefont {Collins}}, \bibinfo {author} {\bibfnamefont
  {W.}~\bibnamefont {Courtney}}, \bibinfo {author} {\bibfnamefont
  {A.}~\bibnamefont {Dunsworth}}, \bibinfo {author} {\bibfnamefont
  {E.}~\bibnamefont {Farhi}}, \bibinfo {author} {\bibfnamefont
  {B.}~\bibnamefont {Foxen}}, \bibinfo {author} {\bibfnamefont
  {A.}~\bibnamefont {Fowler}}, \bibinfo {author} {\bibfnamefont
  {C.}~\bibnamefont {Gidney}}, \bibinfo {author} {\bibfnamefont
  {M.}~\bibnamefont {Giustina}}, \bibinfo {author} {\bibfnamefont
  {R.}~\bibnamefont {Graff}}, \bibinfo {author} {\bibfnamefont
  {K.}~\bibnamefont {Guerin}}, \bibinfo {author} {\bibfnamefont
  {S.}~\bibnamefont {Habegger}}, \bibinfo {author} {\bibfnamefont {M.~P.}\
  \bibnamefont {Harrigan}}, \bibinfo {author} {\bibfnamefont {M.~J.}\
  \bibnamefont {Hartmann}}, \bibinfo {author} {\bibfnamefont {A.}~\bibnamefont
  {Ho}}, \bibinfo {author} {\bibfnamefont {M.}~\bibnamefont {Hoffmann}},
  \bibinfo {author} {\bibfnamefont {T.}~\bibnamefont {Huang}}, \bibinfo
  {author} {\bibfnamefont {T.~S.}\ \bibnamefont {Humble}}, \bibinfo {author}
  {\bibfnamefont {S.~V.}\ \bibnamefont {Isakov}}, \bibinfo {author}
  {\bibfnamefont {E.}~\bibnamefont {Jeffrey}}, \bibinfo {author} {\bibfnamefont
  {Z.}~\bibnamefont {Jiang}}, \bibinfo {author} {\bibfnamefont
  {D.}~\bibnamefont {Kafri}}, \bibinfo {author} {\bibfnamefont
  {K.}~\bibnamefont {Kechedzhi}}, \bibinfo {author} {\bibfnamefont
  {J.}~\bibnamefont {Kelly}}, \bibinfo {author} {\bibfnamefont {P.~V.}\
  \bibnamefont {Klimov}}, \bibinfo {author} {\bibfnamefont {S.}~\bibnamefont
  {Knysh}}, \bibinfo {author} {\bibfnamefont {A.}~\bibnamefont {Korotkov}},
  \bibinfo {author} {\bibfnamefont {F.}~\bibnamefont {Kostritsa}}, \bibinfo
  {author} {\bibfnamefont {D.}~\bibnamefont {Landhuis}}, \bibinfo {author}
  {\bibfnamefont {M.}~\bibnamefont {Lindmark}}, \bibinfo {author}
  {\bibfnamefont {E.}~\bibnamefont {Lucero}}, \bibinfo {author} {\bibfnamefont
  {D.}~\bibnamefont {Lyakh}}, \bibinfo {author} {\bibfnamefont
  {S.}~\bibnamefont {Mandrà}}, \bibinfo {author} {\bibfnamefont {J.~R.}\
  \bibnamefont {McClean}}, \bibinfo {author} {\bibfnamefont {M.}~\bibnamefont
  {McEwen}}, \bibinfo {author} {\bibfnamefont {A.}~\bibnamefont {Megrant}},
  \bibinfo {author} {\bibfnamefont {X.}~\bibnamefont {Mi}}, \bibinfo {author}
  {\bibfnamefont {K.}~\bibnamefont {Michielsen}}, \bibinfo {author}
  {\bibfnamefont {M.}~\bibnamefont {Mohseni}}, \bibinfo {author} {\bibfnamefont
  {J.}~\bibnamefont {Mutus}}, \bibinfo {author} {\bibfnamefont
  {O.}~\bibnamefont {Naaman}}, \bibinfo {author} {\bibfnamefont
  {M.}~\bibnamefont {Neeley}}, \bibinfo {author} {\bibfnamefont
  {C.}~\bibnamefont {Neill}}, \bibinfo {author} {\bibfnamefont {M.~Y.}\
  \bibnamefont {Niu}}, \bibinfo {author} {\bibfnamefont {E.}~\bibnamefont
  {Ostby}}, \bibinfo {author} {\bibfnamefont {A.}~\bibnamefont {Petukhov}},
  \bibinfo {author} {\bibfnamefont {J.~C.}\ \bibnamefont {Platt}}, \bibinfo
  {author} {\bibfnamefont {C.}~\bibnamefont {Quintana}}, \bibinfo {author}
  {\bibfnamefont {E.~G.}\ \bibnamefont {Rieffel}}, \bibinfo {author}
  {\bibfnamefont {P.}~\bibnamefont {Roushan}}, \bibinfo {author} {\bibfnamefont
  {N.~C.}\ \bibnamefont {Rubin}}, \bibinfo {author} {\bibfnamefont
  {D.}~\bibnamefont {Sank}}, \bibinfo {author} {\bibfnamefont {K.~J.}\
  \bibnamefont {Satzinger}}, \bibinfo {author} {\bibfnamefont {V.}~\bibnamefont
  {Smelyanskiy}}, \bibinfo {author} {\bibfnamefont {K.~J.}\ \bibnamefont
  {Sung}}, \bibinfo {author} {\bibfnamefont {M.~D.}\ \bibnamefont
  {Trevithick}}, \bibinfo {author} {\bibfnamefont {A.}~\bibnamefont
  {Vainsencher}}, \bibinfo {author} {\bibfnamefont {B.}~\bibnamefont
  {Villalonga}}, \bibinfo {author} {\bibfnamefont {T.}~\bibnamefont {White}},
  \bibinfo {author} {\bibfnamefont {Z.~J.}\ \bibnamefont {Yao}}, \bibinfo
  {author} {\bibfnamefont {P.}~\bibnamefont {Yeh}}, \bibinfo {author}
  {\bibfnamefont {A.}~\bibnamefont {Zalcman}}, \bibinfo {author} {\bibfnamefont
  {H.}~\bibnamefont {Neven}},\ and\ \bibinfo {author} {\bibfnamefont {J.~M.}\
  \bibnamefont {Martinis}},\ }\bibfield  {title} {\bibinfo {title} {Quantum
  supremacy using a programmable superconducting processor},\ }\href
  {https://doi.org/10.1038/s41586-019-1666-5} {\bibfield  {journal} {\bibinfo
  {journal} {Nature}\ }\textbf {\bibinfo {volume} {574}},\ \bibinfo {pages}
  {505} (\bibinfo {year} {2019})}\BibitemShut {NoStop}%
\bibitem [{\citenamefont {Jurcevic}\ \emph {et~al.}(2021)\citenamefont
  {Jurcevic}, \citenamefont {Javadi-Abhari}, \citenamefont {Bishop},
  \citenamefont {Lauer}, \citenamefont {Bogorin}, \citenamefont {Brink},
  \citenamefont {Capelluto}, \citenamefont {Günlük}, \citenamefont {Itoko},
  \citenamefont {Kanazawa}, \citenamefont {Kandala}, \citenamefont {Keefe},
  \citenamefont {Krsulich}, \citenamefont {Landers}, \citenamefont
  {Lewandowski}, \citenamefont {McClure}, \citenamefont {Nannicini},
  \citenamefont {Narasgond}, \citenamefont {Nayfeh}, \citenamefont {Pritchett},
  \citenamefont {Rothwell}, \citenamefont {Srinivasan}, \citenamefont
  {Sundaresan}, \citenamefont {Wang}, \citenamefont {Wei}, \citenamefont
  {Wood}, \citenamefont {Yau}, \citenamefont {Zhang}, \citenamefont {Dial},
  \citenamefont {Chow},\ and\ \citenamefont
  {Gambetta}}]{jurcevic_demonstration_2021}%
  \BibitemOpen
  \bibfield  {author} {\bibinfo {author} {\bibfnamefont {P.}~\bibnamefont
  {Jurcevic}}, \bibinfo {author} {\bibfnamefont {A.}~\bibnamefont
  {Javadi-Abhari}}, \bibinfo {author} {\bibfnamefont {L.~S.}\ \bibnamefont
  {Bishop}}, \bibinfo {author} {\bibfnamefont {I.}~\bibnamefont {Lauer}},
  \bibinfo {author} {\bibfnamefont {D.~F.}\ \bibnamefont {Bogorin}}, \bibinfo
  {author} {\bibfnamefont {M.}~\bibnamefont {Brink}}, \bibinfo {author}
  {\bibfnamefont {L.}~\bibnamefont {Capelluto}}, \bibinfo {author}
  {\bibfnamefont {O.}~\bibnamefont {Günlük}}, \bibinfo {author}
  {\bibfnamefont {T.}~\bibnamefont {Itoko}}, \bibinfo {author} {\bibfnamefont
  {N.}~\bibnamefont {Kanazawa}}, \bibinfo {author} {\bibfnamefont
  {A.}~\bibnamefont {Kandala}}, \bibinfo {author} {\bibfnamefont {G.~A.}\
  \bibnamefont {Keefe}}, \bibinfo {author} {\bibfnamefont {K.}~\bibnamefont
  {Krsulich}}, \bibinfo {author} {\bibfnamefont {W.}~\bibnamefont {Landers}},
  \bibinfo {author} {\bibfnamefont {E.~P.}\ \bibnamefont {Lewandowski}},
  \bibinfo {author} {\bibfnamefont {D.~T.}\ \bibnamefont {McClure}}, \bibinfo
  {author} {\bibfnamefont {G.}~\bibnamefont {Nannicini}}, \bibinfo {author}
  {\bibfnamefont {A.}~\bibnamefont {Narasgond}}, \bibinfo {author}
  {\bibfnamefont {H.~M.}\ \bibnamefont {Nayfeh}}, \bibinfo {author}
  {\bibfnamefont {E.}~\bibnamefont {Pritchett}}, \bibinfo {author}
  {\bibfnamefont {M.~B.}\ \bibnamefont {Rothwell}}, \bibinfo {author}
  {\bibfnamefont {S.}~\bibnamefont {Srinivasan}}, \bibinfo {author}
  {\bibfnamefont {N.}~\bibnamefont {Sundaresan}}, \bibinfo {author}
  {\bibfnamefont {C.}~\bibnamefont {Wang}}, \bibinfo {author} {\bibfnamefont
  {K.~X.}\ \bibnamefont {Wei}}, \bibinfo {author} {\bibfnamefont {C.~J.}\
  \bibnamefont {Wood}}, \bibinfo {author} {\bibfnamefont {J.-B.}\ \bibnamefont
  {Yau}}, \bibinfo {author} {\bibfnamefont {E.~J.}\ \bibnamefont {Zhang}},
  \bibinfo {author} {\bibfnamefont {O.~E.}\ \bibnamefont {Dial}}, \bibinfo
  {author} {\bibfnamefont {J.~M.}\ \bibnamefont {Chow}},\ and\ \bibinfo
  {author} {\bibfnamefont {J.~M.}\ \bibnamefont {Gambetta}},\ }\bibfield
  {title} {\bibinfo {title} {Demonstration of quantum volume 64 on a
  superconducting quantum computing system},\ }\href
  {https://doi.org/10.1088/2058-9565/abe519} {\bibfield  {journal} {\bibinfo
  {journal} {Quantum Science and Technology}\ }\textbf {\bibinfo {volume}
  {6}},\ \bibinfo {pages} {025020} (\bibinfo {year} {2021})}\BibitemShut
  {NoStop}%
\bibitem [{\citenamefont {Elder}\ \emph {et~al.}(2020)\citenamefont {Elder},
  \citenamefont {Wang}, \citenamefont {Reinhold}, \citenamefont {Hann},
  \citenamefont {Chou}, \citenamefont {Lester}, \citenamefont {Rosenblum},
  \citenamefont {Frunzio}, \citenamefont {Jiang},\ and\ \citenamefont
  {Schoelkopf}}]{elder_high-fidelity_2020}%
  \BibitemOpen
  \bibfield  {author} {\bibinfo {author} {\bibfnamefont {S.~S.}\ \bibnamefont
  {Elder}}, \bibinfo {author} {\bibfnamefont {C.~S.}\ \bibnamefont {Wang}},
  \bibinfo {author} {\bibfnamefont {P.}~\bibnamefont {Reinhold}}, \bibinfo
  {author} {\bibfnamefont {C.~T.}\ \bibnamefont {Hann}}, \bibinfo {author}
  {\bibfnamefont {K.~S.}\ \bibnamefont {Chou}}, \bibinfo {author}
  {\bibfnamefont {B.~J.}\ \bibnamefont {Lester}}, \bibinfo {author}
  {\bibfnamefont {S.}~\bibnamefont {Rosenblum}}, \bibinfo {author}
  {\bibfnamefont {L.}~\bibnamefont {Frunzio}}, \bibinfo {author} {\bibfnamefont
  {L.}~\bibnamefont {Jiang}},\ and\ \bibinfo {author} {\bibfnamefont {R.~J.}\
  \bibnamefont {Schoelkopf}},\ }\bibfield  {title} {\bibinfo {title}
  {High-{Fidelity} {Measurement} of {Qubits} {Encoded} in {Multilevel}
  {Superconducting} {Circuits}},\ }\href
  {https://doi.org/10.1103/PhysRevX.10.011001} {\bibfield  {journal} {\bibinfo
  {journal} {Physical Review X}\ }\textbf {\bibinfo {volume} {10}},\ \bibinfo
  {pages} {011001} (\bibinfo {year} {2020})}\BibitemShut {NoStop}%
\bibitem [{\citenamefont {Wu}\ \emph {et~al.}(2022)\citenamefont {Wu},
  \citenamefont {Kolkowitz}, \citenamefont {Puri},\ and\ \citenamefont
  {Thompson}}]{wu_erasure_2022}%
  \BibitemOpen
  \bibfield  {author} {\bibinfo {author} {\bibfnamefont {Y.}~\bibnamefont
  {Wu}}, \bibinfo {author} {\bibfnamefont {S.}~\bibnamefont {Kolkowitz}},
  \bibinfo {author} {\bibfnamefont {S.}~\bibnamefont {Puri}},\ and\ \bibinfo
  {author} {\bibfnamefont {J.~D.}\ \bibnamefont {Thompson}},\ }\bibfield
  {title} {\bibinfo {title} {Erasure conversion for fault-tolerant quantum
  computing in alkaline earth {Rydberg} atom arrays},\ }\href
  {https://doi.org/10.1038/s41467-022-32094-6} {\bibfield  {journal} {\bibinfo
  {journal} {Nature Communications}\ }\textbf {\bibinfo {volume} {13}},\
  \bibinfo {pages} {4657} (\bibinfo {year} {2022})}\BibitemShut {NoStop}%
\bibitem [{\citenamefont {Kang}\ \emph {et~al.}(2022)\citenamefont {Kang},
  \citenamefont {Campbell},\ and\ \citenamefont {Brown}}]{kang_quantum_2022}%
  \BibitemOpen
  \bibfield  {author} {\bibinfo {author} {\bibfnamefont {M.}~\bibnamefont
  {Kang}}, \bibinfo {author} {\bibfnamefont {W.~C.}\ \bibnamefont {Campbell}},\
  and\ \bibinfo {author} {\bibfnamefont {K.~R.}\ \bibnamefont {Brown}},\
  }\bibfield  {title} {\bibinfo {title} {Quantum error correction with
  metastable states of trapped ions using erasure conversion},\ }\href
  {http://arxiv.org/abs/2210.15024} {\bibfield  {journal} {\bibinfo  {journal}
  {arXiv:2210.15024 [quant-ph]}\ } (\bibinfo {year} {2022})}\BibitemShut
  {NoStop}%
\bibitem [{\citenamefont {Preskill}(2018)}]{preskill_quantum_2018}%
  \BibitemOpen
  \bibfield  {author} {\bibinfo {author} {\bibfnamefont {J.}~\bibnamefont
  {Preskill}},\ }\bibfield  {title} {\bibinfo {title} {Quantum {Computing} in
  the {NISQ} era and beyond},\ }\href
  {https://doi.org/10.22331/q-2018-08-06-79} {\bibfield  {journal} {\bibinfo
  {journal} {Quantum}\ }\textbf {\bibinfo {volume} {2}},\ \bibinfo {pages} {79}
  (\bibinfo {year} {2018})},\ \bibinfo {note} {arXiv:1801.00862 [cond-mat,
  physics:quant-ph]}\BibitemShut {NoStop}%
\bibitem [{\citenamefont {Wallraff}\ \emph {et~al.}(2004)\citenamefont
  {Wallraff}, \citenamefont {Schuster}, \citenamefont {Blais}, \citenamefont
  {Frunzio}, \citenamefont {Huang}, \citenamefont {Majer}, \citenamefont
  {Kumar}, \citenamefont {Girvin},\ and\ \citenamefont
  {Schoelkopf}}]{wallraff_strong_2004}%
  \BibitemOpen
  \bibfield  {author} {\bibinfo {author} {\bibfnamefont {A.}~\bibnamefont
  {Wallraff}}, \bibinfo {author} {\bibfnamefont {D.~I.}\ \bibnamefont
  {Schuster}}, \bibinfo {author} {\bibfnamefont {A.}~\bibnamefont {Blais}},
  \bibinfo {author} {\bibfnamefont {L.}~\bibnamefont {Frunzio}}, \bibinfo
  {author} {\bibfnamefont {R.-S.}\ \bibnamefont {Huang}}, \bibinfo {author}
  {\bibfnamefont {J.}~\bibnamefont {Majer}}, \bibinfo {author} {\bibfnamefont
  {S.}~\bibnamefont {Kumar}}, \bibinfo {author} {\bibfnamefont {S.~M.}\
  \bibnamefont {Girvin}},\ and\ \bibinfo {author} {\bibfnamefont {R.~J.}\
  \bibnamefont {Schoelkopf}},\ }\bibfield  {title} {\bibinfo {title} {Strong
  coupling of a single photon to a superconducting qubit using circuit quantum
  electrodynamics},\ }\href {https://doi.org/10.1038/nature02851} {\bibfield
  {journal} {\bibinfo  {journal} {Nature}\ }\textbf {\bibinfo {volume} {431}},\
  \bibinfo {pages} {162} (\bibinfo {year} {2004})}\BibitemShut {NoStop}%
\bibitem [{\citenamefont {Reagor}\ \emph {et~al.}(2016)\citenamefont {Reagor},
  \citenamefont {Pfaff}, \citenamefont {Axline}, \citenamefont {Heeres},
  \citenamefont {Ofek}, \citenamefont {Sliwa}, \citenamefont {Holland},
  \citenamefont {Wang}, \citenamefont {Blumoff}, \citenamefont {Chou},
  \citenamefont {Hatridge}, \citenamefont {Frunzio}, \citenamefont {Devoret},
  \citenamefont {Jiang},\ and\ \citenamefont
  {Schoelkopf}}]{reagor_quantum_2016}%
  \BibitemOpen
  \bibfield  {author} {\bibinfo {author} {\bibfnamefont {M.}~\bibnamefont
  {Reagor}}, \bibinfo {author} {\bibfnamefont {W.}~\bibnamefont {Pfaff}},
  \bibinfo {author} {\bibfnamefont {C.}~\bibnamefont {Axline}}, \bibinfo
  {author} {\bibfnamefont {R.~W.}\ \bibnamefont {Heeres}}, \bibinfo {author}
  {\bibfnamefont {N.}~\bibnamefont {Ofek}}, \bibinfo {author} {\bibfnamefont
  {K.}~\bibnamefont {Sliwa}}, \bibinfo {author} {\bibfnamefont
  {E.}~\bibnamefont {Holland}}, \bibinfo {author} {\bibfnamefont
  {C.}~\bibnamefont {Wang}}, \bibinfo {author} {\bibfnamefont {J.}~\bibnamefont
  {Blumoff}}, \bibinfo {author} {\bibfnamefont {K.}~\bibnamefont {Chou}},
  \bibinfo {author} {\bibfnamefont {M.~J.}\ \bibnamefont {Hatridge}}, \bibinfo
  {author} {\bibfnamefont {L.}~\bibnamefont {Frunzio}}, \bibinfo {author}
  {\bibfnamefont {M.~H.}\ \bibnamefont {Devoret}}, \bibinfo {author}
  {\bibfnamefont {L.}~\bibnamefont {Jiang}},\ and\ \bibinfo {author}
  {\bibfnamefont {R.~J.}\ \bibnamefont {Schoelkopf}},\ }\bibfield  {title}
  {\bibinfo {title} {Quantum memory with millisecond coherence in circuit
  {QED}},\ }\href {https://doi.org/10.1103/PhysRevB.94.014506} {\bibfield
  {journal} {\bibinfo  {journal} {Physical Review B}\ }\textbf {\bibinfo
  {volume} {94}},\ \bibinfo {pages} {014506} (\bibinfo {year}
  {2016})}\BibitemShut {NoStop}%
\bibitem [{\citenamefont {Michael}\ \emph {et~al.}(2016)\citenamefont
  {Michael}, \citenamefont {Silveri}, \citenamefont {Brierley}, \citenamefont
  {Albert}, \citenamefont {Salmilehto}, \citenamefont {Jiang},\ and\
  \citenamefont {Girvin}}]{michael_new_2016}%
  \BibitemOpen
  \bibfield  {author} {\bibinfo {author} {\bibfnamefont {M.~H.}\ \bibnamefont
  {Michael}}, \bibinfo {author} {\bibfnamefont {M.}~\bibnamefont {Silveri}},
  \bibinfo {author} {\bibfnamefont {R.}~\bibnamefont {Brierley}}, \bibinfo
  {author} {\bibfnamefont {V.~V.}\ \bibnamefont {Albert}}, \bibinfo {author}
  {\bibfnamefont {J.}~\bibnamefont {Salmilehto}}, \bibinfo {author}
  {\bibfnamefont {L.}~\bibnamefont {Jiang}},\ and\ \bibinfo {author}
  {\bibfnamefont {S.}~\bibnamefont {Girvin}},\ }\bibfield  {title} {\bibinfo
  {title} {New {Class} of {Quantum} {Error}-{Correcting} {Codes} for a
  {Bosonic} {Mode}},\ }\href {https://doi.org/10.1103/PhysRevX.6.031006}
  {\bibfield  {journal} {\bibinfo  {journal} {Physical Review X}\ }\textbf
  {\bibinfo {volume} {6}},\ \bibinfo {pages} {031006} (\bibinfo {year}
  {2016})}\BibitemShut {NoStop}%
\bibitem [{\citenamefont {Cochrane}\ \emph {et~al.}(1999)\citenamefont
  {Cochrane}, \citenamefont {Milburn},\ and\ \citenamefont
  {Munro}}]{Cochrane1999}%
  \BibitemOpen
  \bibfield  {author} {\bibinfo {author} {\bibfnamefont {P.~T.}\ \bibnamefont
  {Cochrane}}, \bibinfo {author} {\bibfnamefont {G.~J.}\ \bibnamefont
  {Milburn}},\ and\ \bibinfo {author} {\bibfnamefont {W.~J.}\ \bibnamefont
  {Munro}},\ }\bibfield  {title} {\bibinfo {title} {Macroscopically distinct
  quantum-superposition states as a bosonic code for amplitude damping},\
  }\href {https://doi.org/10.1103/PhysRevA.59.2631} {\bibfield  {journal}
  {\bibinfo  {journal} {Phys. Rev. A}\ }\textbf {\bibinfo {volume} {59}},\
  \bibinfo {pages} {2631} (\bibinfo {year} {1999})}\BibitemShut {NoStop}%
\bibitem [{\citenamefont {Mirrahimi}\ \emph {et~al.}(2014)\citenamefont
  {Mirrahimi}, \citenamefont {Leghtas}, \citenamefont {Albert}, \citenamefont
  {Touzard}, \citenamefont {Schoelkopf}, \citenamefont {Jiang},\ and\
  \citenamefont {Devoret}}]{mirrahimi_dynamically_2014}%
  \BibitemOpen
  \bibfield  {author} {\bibinfo {author} {\bibfnamefont {M.}~\bibnamefont
  {Mirrahimi}}, \bibinfo {author} {\bibfnamefont {Z.}~\bibnamefont {Leghtas}},
  \bibinfo {author} {\bibfnamefont {V.~V.}\ \bibnamefont {Albert}}, \bibinfo
  {author} {\bibfnamefont {S.}~\bibnamefont {Touzard}}, \bibinfo {author}
  {\bibfnamefont {R.~J.}\ \bibnamefont {Schoelkopf}}, \bibinfo {author}
  {\bibfnamefont {L.}~\bibnamefont {Jiang}},\ and\ \bibinfo {author}
  {\bibfnamefont {M.~H.}\ \bibnamefont {Devoret}},\ }\bibfield  {title}
  {\bibinfo {title} {Dynamically protected cat-qubits: a new paradigm for
  universal quantum computation},\ }\href
  {https://doi.org/10.1088/1367-2630/16/4/045014} {\bibfield  {journal}
  {\bibinfo  {journal} {New Journal of Physics}\ }\textbf {\bibinfo {volume}
  {16}},\ \bibinfo {pages} {045014} (\bibinfo {year} {2014})}\BibitemShut
  {NoStop}%
\bibitem [{\citenamefont {Grimsmo}\ \emph {et~al.}(2020)\citenamefont
  {Grimsmo}, \citenamefont {Combes},\ and\ \citenamefont
  {Baragiola}}]{grimsmo_quantum_2020}%
  \BibitemOpen
  \bibfield  {author} {\bibinfo {author} {\bibfnamefont {A.~L.}\ \bibnamefont
  {Grimsmo}}, \bibinfo {author} {\bibfnamefont {J.}~\bibnamefont {Combes}},\
  and\ \bibinfo {author} {\bibfnamefont {B.~Q.}\ \bibnamefont {Baragiola}},\
  }\bibfield  {title} {\bibinfo {title} {Quantum {Computing} with
  {Rotation}-{Symmetric} {Bosonic} {Codes}},\ }\href
  {https://doi.org/10.1103/PhysRevX.10.011058} {\bibfield  {journal} {\bibinfo
  {journal} {Physical Review X}\ }\textbf {\bibinfo {volume} {10}},\ \bibinfo
  {pages} {011058} (\bibinfo {year} {2020})}\BibitemShut {NoStop}%
\bibitem [{\citenamefont {Chuang}\ and\ \citenamefont
  {Yamamoto}(1995)}]{chuang_simple_1995}%
  \BibitemOpen
  \bibfield  {author} {\bibinfo {author} {\bibfnamefont {I.~L.}\ \bibnamefont
  {Chuang}}\ and\ \bibinfo {author} {\bibfnamefont {Y.}~\bibnamefont
  {Yamamoto}},\ }\bibfield  {title} {\bibinfo {title} {Simple quantum
  computer},\ }\href {https://doi.org/10.1103/PhysRevA.52.3489} {\bibfield
  {journal} {\bibinfo  {journal} {Physical Review A}\ }\textbf {\bibinfo
  {volume} {52}},\ \bibinfo {pages} {3489} (\bibinfo {year}
  {1995})}\BibitemShut {NoStop}%
\bibitem [{\citenamefont {Teoh}\ \emph {et~al.}()\citenamefont {Teoh},
  \citenamefont {Winkel}, \citenamefont {Babla}, \citenamefont {Chapman},
  \citenamefont {Claes}, \citenamefont {de~Graaf}, \citenamefont {Garmon},
  \citenamefont {Kalfus}, \citenamefont {Lu}, \citenamefont {Maiti},
  \citenamefont {Sahay}, \citenamefont {Thakur}, \citenamefont {Tsunoda},
  \citenamefont {Xue}, \citenamefont {Frunzio}, \citenamefont {Girvin},
  \citenamefont {Puri},\ and\ \citenamefont {Schoelkopf}}]{Teoh2022}%
  \BibitemOpen
  \bibfield  {author} {\bibinfo {author} {\bibfnamefont {J.}~\bibnamefont
  {Teoh}}, \bibinfo {author} {\bibfnamefont {P.}~\bibnamefont {Winkel}},
  \bibinfo {author} {\bibfnamefont {H.~K.}\ \bibnamefont {Babla}}, \bibinfo
  {author} {\bibfnamefont {B.~J.}\ \bibnamefont {Chapman}}, \bibinfo {author}
  {\bibfnamefont {J.}~\bibnamefont {Claes}}, \bibinfo {author} {\bibfnamefont
  {S.~J.}\ \bibnamefont {de~Graaf}}, \bibinfo {author} {\bibfnamefont {J.~W.}\
  \bibnamefont {Garmon}}, \bibinfo {author} {\bibfnamefont {W.~D.}\
  \bibnamefont {Kalfus}}, \bibinfo {author} {\bibfnamefont {Y.}~\bibnamefont
  {Lu}}, \bibinfo {author} {\bibfnamefont {A.}~\bibnamefont {Maiti}}, \bibinfo
  {author} {\bibfnamefont {K.}~\bibnamefont {Sahay}}, \bibinfo {author}
  {\bibfnamefont {N.}~\bibnamefont {Thakur}}, \bibinfo {author} {\bibfnamefont
  {T.}~\bibnamefont {Tsunoda}}, \bibinfo {author} {\bibfnamefont
  {S.}~\bibnamefont {Xue}}, \bibinfo {author} {\bibfnamefont {L.}~\bibnamefont
  {Frunzio}}, \bibinfo {author} {\bibfnamefont {S.~M.}\ \bibnamefont {Girvin}},
  \bibinfo {author} {\bibfnamefont {S.}~\bibnamefont {Puri}},\ and\ \bibinfo
  {author} {\bibfnamefont {R.~J.}\ \bibnamefont {Schoelkopf}},\ }\bibfield
  {title} {\bibinfo {title} {Dual-rail encoding with superconducting cavities
  (in prep.)},\ }\href@noop {} {\ }\BibitemShut {NoStop}%
\bibitem [{\citenamefont {Gao}\ \emph {et~al.}(2019)\citenamefont {Gao},
  \citenamefont {Lester}, \citenamefont {Chou}, \citenamefont {Frunzio},
  \citenamefont {Devoret}, \citenamefont {Jiang}, \citenamefont {Girvin},\ and\
  \citenamefont {Schoelkopf}}]{gao_entanglement_2019}%
  \BibitemOpen
  \bibfield  {author} {\bibinfo {author} {\bibfnamefont {Y.~Y.}\ \bibnamefont
  {Gao}}, \bibinfo {author} {\bibfnamefont {B.~J.}\ \bibnamefont {Lester}},
  \bibinfo {author} {\bibfnamefont {K.~S.}\ \bibnamefont {Chou}}, \bibinfo
  {author} {\bibfnamefont {L.}~\bibnamefont {Frunzio}}, \bibinfo {author}
  {\bibfnamefont {M.~H.}\ \bibnamefont {Devoret}}, \bibinfo {author}
  {\bibfnamefont {L.}~\bibnamefont {Jiang}}, \bibinfo {author} {\bibfnamefont
  {S.~M.}\ \bibnamefont {Girvin}},\ and\ \bibinfo {author} {\bibfnamefont
  {R.~J.}\ \bibnamefont {Schoelkopf}},\ }\bibfield  {title} {\bibinfo {title}
  {Entanglement of bosonic modes through an engineered exchange interaction},\
  }\href {https://doi.org/10.1038/s41586-019-0970-4} {\bibfield  {journal}
  {\bibinfo  {journal} {Nature}\ }\textbf {\bibinfo {volume} {566}},\ \bibinfo
  {pages} {509} (\bibinfo {year} {2019})}\BibitemShut {NoStop}%
\bibitem [{\citenamefont {Ma}\ \emph {et~al.}(2020{\natexlab{a}})\citenamefont
  {Ma}, \citenamefont {Zhang}, \citenamefont {Wong}, \citenamefont {Noh},
  \citenamefont {Rosenblum}, \citenamefont {Reinhold}, \citenamefont
  {Schoelkopf},\ and\ \citenamefont {Jiang}}]{ma_path-independent_2020}%
  \BibitemOpen
  \bibfield  {author} {\bibinfo {author} {\bibfnamefont {W.-L.}\ \bibnamefont
  {Ma}}, \bibinfo {author} {\bibfnamefont {M.}~\bibnamefont {Zhang}}, \bibinfo
  {author} {\bibfnamefont {Y.}~\bibnamefont {Wong}}, \bibinfo {author}
  {\bibfnamefont {K.}~\bibnamefont {Noh}}, \bibinfo {author} {\bibfnamefont
  {S.}~\bibnamefont {Rosenblum}}, \bibinfo {author} {\bibfnamefont
  {P.}~\bibnamefont {Reinhold}}, \bibinfo {author} {\bibfnamefont {R.~J.}\
  \bibnamefont {Schoelkopf}},\ and\ \bibinfo {author} {\bibfnamefont
  {L.}~\bibnamefont {Jiang}},\ }\bibfield  {title} {\bibinfo {title}
  {Path-{Independent} {Quantum} {Gates} with {Noisy} {Ancilla}},\ }\href
  {https://doi.org/10.1103/PhysRevLett.125.110503} {\bibfield  {journal}
  {\bibinfo  {journal} {Physical Review Letters}\ }\textbf {\bibinfo {volume}
  {125}},\ \bibinfo {pages} {110503} (\bibinfo {year}
  {2020}{\natexlab{a}})}\BibitemShut {NoStop}%
\bibitem [{\citenamefont {Reinhold}\ \emph {et~al.}(2020)\citenamefont
  {Reinhold}, \citenamefont {Rosenblum}, \citenamefont {Ma}, \citenamefont
  {Frunzio}, \citenamefont {Jiang},\ and\ \citenamefont
  {Schoelkopf}}]{reinhold_error-corrected_2020}%
  \BibitemOpen
  \bibfield  {author} {\bibinfo {author} {\bibfnamefont {P.}~\bibnamefont
  {Reinhold}}, \bibinfo {author} {\bibfnamefont {S.}~\bibnamefont {Rosenblum}},
  \bibinfo {author} {\bibfnamefont {W.-L.}\ \bibnamefont {Ma}}, \bibinfo
  {author} {\bibfnamefont {L.}~\bibnamefont {Frunzio}}, \bibinfo {author}
  {\bibfnamefont {L.}~\bibnamefont {Jiang}},\ and\ \bibinfo {author}
  {\bibfnamefont {R.~J.}\ \bibnamefont {Schoelkopf}},\ }\bibfield  {title}
  {\bibinfo {title} {Error-corrected gates on an encoded qubit},\ }\href
  {https://doi.org/10.1038/s41567-020-0931-8} {\bibfield  {journal} {\bibinfo
  {journal} {Nature Physics}\ }\textbf {\bibinfo {volume} {16}},\ \bibinfo
  {pages} {822} (\bibinfo {year} {2020})}\BibitemShut {NoStop}%
\bibitem [{\citenamefont {Kirchmair}\ \emph {et~al.}(2013)\citenamefont
  {Kirchmair}, \citenamefont {Vlastakis}, \citenamefont {Leghtas},
  \citenamefont {Nigg}, \citenamefont {Paik}, \citenamefont {Ginossar},
  \citenamefont {Mirrahimi}, \citenamefont {Frunzio}, \citenamefont {Girvin},\
  and\ \citenamefont {Schoelkopf}}]{kirchmair_observation_2013}%
  \BibitemOpen
  \bibfield  {author} {\bibinfo {author} {\bibfnamefont {G.}~\bibnamefont
  {Kirchmair}}, \bibinfo {author} {\bibfnamefont {B.}~\bibnamefont
  {Vlastakis}}, \bibinfo {author} {\bibfnamefont {Z.}~\bibnamefont {Leghtas}},
  \bibinfo {author} {\bibfnamefont {S.~E.}\ \bibnamefont {Nigg}}, \bibinfo
  {author} {\bibfnamefont {H.}~\bibnamefont {Paik}}, \bibinfo {author}
  {\bibfnamefont {E.}~\bibnamefont {Ginossar}}, \bibinfo {author}
  {\bibfnamefont {M.}~\bibnamefont {Mirrahimi}}, \bibinfo {author}
  {\bibfnamefont {L.}~\bibnamefont {Frunzio}}, \bibinfo {author} {\bibfnamefont
  {S.~M.}\ \bibnamefont {Girvin}},\ and\ \bibinfo {author} {\bibfnamefont
  {R.~J.}\ \bibnamefont {Schoelkopf}},\ }\bibfield  {title} {\bibinfo {title}
  {Observation of quantum state collapse and revival due to the single-photon
  {Kerr} effect},\ }\href {https://doi.org/10.1038/nature11902} {\bibfield
  {journal} {\bibinfo  {journal} {Nature}\ }\textbf {\bibinfo {volume} {495}},\
  \bibinfo {pages} {205} (\bibinfo {year} {2013})}\BibitemShut {NoStop}%
\bibitem [{\citenamefont {Blais}\ \emph {et~al.}(2021)\citenamefont {Blais},
  \citenamefont {Grimsmo}, \citenamefont {Girvin},\ and\ \citenamefont
  {Wallraff}}]{blais_circuit_2021}%
  \BibitemOpen
  \bibfield  {author} {\bibinfo {author} {\bibfnamefont {A.}~\bibnamefont
  {Blais}}, \bibinfo {author} {\bibfnamefont {A.~L.}\ \bibnamefont {Grimsmo}},
  \bibinfo {author} {\bibfnamefont {S.}~\bibnamefont {Girvin}},\ and\ \bibinfo
  {author} {\bibfnamefont {A.}~\bibnamefont {Wallraff}},\ }\bibfield  {title}
  {\bibinfo {title} {Circuit quantum electrodynamics},\ }\href
  {https://doi.org/10.1103/RevModPhys.93.025005} {\bibfield  {journal}
  {\bibinfo  {journal} {Reviews of Modern Physics}\ }\textbf {\bibinfo {volume}
  {93}},\ \bibinfo {pages} {025005} (\bibinfo {year} {2021})}\BibitemShut
  {NoStop}%
\bibitem [{\citenamefont {Zhang}\ \emph {et~al.}(2019)\citenamefont {Zhang},
  \citenamefont {Lester}, \citenamefont {Gao}, \citenamefont {Jiang},
  \citenamefont {Schoelkopf},\ and\ \citenamefont
  {Girvin}}]{zhang_engineering_2019}%
  \BibitemOpen
  \bibfield  {author} {\bibinfo {author} {\bibfnamefont {Y.}~\bibnamefont
  {Zhang}}, \bibinfo {author} {\bibfnamefont {B.~J.}\ \bibnamefont {Lester}},
  \bibinfo {author} {\bibfnamefont {Y.~Y.}\ \bibnamefont {Gao}}, \bibinfo
  {author} {\bibfnamefont {L.}~\bibnamefont {Jiang}}, \bibinfo {author}
  {\bibfnamefont {R.~J.}\ \bibnamefont {Schoelkopf}},\ and\ \bibinfo {author}
  {\bibfnamefont {S.~M.}\ \bibnamefont {Girvin}},\ }\bibfield  {title}
  {\bibinfo {title} {Engineering bilinear mode coupling in circuit {QED}:
  {Theory} and experiment},\ }\href
  {https://doi.org/10.1103/PhysRevA.99.012314} {\bibfield  {journal} {\bibinfo
  {journal} {Physical Review A}\ }\textbf {\bibinfo {volume} {99}},\ \bibinfo
  {pages} {012314} (\bibinfo {year} {2019})}\BibitemShut {NoStop}%
\bibitem [{\citenamefont {Chapman}\ \emph {et~al.}()\citenamefont {Chapman},
  \citenamefont {de~Graaf}, \citenamefont {Xue}, \citenamefont {Zhang},
  \citenamefont {Teoh}, \citenamefont {Curtis}, \citenamefont {Tsunoda},
  \citenamefont {Eickbusch}, \citenamefont {Read}, \citenamefont
  {Kootandavida}, \citenamefont {Mundhaha}, \citenamefont {Frunzio},
  \citenamefont {Devoret}, \citenamefont {Girvin},\ and\ \citenamefont
  {Schoelkopf}}]{StijnBS2022}%
  \BibitemOpen
  \bibfield  {author} {\bibinfo {author} {\bibfnamefont {B.~J.}\ \bibnamefont
  {Chapman}}, \bibinfo {author} {\bibfnamefont {S.~J.}\ \bibnamefont
  {de~Graaf}}, \bibinfo {author} {\bibfnamefont {S.~X.}\ \bibnamefont {Xue}},
  \bibinfo {author} {\bibfnamefont {Y.}~\bibnamefont {Zhang}}, \bibinfo
  {author} {\bibfnamefont {J.}~\bibnamefont {Teoh}}, \bibinfo {author}
  {\bibfnamefont {J.~C.}\ \bibnamefont {Curtis}}, \bibinfo {author}
  {\bibfnamefont {T.}~\bibnamefont {Tsunoda}}, \bibinfo {author} {\bibfnamefont
  {A.}~\bibnamefont {Eickbusch}}, \bibinfo {author} {\bibfnamefont {A.~P.}\
  \bibnamefont {Read}}, \bibinfo {author} {\bibfnamefont {A.}~\bibnamefont
  {Kootandavida}}, \bibinfo {author} {\bibfnamefont {S.}~\bibnamefont
  {Mundhaha}}, \bibinfo {author} {\bibfnamefont {L.}~\bibnamefont {Frunzio}},
  \bibinfo {author} {\bibfnamefont {M.~H.}\ \bibnamefont {Devoret}}, \bibinfo
  {author} {\bibfnamefont {S.~M.}\ \bibnamefont {Girvin}},\ and\ \bibinfo
  {author} {\bibfnamefont {R.~J.}\ \bibnamefont {Schoelkopf}},\ }\bibfield
  {title} {\bibinfo {title} {A high on-off ratio beamsplitter interaction for
  gates on bosonically encoded qubits (in prep.)},\ }\href@noop {} {\
  }\BibitemShut {NoStop}%
\bibitem [{\citenamefont {Zhou}\ \emph {et~al.}(2022)\citenamefont {Zhou},
  \citenamefont {Lu}, \citenamefont {Praquin}, \citenamefont {Chien},
  \citenamefont {Kaufman}, \citenamefont {Cao}, \citenamefont {Xia},
  \citenamefont {Mong}, \citenamefont {Pfaff}, \citenamefont {Pekker},\ and\
  \citenamefont {Hatridge}}]{zhou_modular_2022}%
  \BibitemOpen
  \bibfield  {author} {\bibinfo {author} {\bibfnamefont {C.}~\bibnamefont
  {Zhou}}, \bibinfo {author} {\bibfnamefont {P.}~\bibnamefont {Lu}}, \bibinfo
  {author} {\bibfnamefont {M.}~\bibnamefont {Praquin}}, \bibinfo {author}
  {\bibfnamefont {T.-C.}\ \bibnamefont {Chien}}, \bibinfo {author}
  {\bibfnamefont {R.}~\bibnamefont {Kaufman}}, \bibinfo {author} {\bibfnamefont
  {X.}~\bibnamefont {Cao}}, \bibinfo {author} {\bibfnamefont {M.}~\bibnamefont
  {Xia}}, \bibinfo {author} {\bibfnamefont {R.}~\bibnamefont {Mong}}, \bibinfo
  {author} {\bibfnamefont {W.}~\bibnamefont {Pfaff}}, \bibinfo {author}
  {\bibfnamefont {D.}~\bibnamefont {Pekker}},\ and\ \bibinfo {author}
  {\bibfnamefont {M.}~\bibnamefont {Hatridge}},\ }\bibfield  {title} {\bibinfo
  {title} {A modular quantum computer based on a quantum state router},\ }\href
  {http://arxiv.org/abs/2109.06848} {\bibfield  {journal} {\bibinfo  {journal}
  {arXiv:2109.06848 [quant-ph]}\ } (\bibinfo {year} {2022})}\BibitemShut
  {NoStop}%
\bibitem [{\citenamefont {Lu}\ \emph {et~al.}()\citenamefont {Lu},
  \citenamefont {Maiti}, \citenamefont {Garmon}, \citenamefont {Ganjam},
  \citenamefont {Zhang}, \citenamefont {Claes}, \citenamefont {Frunzio},\ and\
  \citenamefont {Schoelkopf}}]{YaoBS2022}%
  \BibitemOpen
  \bibfield  {author} {\bibinfo {author} {\bibfnamefont {Y.}~\bibnamefont
  {Lu}}, \bibinfo {author} {\bibfnamefont {A.}~\bibnamefont {Maiti}}, \bibinfo
  {author} {\bibfnamefont {J.}~\bibnamefont {Garmon}}, \bibinfo {author}
  {\bibfnamefont {S.}~\bibnamefont {Ganjam}}, \bibinfo {author} {\bibfnamefont
  {Y.}~\bibnamefont {Zhang}}, \bibinfo {author} {\bibfnamefont
  {J.}~\bibnamefont {Claes}}, \bibinfo {author} {\bibfnamefont
  {L.}~\bibnamefont {Frunzio}},\ and\ \bibinfo {author} {\bibfnamefont {R.~J.}\
  \bibnamefont {Schoelkopf}},\ }\bibfield  {title} {\bibinfo {title} {High
  fidelity bosonic swaps with a parity-protected mixer (in prep.)},\
  }\href@noop {} {\ }\BibitemShut {NoStop}%
\bibitem [{\citenamefont {Sirois}\ \emph {et~al.}(2015)\citenamefont {Sirois},
  \citenamefont {Castellanos-Beltran}, \citenamefont {DeFeo}, \citenamefont
  {Ranzani}, \citenamefont {Lecocq}, \citenamefont {Simmonds}, \citenamefont
  {Teufel},\ and\ \citenamefont {Aumentado}}]{sirois_coherent-state_2015}%
  \BibitemOpen
  \bibfield  {author} {\bibinfo {author} {\bibfnamefont {A.~J.}\ \bibnamefont
  {Sirois}}, \bibinfo {author} {\bibfnamefont {M.~A.}\ \bibnamefont
  {Castellanos-Beltran}}, \bibinfo {author} {\bibfnamefont {M.~P.}\
  \bibnamefont {DeFeo}}, \bibinfo {author} {\bibfnamefont {L.}~\bibnamefont
  {Ranzani}}, \bibinfo {author} {\bibfnamefont {F.}~\bibnamefont {Lecocq}},
  \bibinfo {author} {\bibfnamefont {R.~W.}\ \bibnamefont {Simmonds}}, \bibinfo
  {author} {\bibfnamefont {J.~D.}\ \bibnamefont {Teufel}},\ and\ \bibinfo
  {author} {\bibfnamefont {J.}~\bibnamefont {Aumentado}},\ }\bibfield  {title}
  {\bibinfo {title} {Coherent-state storage and retrieval between
  superconducting cavities using parametric frequency conversion},\ }\href
  {https://doi.org/10.1063/1.4919759} {\bibfield  {journal} {\bibinfo
  {journal} {Applied Physics Letters}\ }\textbf {\bibinfo {volume} {106}},\
  \bibinfo {pages} {172603} (\bibinfo {year} {2015})}\BibitemShut {NoStop}%
\bibitem [{\citenamefont {Rigetti}\ and\ \citenamefont
  {Devoret}(2010)}]{rigetti_fully_2010}%
  \BibitemOpen
  \bibfield  {author} {\bibinfo {author} {\bibfnamefont {C.}~\bibnamefont
  {Rigetti}}\ and\ \bibinfo {author} {\bibfnamefont {M.}~\bibnamefont
  {Devoret}},\ }\bibfield  {title} {\bibinfo {title} {Fully microwave-tunable
  universal gates in superconducting qubits with linear couplings and fixed
  transition frequencies},\ }\href {https://doi.org/10.1103/PhysRevB.81.134507}
  {\bibfield  {journal} {\bibinfo  {journal} {Physical Review B}\ }\textbf
  {\bibinfo {volume} {81}},\ \bibinfo {pages} {134507} (\bibinfo {year}
  {2010})}\BibitemShut {NoStop}%
\bibitem [{\citenamefont {Schwinger}(1952)}]{schwinger_angular_1952}%
  \BibitemOpen
  \bibfield  {author} {\bibinfo {author} {\bibfnamefont {J.}~\bibnamefont
  {Schwinger}},\ }\href {https://doi.org/10.2172/4389568} {\emph {\bibinfo
  {title} {{ON} {ANGULAR} {MOMENTUM}}}},\ \bibinfo {type} {Tech. Rep.}\
  \bibinfo {number} {NYO-3071}\ (\bibinfo  {institution} {Harvard Univ.,
  Cambridge, MA (United States); Nuclear Development Associates, Inc. (US)},\
  \bibinfo {year} {1952})\BibitemShut {NoStop}%
\bibitem [{\citenamefont {Reck}\ \emph {et~al.}(1994)\citenamefont {Reck},
  \citenamefont {Zeilinger}, \citenamefont {Bernstein},\ and\ \citenamefont
  {Bertani}}]{reck_experimental_1994}%
  \BibitemOpen
  \bibfield  {author} {\bibinfo {author} {\bibfnamefont {M.}~\bibnamefont
  {Reck}}, \bibinfo {author} {\bibfnamefont {A.}~\bibnamefont {Zeilinger}},
  \bibinfo {author} {\bibfnamefont {H.~J.}\ \bibnamefont {Bernstein}},\ and\
  \bibinfo {author} {\bibfnamefont {P.}~\bibnamefont {Bertani}},\ }\bibfield
  {title} {\bibinfo {title} {Experimental realization of any discrete unitary
  operator},\ }\href {https://doi.org/10.1103/PhysRevLett.73.58} {\bibfield
  {journal} {\bibinfo  {journal} {Physical Review Letters}\ }\textbf {\bibinfo
  {volume} {73}},\ \bibinfo {pages} {58} (\bibinfo {year} {1994})}\BibitemShut
  {NoStop}%
\bibitem [{\citenamefont {Shor}(1997)}]{shor_fault-tolerant_1997}%
  \BibitemOpen
  \bibfield  {author} {\bibinfo {author} {\bibfnamefont {P.~W.}\ \bibnamefont
  {Shor}},\ }\bibfield  {title} {\bibinfo {title} {Fault-tolerant quantum
  computation},\ }\href {http://arxiv.org/abs/quant-ph/9605011} {\bibfield
  {journal} {\bibinfo  {journal} {arXiv:quant-ph/9605011}\ } (\bibinfo {year}
  {1997})}\BibitemShut {NoStop}%
\bibitem [{\citenamefont {Gottesman}\ and\ \citenamefont
  {Chuang}(1999)}]{gottesman_demonstrating_1999}%
  \BibitemOpen
  \bibfield  {author} {\bibinfo {author} {\bibfnamefont {D.}~\bibnamefont
  {Gottesman}}\ and\ \bibinfo {author} {\bibfnamefont {I.~L.}\ \bibnamefont
  {Chuang}},\ }\bibfield  {title} {\bibinfo {title} {Demonstrating the
  viability of universal quantum computation using teleportation and
  single-qubit operations},\ }\href {https://doi.org/10.1038/46503} {\bibfield
  {journal} {\bibinfo  {journal} {Nature}\ }\textbf {\bibinfo {volume} {402}},\
  \bibinfo {pages} {390} (\bibinfo {year} {1999})}\BibitemShut {NoStop}%
\bibitem [{\citenamefont {Zhou}\ \emph {et~al.}(2000)\citenamefont {Zhou},
  \citenamefont {Leung},\ and\ \citenamefont {Chuang}}]{zhou_methodology_2000}%
  \BibitemOpen
  \bibfield  {author} {\bibinfo {author} {\bibfnamefont {X.}~\bibnamefont
  {Zhou}}, \bibinfo {author} {\bibfnamefont {D.~W.}\ \bibnamefont {Leung}},\
  and\ \bibinfo {author} {\bibfnamefont {I.~L.}\ \bibnamefont {Chuang}},\
  }\bibfield  {title} {\bibinfo {title} {Methodology for quantum logic gate
  construction},\ }\href {https://doi.org/10.1103/PhysRevA.62.052316}
  {\bibfield  {journal} {\bibinfo  {journal} {Physical Review A}\ }\textbf
  {\bibinfo {volume} {62}},\ \bibinfo {pages} {052316} (\bibinfo {year}
  {2000})}\BibitemShut {NoStop}%
\bibitem [{\citenamefont {Bravyi}\ and\ \citenamefont
  {Kitaev}(2005)}]{bravyi_universal_2005}%
  \BibitemOpen
  \bibfield  {author} {\bibinfo {author} {\bibfnamefont {S.}~\bibnamefont
  {Bravyi}}\ and\ \bibinfo {author} {\bibfnamefont {A.}~\bibnamefont
  {Kitaev}},\ }\bibfield  {title} {\bibinfo {title} {Universal quantum
  computation with ideal {Clifford} gates and noisy ancillas},\ }\href
  {https://doi.org/10.1103/PhysRevA.71.022316} {\bibfield  {journal} {\bibinfo
  {journal} {Physical Review A}\ }\textbf {\bibinfo {volume} {71}},\ \bibinfo
  {pages} {022316} (\bibinfo {year} {2005})}\BibitemShut {NoStop}%
\bibitem [{\citenamefont {Nielsen}\ and\ \citenamefont
  {Chuang}(2010)}]{nielsen_chuang_2010}%
  \BibitemOpen
  \bibfield  {author} {\bibinfo {author} {\bibfnamefont {M.~A.}\ \bibnamefont
  {Nielsen}}\ and\ \bibinfo {author} {\bibfnamefont {I.~L.}\ \bibnamefont
  {Chuang}},\ }\href {https://doi.org/10.1017/CBO9780511976667} {\emph
  {\bibinfo {title} {Quantum Computation and Quantum Information: 10th
  Anniversary Edition}}}\ (\bibinfo  {publisher} {Cambridge University Press},\
  \bibinfo {year} {2010})\BibitemShut {NoStop}%
\bibitem [{\citenamefont {Lau}\ and\ \citenamefont
  {Plenio}(2016)}]{lau_universal_2016}%
  \BibitemOpen
  \bibfield  {author} {\bibinfo {author} {\bibfnamefont {H.-K.}\ \bibnamefont
  {Lau}}\ and\ \bibinfo {author} {\bibfnamefont {M.~B.}\ \bibnamefont
  {Plenio}},\ }\bibfield  {title} {\bibinfo {title} {Universal {Quantum}
  {Computing} with {Arbitrary} {Continuous}-{Variable} {Encoding}}\ }\href
  {https://doi.org/10.1103/PhysRevLett.117.100501}
  {10.1103/PhysRevLett.117.100501} (\bibinfo {year} {2016}),\ \bibinfo {note}
  {arXiv:1605.09278 [quant-ph]}\BibitemShut {NoStop}%
\bibitem [{\citenamefont {Heeres}\ \emph {et~al.}(2015)\citenamefont {Heeres},
  \citenamefont {Vlastakis}, \citenamefont {Holland}, \citenamefont
  {Krastanov}, \citenamefont {Albert}, \citenamefont {Frunzio}, \citenamefont
  {Jiang},\ and\ \citenamefont {Schoelkopf}}]{heeres_cavity_2015}%
  \BibitemOpen
  \bibfield  {author} {\bibinfo {author} {\bibfnamefont {R.~W.}\ \bibnamefont
  {Heeres}}, \bibinfo {author} {\bibfnamefont {B.}~\bibnamefont {Vlastakis}},
  \bibinfo {author} {\bibfnamefont {E.}~\bibnamefont {Holland}}, \bibinfo
  {author} {\bibfnamefont {S.}~\bibnamefont {Krastanov}}, \bibinfo {author}
  {\bibfnamefont {V.~V.}\ \bibnamefont {Albert}}, \bibinfo {author}
  {\bibfnamefont {L.}~\bibnamefont {Frunzio}}, \bibinfo {author} {\bibfnamefont
  {L.}~\bibnamefont {Jiang}},\ and\ \bibinfo {author} {\bibfnamefont {R.~J.}\
  \bibnamefont {Schoelkopf}},\ }\bibfield  {title} {\bibinfo {title} {Cavity
  {State} {Manipulation} {Using} {Photon}-{Number} {Selective} {Phase}
  {Gates}},\ }\href {https://doi.org/10.1103/PhysRevLett.115.137002} {\bibfield
   {journal} {\bibinfo  {journal} {Physical Review Letters}\ }\textbf {\bibinfo
  {volume} {115}},\ \bibinfo {pages} {137002} (\bibinfo {year}
  {2015})}\BibitemShut {NoStop}%
\bibitem [{\citenamefont {Briegel}\ \emph {et~al.}(2009)\citenamefont
  {Briegel}, \citenamefont {Browne}, \citenamefont {Dür}, \citenamefont
  {Raussendorf},\ and\ \citenamefont {Van~den
  Nest}}]{briegel_measurement-based_2009}%
  \BibitemOpen
  \bibfield  {author} {\bibinfo {author} {\bibfnamefont {H.~J.}\ \bibnamefont
  {Briegel}}, \bibinfo {author} {\bibfnamefont {D.~E.}\ \bibnamefont {Browne}},
  \bibinfo {author} {\bibfnamefont {W.}~\bibnamefont {Dür}}, \bibinfo {author}
  {\bibfnamefont {R.}~\bibnamefont {Raussendorf}},\ and\ \bibinfo {author}
  {\bibfnamefont {M.}~\bibnamefont {Van~den Nest}},\ }\bibfield  {title}
  {\bibinfo {title} {Measurement-based quantum computation},\ }\href
  {https://doi.org/10.1038/nphys1157} {\bibfield  {journal} {\bibinfo
  {journal} {Nature Physics}\ }\textbf {\bibinfo {volume} {5}},\ \bibinfo
  {pages} {19} (\bibinfo {year} {2009})}\BibitemShut {NoStop}%
\bibitem [{\citenamefont {Puri}\ \emph {et~al.}(2019)\citenamefont {Puri},
  \citenamefont {Grimm}, \citenamefont {Campagne-Ibarcq}, \citenamefont
  {Eickbusch}, \citenamefont {Noh}, \citenamefont {Roberts}, \citenamefont
  {Jiang}, \citenamefont {Mirrahimi}, \citenamefont {Devoret},\ and\
  \citenamefont {Girvin}}]{puri_stabilized_2019}%
  \BibitemOpen
  \bibfield  {author} {\bibinfo {author} {\bibfnamefont {S.}~\bibnamefont
  {Puri}}, \bibinfo {author} {\bibfnamefont {A.}~\bibnamefont {Grimm}},
  \bibinfo {author} {\bibfnamefont {P.}~\bibnamefont {Campagne-Ibarcq}},
  \bibinfo {author} {\bibfnamefont {A.}~\bibnamefont {Eickbusch}}, \bibinfo
  {author} {\bibfnamefont {K.}~\bibnamefont {Noh}}, \bibinfo {author}
  {\bibfnamefont {G.}~\bibnamefont {Roberts}}, \bibinfo {author} {\bibfnamefont
  {L.}~\bibnamefont {Jiang}}, \bibinfo {author} {\bibfnamefont
  {M.}~\bibnamefont {Mirrahimi}}, \bibinfo {author} {\bibfnamefont {M.~H.}\
  \bibnamefont {Devoret}},\ and\ \bibinfo {author} {\bibfnamefont
  {S.}~\bibnamefont {Girvin}},\ }\bibfield  {title} {\bibinfo {title}
  {Stabilized {Cat} in a {Driven} {Nonlinear} {Cavity}: {A} {Fault}-{Tolerant}
  {Error} {Syndrome} {Detector}},\ }\href
  {https://doi.org/10.1103/PhysRevX.9.041009} {\bibfield  {journal} {\bibinfo
  {journal} {Physical Review X}\ }\textbf {\bibinfo {volume} {9}},\ \bibinfo
  {pages} {041009} (\bibinfo {year} {2019})}\BibitemShut {NoStop}%
\bibitem [{\citenamefont {Rosenblum}\ \emph {et~al.}(2018)\citenamefont
  {Rosenblum}, \citenamefont {Reinhold}, \citenamefont {Mirrahimi},
  \citenamefont {Jiang}, \citenamefont {Frunzio},\ and\ \citenamefont
  {Schoelkopf}}]{rosenblum_fault-tolerant_2018}%
  \BibitemOpen
  \bibfield  {author} {\bibinfo {author} {\bibfnamefont {S.}~\bibnamefont
  {Rosenblum}}, \bibinfo {author} {\bibfnamefont {P.}~\bibnamefont {Reinhold}},
  \bibinfo {author} {\bibfnamefont {M.}~\bibnamefont {Mirrahimi}}, \bibinfo
  {author} {\bibfnamefont {L.}~\bibnamefont {Jiang}}, \bibinfo {author}
  {\bibfnamefont {L.}~\bibnamefont {Frunzio}},\ and\ \bibinfo {author}
  {\bibfnamefont {R.~J.}\ \bibnamefont {Schoelkopf}},\ }\bibfield  {title}
  {\bibinfo {title} {Fault-tolerant detection of a quantum error},\ }\href
  {https://doi.org/10.1126/science.aat3996} {\bibfield  {journal} {\bibinfo
  {journal} {Science}\ }\textbf {\bibinfo {volume} {361}},\ \bibinfo {pages}
  {266} (\bibinfo {year} {2018})}\BibitemShut {NoStop}%
\bibitem [{\citenamefont {Nielsen}(2002)}]{nielsen_simple_2002}%
  \BibitemOpen
  \bibfield  {author} {\bibinfo {author} {\bibfnamefont {M.~A.}\ \bibnamefont
  {Nielsen}},\ }\bibfield  {title} {\bibinfo {title} {A simple formula for the
  average gate fidelity of a quantum dynamical operation},\ }\href
  {https://doi.org/10.1016/S0375-9601(02)01272-0} {\bibfield  {journal}
  {\bibinfo  {journal} {Physics Letters A}\ }\textbf {\bibinfo {volume}
  {303}},\ \bibinfo {pages} {249} (\bibinfo {year} {2002})}\BibitemShut
  {NoStop}%
\bibitem [{\citenamefont {Johansson}\ \emph {et~al.}(2013)\citenamefont
  {Johansson}, \citenamefont {Nation},\ and\ \citenamefont
  {Nori}}]{johansson_qutip_2013}%
  \BibitemOpen
  \bibfield  {author} {\bibinfo {author} {\bibfnamefont {J.~R.}\ \bibnamefont
  {Johansson}}, \bibinfo {author} {\bibfnamefont {P.~D.}\ \bibnamefont
  {Nation}},\ and\ \bibinfo {author} {\bibfnamefont {F.}~\bibnamefont {Nori}},\
  }\bibfield  {title} {\bibinfo {title} {{QuTiP} 2: {A} {Python} framework for
  the dynamics of open quantum systems},\ }\href
  {https://doi.org/10.1016/j.cpc.2012.11.019} {\bibfield  {journal} {\bibinfo
  {journal} {Computer Physics Communications}\ }\textbf {\bibinfo {volume}
  {184}},\ \bibinfo {pages} {1234} (\bibinfo {year} {2013})}\BibitemShut
  {NoStop}%
\bibitem [{\citenamefont {Gao}\ \emph {et~al.}(2018)\citenamefont {Gao},
  \citenamefont {Lester}, \citenamefont {Zhang}, \citenamefont {Wang},
  \citenamefont {Rosenblum}, \citenamefont {Frunzio}, \citenamefont {Jiang},
  \citenamefont {Girvin},\ and\ \citenamefont
  {Schoelkopf}}]{gao_programmable_2018}%
  \BibitemOpen
  \bibfield  {author} {\bibinfo {author} {\bibfnamefont {Y.~Y.}\ \bibnamefont
  {Gao}}, \bibinfo {author} {\bibfnamefont {B.~J.}\ \bibnamefont {Lester}},
  \bibinfo {author} {\bibfnamefont {Y.}~\bibnamefont {Zhang}}, \bibinfo
  {author} {\bibfnamefont {C.}~\bibnamefont {Wang}}, \bibinfo {author}
  {\bibfnamefont {S.}~\bibnamefont {Rosenblum}}, \bibinfo {author}
  {\bibfnamefont {L.}~\bibnamefont {Frunzio}}, \bibinfo {author} {\bibfnamefont
  {L.}~\bibnamefont {Jiang}}, \bibinfo {author} {\bibfnamefont
  {S.}~\bibnamefont {Girvin}},\ and\ \bibinfo {author} {\bibfnamefont {R.~J.}\
  \bibnamefont {Schoelkopf}},\ }\bibfield  {title} {\bibinfo {title}
  {Programmable {Interference} between {Two} {Microwave} {Quantum}
  {Memories}},\ }\href {https://doi.org/10.1103/PhysRevX.8.021073} {\bibfield
  {journal} {\bibinfo  {journal} {Physical Review X}\ }\textbf {\bibinfo
  {volume} {8}},\ \bibinfo {pages} {021073} (\bibinfo {year}
  {2018})}\BibitemShut {NoStop}%
\bibitem [{\citenamefont {Hou}\ \emph {et~al.}(2022)\citenamefont {Hou},
  \citenamefont {Wu}, \citenamefont {Erickson}, \citenamefont {Cole},
  \citenamefont {Zarantonello}, \citenamefont {Brandt}, \citenamefont {Wilson},
  \citenamefont {Slichter},\ and\ \citenamefont
  {Leibfried}}]{hou_coherently_2022}%
  \BibitemOpen
  \bibfield  {author} {\bibinfo {author} {\bibfnamefont {P.-Y.}\ \bibnamefont
  {Hou}}, \bibinfo {author} {\bibfnamefont {J.~J.}\ \bibnamefont {Wu}},
  \bibinfo {author} {\bibfnamefont {S.~D.}\ \bibnamefont {Erickson}}, \bibinfo
  {author} {\bibfnamefont {D.~C.}\ \bibnamefont {Cole}}, \bibinfo {author}
  {\bibfnamefont {G.}~\bibnamefont {Zarantonello}}, \bibinfo {author}
  {\bibfnamefont {A.~D.}\ \bibnamefont {Brandt}}, \bibinfo {author}
  {\bibfnamefont {A.~C.}\ \bibnamefont {Wilson}}, \bibinfo {author}
  {\bibfnamefont {D.~H.}\ \bibnamefont {Slichter}},\ and\ \bibinfo {author}
  {\bibfnamefont {D.}~\bibnamefont {Leibfried}},\ }\bibfield  {title} {\bibinfo
  {title} {Coherently {Coupled} {Mechanical} {Oscillators} in the {Quantum}
  {Regime}},\ }\href {http://arxiv.org/abs/2205.14841} {\bibfield  {journal}
  {\bibinfo  {journal} {arXiv:2205.14841 [quant-ph]}\ } (\bibinfo {year}
  {2022})}\BibitemShut {NoStop}%
\bibitem [{\citenamefont {Ortiz-Gutiérrez}\ \emph {et~al.}(2017)\citenamefont
  {Ortiz-Gutiérrez}, \citenamefont {Gabrielly}, \citenamefont {Muñoz},
  \citenamefont {Pereira}, \citenamefont {Filgueiras},\ and\ \citenamefont
  {Villar}}]{ortiz-gutierrez_continuous_2017}%
  \BibitemOpen
  \bibfield  {author} {\bibinfo {author} {\bibfnamefont {L.}~\bibnamefont
  {Ortiz-Gutiérrez}}, \bibinfo {author} {\bibfnamefont {B.}~\bibnamefont
  {Gabrielly}}, \bibinfo {author} {\bibfnamefont {L.~F.}\ \bibnamefont
  {Muñoz}}, \bibinfo {author} {\bibfnamefont {K.~T.}\ \bibnamefont {Pereira}},
  \bibinfo {author} {\bibfnamefont {J.~G.}\ \bibnamefont {Filgueiras}},\ and\
  \bibinfo {author} {\bibfnamefont {A.~S.}\ \bibnamefont {Villar}},\ }\bibfield
   {title} {\bibinfo {title} {Continuous variables quantum computation over the
  vibrational modes of a single trapped ion},\ }\href
  {https://doi.org/10.1016/j.optcom.2017.04.011} {\bibfield  {journal}
  {\bibinfo  {journal} {Optics Communications}\ }\textbf {\bibinfo {volume}
  {397}},\ \bibinfo {pages} {166} (\bibinfo {year} {2017})}\BibitemShut
  {NoStop}%
\bibitem [{\citenamefont {Gan}\ \emph {et~al.}(2020)\citenamefont {Gan},
  \citenamefont {Maslennikov}, \citenamefont {Tseng}, \citenamefont {Nguyen},\
  and\ \citenamefont {Matsukevich}}]{gan_hybrid_2020}%
  \BibitemOpen
  \bibfield  {author} {\bibinfo {author} {\bibfnamefont {H.}~\bibnamefont
  {Gan}}, \bibinfo {author} {\bibfnamefont {G.}~\bibnamefont {Maslennikov}},
  \bibinfo {author} {\bibfnamefont {K.-W.}\ \bibnamefont {Tseng}}, \bibinfo
  {author} {\bibfnamefont {C.}~\bibnamefont {Nguyen}},\ and\ \bibinfo {author}
  {\bibfnamefont {D.}~\bibnamefont {Matsukevich}},\ }\bibfield  {title}
  {\bibinfo {title} {Hybrid {Quantum} {Computing} with {Conditional} {Beam}
  {Splitter} {Gate} in {Trapped} {Ion} {System}},\ }\href
  {https://doi.org/10.1103/PhysRevLett.124.170502} {\bibfield  {journal}
  {\bibinfo  {journal} {Physical Review Letters}\ }\textbf {\bibinfo {volume}
  {124}},\ \bibinfo {pages} {170502} (\bibinfo {year} {2020})}\BibitemShut
  {NoStop}%
\bibitem [{\citenamefont {Katz}\ and\ \citenamefont
  {Monroe}(2022)}]{katz_programmable_2022}%
  \BibitemOpen
  \bibfield  {author} {\bibinfo {author} {\bibfnamefont {O.}~\bibnamefont
  {Katz}}\ and\ \bibinfo {author} {\bibfnamefont {C.}~\bibnamefont {Monroe}},\
  }\bibfield  {title} {\bibinfo {title} {Programmable quantum simulations of
  bosonic systems with trapped ions},\ }\href {http://arxiv.org/abs/2207.13653}
  {\bibfield  {journal} {\bibinfo  {journal} {arXiv:2207.13653 [physics,
  physics:quant-ph]}\ } (\bibinfo {year} {2022})}\BibitemShut {NoStop}%
\bibitem [{\citenamefont {Kivlichan}\ \emph {et~al.}(2018)\citenamefont
  {Kivlichan}, \citenamefont {McClean}, \citenamefont {Wiebe}, \citenamefont
  {Gidney}, \citenamefont {Aspuru-Guzik}, \citenamefont {Chan},\ and\
  \citenamefont {Babbush}}]{kivlichan_quantum_2018}%
  \BibitemOpen
  \bibfield  {author} {\bibinfo {author} {\bibfnamefont {I.~D.}\ \bibnamefont
  {Kivlichan}}, \bibinfo {author} {\bibfnamefont {J.}~\bibnamefont {McClean}},
  \bibinfo {author} {\bibfnamefont {N.}~\bibnamefont {Wiebe}}, \bibinfo
  {author} {\bibfnamefont {C.}~\bibnamefont {Gidney}}, \bibinfo {author}
  {\bibfnamefont {A.}~\bibnamefont {Aspuru-Guzik}}, \bibinfo {author}
  {\bibfnamefont {G.~K.-L.}\ \bibnamefont {Chan}},\ and\ \bibinfo {author}
  {\bibfnamefont {R.}~\bibnamefont {Babbush}},\ }\bibfield  {title} {\bibinfo
  {title} {Quantum {Simulation} of {Electronic} {Structure} with {Linear}
  {Depth} and {Connectivity}},\ }\href
  {https://doi.org/10.1103/PhysRevLett.120.110501} {\bibfield  {journal}
  {\bibinfo  {journal} {Physical Review Letters}\ }\textbf {\bibinfo {volume}
  {120}},\ \bibinfo {pages} {110501} (\bibinfo {year} {2018})}\BibitemShut
  {NoStop}%
\bibitem [{\citenamefont {Ma}\ \emph {et~al.}(2020{\natexlab{b}})\citenamefont
  {Ma}, \citenamefont {Xu}, \citenamefont {Mu}, \citenamefont {Cai},
  \citenamefont {Hu}, \citenamefont {Wang}, \citenamefont {Pan}, \citenamefont
  {Wang}, \citenamefont {Song}, \citenamefont {Zou},\ and\ \citenamefont
  {Sun}}]{ma_error-transparent_2020}%
  \BibitemOpen
  \bibfield  {author} {\bibinfo {author} {\bibfnamefont {Y.}~\bibnamefont
  {Ma}}, \bibinfo {author} {\bibfnamefont {Y.}~\bibnamefont {Xu}}, \bibinfo
  {author} {\bibfnamefont {X.}~\bibnamefont {Mu}}, \bibinfo {author}
  {\bibfnamefont {W.}~\bibnamefont {Cai}}, \bibinfo {author} {\bibfnamefont
  {L.}~\bibnamefont {Hu}}, \bibinfo {author} {\bibfnamefont {W.}~\bibnamefont
  {Wang}}, \bibinfo {author} {\bibfnamefont {X.}~\bibnamefont {Pan}}, \bibinfo
  {author} {\bibfnamefont {H.}~\bibnamefont {Wang}}, \bibinfo {author}
  {\bibfnamefont {Y.~P.}\ \bibnamefont {Song}}, \bibinfo {author}
  {\bibfnamefont {C.-L.}\ \bibnamefont {Zou}},\ and\ \bibinfo {author}
  {\bibfnamefont {L.}~\bibnamefont {Sun}},\ }\bibfield  {title} {\bibinfo
  {title} {Error-transparent operations on a logical qubit protected by quantum
  error correction},\ }\href {https://doi.org/10.1038/s41567-020-0893-x}
  {\bibfield  {journal} {\bibinfo  {journal} {Nature Physics}\ }\textbf
  {\bibinfo {volume} {16}},\ \bibinfo {pages} {827} (\bibinfo {year}
  {2020}{\natexlab{b}})},\ \bibinfo {note} {number: 8}\BibitemShut {NoStop}%
\bibitem [{\citenamefont {Ofek}\ \emph {et~al.}(2016)\citenamefont {Ofek},
  \citenamefont {Petrenko}, \citenamefont {Heeres}, \citenamefont {Reinhold},
  \citenamefont {Leghtas}, \citenamefont {Vlastakis}, \citenamefont {Liu},
  \citenamefont {Frunzio}, \citenamefont {Girvin}, \citenamefont {Jiang},
  \citenamefont {Mirrahimi}, \citenamefont {Devoret},\ and\ \citenamefont
  {Schoelkopf}}]{ofek_extending_2016}%
  \BibitemOpen
  \bibfield  {author} {\bibinfo {author} {\bibfnamefont {N.}~\bibnamefont
  {Ofek}}, \bibinfo {author} {\bibfnamefont {A.}~\bibnamefont {Petrenko}},
  \bibinfo {author} {\bibfnamefont {R.}~\bibnamefont {Heeres}}, \bibinfo
  {author} {\bibfnamefont {P.}~\bibnamefont {Reinhold}}, \bibinfo {author}
  {\bibfnamefont {Z.}~\bibnamefont {Leghtas}}, \bibinfo {author} {\bibfnamefont
  {B.}~\bibnamefont {Vlastakis}}, \bibinfo {author} {\bibfnamefont
  {Y.}~\bibnamefont {Liu}}, \bibinfo {author} {\bibfnamefont {L.}~\bibnamefont
  {Frunzio}}, \bibinfo {author} {\bibfnamefont {S.~M.}\ \bibnamefont {Girvin}},
  \bibinfo {author} {\bibfnamefont {L.}~\bibnamefont {Jiang}}, \bibinfo
  {author} {\bibfnamefont {M.}~\bibnamefont {Mirrahimi}}, \bibinfo {author}
  {\bibfnamefont {M.~H.}\ \bibnamefont {Devoret}},\ and\ \bibinfo {author}
  {\bibfnamefont {R.~J.}\ \bibnamefont {Schoelkopf}},\ }\bibfield  {title}
  {\bibinfo {title} {Extending the lifetime of a quantum bit with error
  correction in superconducting circuits},\ }\href
  {https://doi.org/10.1038/nature18949} {\bibfield  {journal} {\bibinfo
  {journal} {Nature}\ }\textbf {\bibinfo {volume} {536}},\ \bibinfo {pages}
  {441} (\bibinfo {year} {2016})}\BibitemShut {NoStop}%
\bibitem [{\citenamefont {Campagne-Ibarcq}\ \emph {et~al.}(2020)\citenamefont
  {Campagne-Ibarcq}, \citenamefont {Eickbusch}, \citenamefont {Touzard},
  \citenamefont {Zalys-Geller}, \citenamefont {Frattini}, \citenamefont
  {Sivak}, \citenamefont {Reinhold}, \citenamefont {Puri}, \citenamefont
  {Shankar}, \citenamefont {Schoelkopf}, \citenamefont {Frunzio}, \citenamefont
  {Mirrahimi},\ and\ \citenamefont {Devoret}}]{Campagne-Ibarcq2020}%
  \BibitemOpen
  \bibfield  {author} {\bibinfo {author} {\bibfnamefont {P.}~\bibnamefont
  {Campagne-Ibarcq}}, \bibinfo {author} {\bibfnamefont {A.}~\bibnamefont
  {Eickbusch}}, \bibinfo {author} {\bibfnamefont {S.}~\bibnamefont {Touzard}},
  \bibinfo {author} {\bibfnamefont {E.}~\bibnamefont {Zalys-Geller}}, \bibinfo
  {author} {\bibfnamefont {N.~E.}\ \bibnamefont {Frattini}}, \bibinfo {author}
  {\bibfnamefont {V.~V.}\ \bibnamefont {Sivak}}, \bibinfo {author}
  {\bibfnamefont {P.}~\bibnamefont {Reinhold}}, \bibinfo {author}
  {\bibfnamefont {S.}~\bibnamefont {Puri}}, \bibinfo {author} {\bibfnamefont
  {S.}~\bibnamefont {Shankar}}, \bibinfo {author} {\bibfnamefont {R.~J.}\
  \bibnamefont {Schoelkopf}}, \bibinfo {author} {\bibfnamefont
  {L.}~\bibnamefont {Frunzio}}, \bibinfo {author} {\bibfnamefont
  {M.}~\bibnamefont {Mirrahimi}},\ and\ \bibinfo {author} {\bibfnamefont
  {M.~H.}\ \bibnamefont {Devoret}},\ }\bibfield  {title} {\bibinfo {title}
  {Quantum error correction of a qubit encoded in grid states of an
  oscillator},\ }\href {https://doi.org/10.1038/s41586-020-2603-3} {\bibfield
  {journal} {\bibinfo  {journal} {Nature}\ }\textbf {\bibinfo {volume} {584}},\
  \bibinfo {pages} {368} (\bibinfo {year} {2020})}\BibitemShut {NoStop}%
\bibitem [{\citenamefont {Eickbusch}\ \emph {et~al.}(2022)\citenamefont
  {Eickbusch}, \citenamefont {Sivak}, \citenamefont {Ding}, \citenamefont
  {Elder}, \citenamefont {Jha}, \citenamefont {Venkatraman}, \citenamefont
  {Royer}, \citenamefont {Girvin}, \citenamefont {Schoelkopf},\ and\
  \citenamefont {Devoret}}]{eickbusch_fast_2022}%
  \BibitemOpen
  \bibfield  {author} {\bibinfo {author} {\bibfnamefont {A.}~\bibnamefont
  {Eickbusch}}, \bibinfo {author} {\bibfnamefont {V.}~\bibnamefont {Sivak}},
  \bibinfo {author} {\bibfnamefont {A.~Z.}\ \bibnamefont {Ding}}, \bibinfo
  {author} {\bibfnamefont {S.~S.}\ \bibnamefont {Elder}}, \bibinfo {author}
  {\bibfnamefont {S.~R.}\ \bibnamefont {Jha}}, \bibinfo {author} {\bibfnamefont
  {J.}~\bibnamefont {Venkatraman}}, \bibinfo {author} {\bibfnamefont
  {B.}~\bibnamefont {Royer}}, \bibinfo {author} {\bibfnamefont {S.~M.}\
  \bibnamefont {Girvin}}, \bibinfo {author} {\bibfnamefont {R.~J.}\
  \bibnamefont {Schoelkopf}},\ and\ \bibinfo {author} {\bibfnamefont {M.~H.}\
  \bibnamefont {Devoret}},\ }\bibfield  {title} {\bibinfo {title} {Fast
  universal control of an oscillator with weak dispersive coupling to a
  qubit},\ }\href {https://doi.org/10.1038/s41567-022-01776-9} {\bibfield
  {journal} {\bibinfo  {journal} {Nature Physics}\ }\textbf {\bibinfo {volume}
  {18}},\ \bibinfo {pages} {1464} (\bibinfo {year} {2022})}\BibitemShut
  {NoStop}%
\bibitem [{\citenamefont {Walter}\ \emph {et~al.}(2017)\citenamefont {Walter},
  \citenamefont {Kurpiers}, \citenamefont {Gasparinetti}, \citenamefont
  {Magnard}, \citenamefont {Poto\ifmmode~\check{c}\else \v{c}\fi{}nik},
  \citenamefont {Salath\'e}, \citenamefont {Pechal}, \citenamefont {Mondal},
  \citenamefont {Oppliger}, \citenamefont {Eichler},\ and\ \citenamefont
  {Wallraff}}]{WallraffPRA2017readout}%
  \BibitemOpen
  \bibfield  {author} {\bibinfo {author} {\bibfnamefont {T.}~\bibnamefont
  {Walter}}, \bibinfo {author} {\bibfnamefont {P.}~\bibnamefont {Kurpiers}},
  \bibinfo {author} {\bibfnamefont {S.}~\bibnamefont {Gasparinetti}}, \bibinfo
  {author} {\bibfnamefont {P.}~\bibnamefont {Magnard}}, \bibinfo {author}
  {\bibfnamefont {A.}~\bibnamefont {Poto\ifmmode~\check{c}\else
  \v{c}\fi{}nik}}, \bibinfo {author} {\bibfnamefont {Y.}~\bibnamefont
  {Salath\'e}}, \bibinfo {author} {\bibfnamefont {M.}~\bibnamefont {Pechal}},
  \bibinfo {author} {\bibfnamefont {M.}~\bibnamefont {Mondal}}, \bibinfo
  {author} {\bibfnamefont {M.}~\bibnamefont {Oppliger}}, \bibinfo {author}
  {\bibfnamefont {C.}~\bibnamefont {Eichler}},\ and\ \bibinfo {author}
  {\bibfnamefont {A.}~\bibnamefont {Wallraff}},\ }\bibfield  {title} {\bibinfo
  {title} {Rapid high-fidelity single-shot dispersive readout of
  superconducting qubits},\ }\href
  {https://doi.org/10.1103/PhysRevApplied.7.054020} {\bibfield  {journal}
  {\bibinfo  {journal} {Phys. Rev. Appl.}\ }\textbf {\bibinfo {volume} {7}},\
  \bibinfo {pages} {054020} (\bibinfo {year} {2017})}\BibitemShut {NoStop}%
\end{thebibliography}%

\appendix

\section{Dual-rail qubits}
We discuss here how to apply the measurement and gate constructions to dual-rail qubits. 
Dual-rail qubits differ from the other bosonic codewords we discuss because a logical qubit occupies two bosonic modes $(\hat{a}_1,\hat{b}_1)$ with codewords $\ket{0}_L=\ket{01}$ and $\ket{1}_L = \ket{10}_L$ 
Nonetheless, we can perform single qubit logical $Z$ gates by physically interacting with one of the bosonic modes in the dual-rail qubit. 
Explicitly, we can perform a $Z$ gate via $Z_L=e^{i\pi \hat{a}_1^\dagger \hat{a}_1}$  or equivalently via $Z_L=e^{i\pi \hat{b}_1^\dagger \hat{b}_1}$. 
This means that even though two dual-rail qubits comprise four physical modes, only two of them need to interact to perform logical two qubit gates and measurements.
If we define $(\hat{a}_2,\hat{b}_2)$ as the modes in a second dual-rail qubit, we can perform a logical $ZZ_L(\theta)$ gate by using an ancilla coupled to mode $\hat{a}_2$ and setting $\widehat{P} = e^{i\pi (\hat{a}_1^\dagger \hat{a}_1 + \hat{a}_2^\dagger \hat{a}_2)}$. 

Another distinction with the dual-rail code is the fact that all logical gates conserve the total number of excitations in the system. 
Arbitrary single qubit rotations in dual-rail qubits can be realized with the beamsplitter interaction between the modes $\hat{a}_i$ and $\hat{b}_i$. 
When combined with the $ZZ_L(\theta)$ gate this forms a universal gate set. In contrast, any bosonic code that uses only one bosonic mode per logical qubit by necessity requires gates that do not conserve the total number of excitations. 
E.g. an X gate in the Fock 01 code is $\hat{X}_{\text{Fock}}=\ket{0}\bra{1} + \ket{1}\bra{0}$ which involves transitions between states with different photon number whereas $\hat{X}_{\text{Dual-rail}}=\ket{01}\bra{10}+\ket{10}\bra{01}$ does not.

For our gate and measurement constructions to be applied to the Fock 01 or dual-rail code, the modes must still be bosonic, with the ability to support up to two excitations in each mode. 
This is because constructions rely on Hong-Ou-Mandel-like interference when we start in the state $\ket{11}_{a_1,a_2}$. 
The dual-rail code also has the ability to detect photon loss errors after the gate or measurement. One or both of the dual-rail qubits may end in the state $\ket{00}_{a_i,b_i}$

\begin{figure}[t]
\centering
    \begin{tabular}{c c}
    \includegraphics[width=1\linewidth]{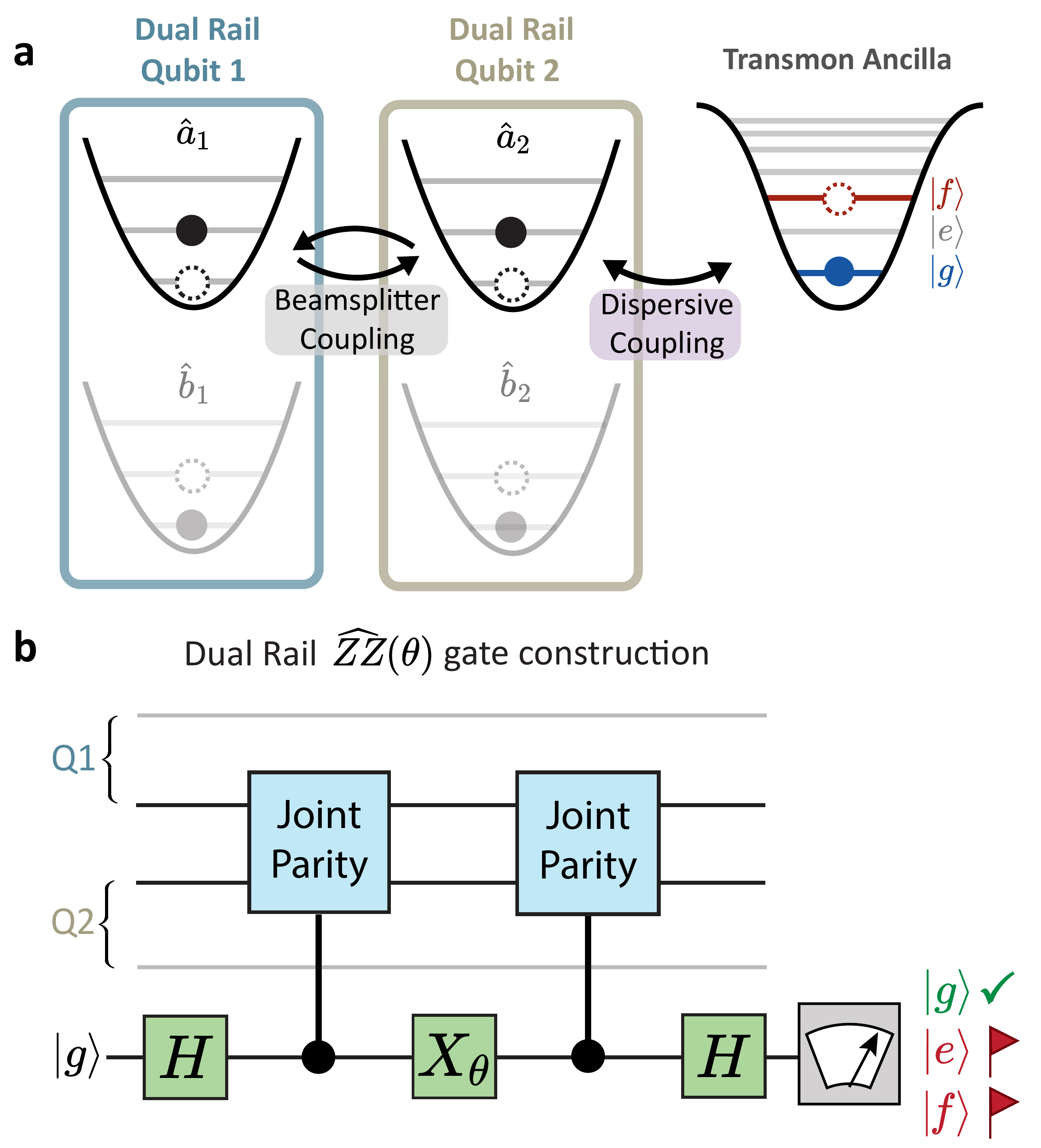}
    \end{tabular}
\caption{\textbf{Error-detected gates for dual-rail bosonic qubits} \textbf{(a)} Required hardware layout. Modes $(\hat{a}_1,\hat{b}_1)$ comprise qubit 1 and $(\hat{a}_2,\hat{b}_2)$ comprise qubit 2. An ancilla operated in the $gf$-manifold is dispersively coupled to $\hat{a}_2$ and we only need a beamsplitter interaction between modes $\hat{a}_1$ and $\hat{a}_2$ \textbf{(b)} Gate construction for a $ZZ_L(\theta)$ gate for dual-rail qubits. We need to engineer the ancilla-controlled joint-parity unitary acting on modes $\hat{a}_1$ and $\hat{a}_2$, where the joint parity operator is $e^{i\pi\left(\hat{a}_1^\dagger \hat{a}_1 + \hat{a}_2^\dagger \hat{a}_2\right)}$.}
\label{fig:DR qubits}
\end{figure}

\section{Interpretation of geometric phases enclosed on the operator Bloch sphere}
\label{app:interpretation_phase}

A powerful resource we use is the geometric phase enclosed by trajectories on the operator Bloch sphere. For a qubit Bloch sphere, this phase is often just an irrelevant global phase but in the operator Bloch sphere, this phase corresponds to performing a unitary of the form $\hat{R}_\phi=e^{i\phi(\hat{a}^\dagger\hat{a}+\hat{b}^\dagger\hat{b})}$. 
The mathematical reason behind this is because the effective angular momentum operator is given by $\hat{L}_I=\hat{a}^\dagger\hat{a}+\hat{b}^\dagger\hat{b}$, whereas for a qubit Bloch sphere, it takes the form $\hat{L}_I=\ket{0}\bra{0}+\ket{1}\bra{1}=\hat{\sigma}_0=\mathds{1}$. Enclosing phase $\phi$ on a ``Bloch sphere'' corresponds to performing the unitary $e^{i\phi \hat{L}_I}$ . This is a trivial global phase for the qubit Bloch sphere but a non-trivial unitary, $\hat{R}_\phi$ on the bosonic modes.
\section{Deriving useful ancilla-controlled unitaries from the dispersive beamsplitter Hamiltonian}
\label{app:family of primitive gates}

The dispersive beamsplitter Hamiltonian (Eq.~\ref{eq:detuned_bs}) allows for a wide variety of ancilla-controlled unitaries to be constructed for different settings of the Hamiltonian parameters $g$, $\Delta$, and $\varphi$. Even when these settings are fixed for the duration of the gate, we can realize several useful operations listed in Table~\ref{tab:primitives} and derived below.

\setlength{\tabcolsep}{8pt} 
\renewcommand{\arraystretch}{2} 

\begin{table*}
    \centering
    \begin{tabular}{C{3cm}C{2cm}C{2cm}C{2.5cm}C{2.5cm}C{2cm}}
        \toprule
        Operation & $g$ & $\Delta$ & Duration & $\vec{n}_{\ket{g}}$ & $\vec{n}_{\ket{f}}$\\
        \hline
        50:50 Beamsplitter & Any & $-\frac{\chi}{2}$ & $\frac{\pi}{2g}$ & $(1,0,0)$ & --- \\
        Control-SWAP & $\frac{1}{\sqrt{3}}|\chi|$ & $+\frac{\chi}{2}$ & $\frac{\sqrt{3}\pi}{|\chi|}$ & $\left(\frac{1}{2}, 0, \frac{\sqrt{3}}{2}\right)$ & $\left(1,0,0\right)$ \\
        \text{Control-joint parity} & $\frac{\sqrt{3}}{2}|\chi|$ & 0 & $\frac{2\pi}{|\chi|}$ & $\left(\frac{\sqrt{3}}{2}, 0, \frac{1}{2}\right)$ & $\left(\frac{\sqrt{3}}{2},0,-\frac{1}{2}\right)$ \\
        Control-joint 4-parity (slow) & $\frac{\sqrt{7}}{6}|\chi|$ & 0 & $\frac{3\pi}{|\chi|}$  & $\left(\frac{\sqrt{7}}{4}, 0, \frac{3}{4}\right)$ & $\left(\frac{\sqrt{7}}{4},0,-\frac{3}{4}\right)$ \\
        Control-joint 4-parity (fast) & $\frac{\sqrt{15}}{2}|\chi|$ & 0 & $\frac{\pi}{|\chi|}$  & $\left(\frac{\sqrt{15}}{4}, 0, \frac{1}{4}\right)$ & $\left(\frac{\sqrt{15}}{4},0,-\frac{1}{4}\right)$ \\
        \toprule
    \end{tabular}
\caption{\textbf{Hamiltonian parameters for useful ancilla-controlled unitaries.} The beamsplitter rate $g$, detuning $\Delta$, and gate duration are easily set by microwave drives in cQED hardware to match the conditions listed for useful operating points. The ancilla-state-dependent precession vectors $\vec{n}_{\ket{g}}$ and $\vec{n}_{\ket{f}}$ for the operator Bloch sphere are also listed. Controlled-joint 4-parity unitaries allow one to implement logical $ZZ_L(\theta)$ gates for bosonic codewords with 2-fold rotational symmetry, such as the binomial and 4-legged cat code.}
\label{tab:primitives}
\end{table*}

\begin{figure*}
    \centering
    \includegraphics[width=0.8\linewidth]{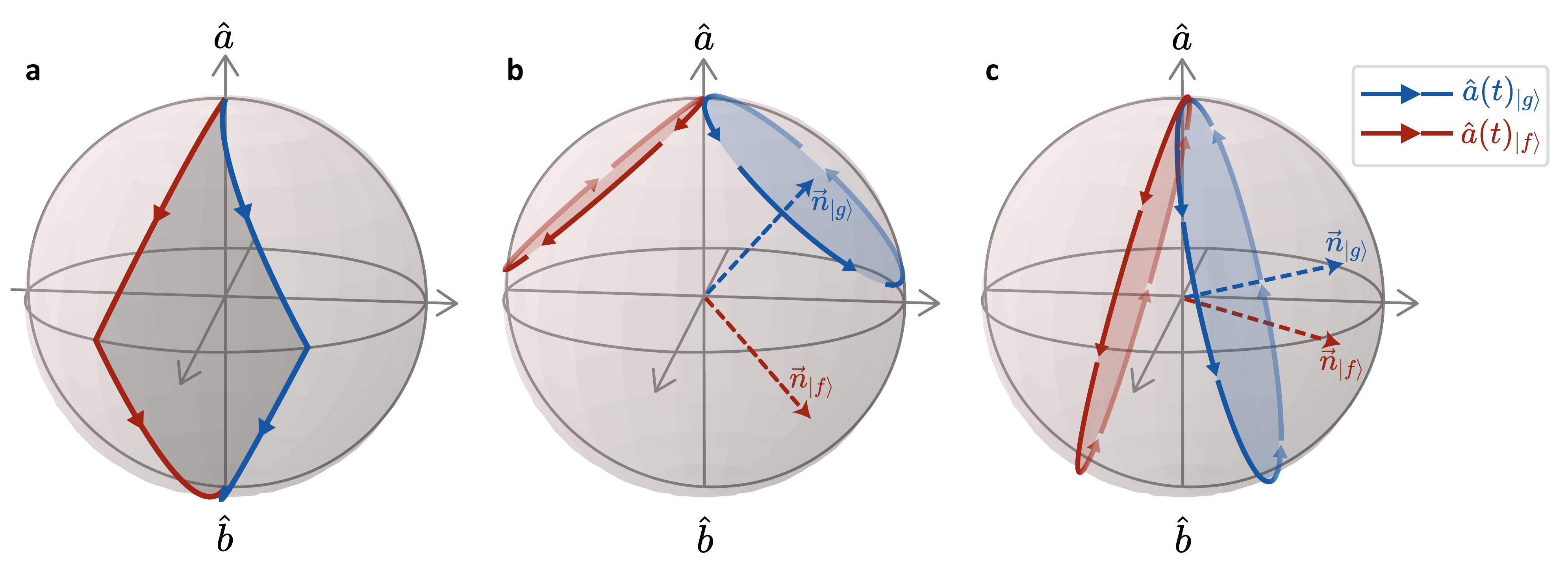}
    \caption{Operator Bloch spheres showing the trajectories for $\hat{a}(t)$ and precession vectors during \textbf{(a)} unconditional SWAP, \textbf{(b)} control-joint 4-parity (slow) and \textbf{(c)} control-joint 4-parity (fast). The trajectory in \textbf{(a)} is constructed piecewise from two different ancilla-controlled unitaries, where we perform a $\pi$-pulse on the ancilla at the halfway point to reverse the $\ket{g}$ and $\ket{f}$ states.}
    \label{fig:bonus_trajs}
\end{figure*}

\setlength{\tabcolsep}{6pt} 
\renewcommand{\arraystretch}{1} 

\subsection{Controlled SWAP}
For the ancilla-controlled SWAP unitary, we wish to realize
\begin{equation}
    \textbf{cSWAP} \equiv \ket{g}\bra{g}\mathbb{1} + \ket{f}\bra{f}\text{SWAP}.
\end{equation}
If the ancilla is in $\ket{f}$, we exchange the states in modes $\hat{a}$ and $\hat{b}$. If the ancilla is in $\ket{g}$, the states should be unaffected at the end of the operation. In the Heisenberg picture, one can write the desired mode transformations as
\begin{eqnarray}
\begin{pmatrix}
\hat{a}_g \\
\hat{b}_g \\
\end{pmatrix}
\rightarrow
\begin{pmatrix}
\hat{a}_g \\
\hat{b}_g \\
\end{pmatrix}
,\ \ \ \  
\begin{pmatrix}
\hat{a}_f \\
\hat{b}_f \\
\end{pmatrix}
\rightarrow
\begin{pmatrix}
\hat{b}_f \\
\hat{a}_f \\
\end{pmatrix}.
\end{eqnarray}
Setting $\Delta=+\frac{|\chi|}{2}$ ensures that $\vec{n}_{\ket{f}}$ lies on the equator and that after time $T=\pi/g$ the mode transformations ($\hat{a}_f\rightarrow -i\hat{b}_f,\hat{b}_f\rightarrow -i\hat{a}_f$) will have been carried out, which is a SWAP up to $90^\circ$ cavity rotations.\\
\\
We are still free to choose the parameter $g$. The goal is to find $g$ such that the state precesses around vector $\vec{n}_g=(g,0,\chi)/\sqrt{g^2+\chi^2}$ and such that $\hat{a}(t)$ returns to the pole after time $T=\frac{2\pi}{\sqrt{g^2+\chi^2}}$. 
\\
Setting these times to be equal and solving for $g$
\begin{equation}
    \frac{2\pi}{\sqrt{g^2+\chi^2}}=\frac{\pi}{g},
\end{equation}
gives $g=\chi/\sqrt{3}$ and duration $\sqrt{3}\pi/\chi$.
The resulting mode transformations are
\begin{eqnarray}
\begin{pmatrix}
\hat{a}_g \\
\hat{b}_g \\
\end{pmatrix}
\rightarrow
\begin{pmatrix}
e^{i\pi\left(1-\frac{\sqrt{3}}{2}\right)}\hat{a}_g\\
e^{i\pi\left(1-\frac{\sqrt{3}}{2}\right)}\hat{b}_g\\
\end{pmatrix}
,\ \ \ \  
\begin{pmatrix}
\hat{a}_f \\
\hat{b}_f \\
\end{pmatrix}
\rightarrow
\begin{pmatrix}
-i\hat{b}_f \\
-i\hat{a}_f \\
\end{pmatrix}
\end{eqnarray}
which is almost the desired \textbf{cSWAP}. To fix the unwanted phase accumulations, one can add delay times before and after the unitary. The dispersive interaction acting for time $t$ between mode $\hat{a}$ and the ancilla gives the mode transformations,
\begin{eqnarray}
\begin{pmatrix}
\hat{a}_g \\
\hat{b}_g \\
\end{pmatrix}
\rightarrow
\begin{pmatrix}
\hat{a}_g \\
\hat{b}_g \\
\end{pmatrix}
,\ \ \ \  
\begin{pmatrix}
\hat{a}_f \\
\hat{b}_f \\
\end{pmatrix}
\rightarrow
\begin{pmatrix}
e^{-i\chi t}\hat{a}_f \\
\hat{b}_f \\
\end{pmatrix}.
\end{eqnarray}
Adding a second delay after the ancilla-controlled unitary effectively implements a dispersive interaction between $\hat{b}_f$ and the ancilla since the modes have swapped. When one chooses $T=\frac{\pi}{2|\chi|}(3-\sqrt{3})$ the overall mode transformations are
\begin{eqnarray}
\begin{pmatrix}
\hat{a}_g \\
\hat{b}_g \\
\end{pmatrix}
\rightarrow
\begin{pmatrix}
e^{i\phi}\hat{a}_g \\
e^{i\phi}\hat{b}_g \\
\end{pmatrix}
,\ \ \ \  
\begin{pmatrix}
\hat{a}_f \\
\hat{b}_f \\
\end{pmatrix}
\rightarrow
\begin{pmatrix}
e^{i\phi}\hat{b}_f \\
e^{i\phi}\hat{a}_f \\
\end{pmatrix},
\end{eqnarray}
where $\phi = \pi\left(1-\frac{\sqrt{3}}{2}\right)$, which are the desired mode transformations for the \textbf{cSWAP} unitary up to the deterministic rotation  $\hat{R}_\phi=e^{i\phi\left(\hat{a}^\dagger\hat{a}+\hat{b}^\dagger\hat{b}\right)}$ (which can be easily tracked in software).
This trick can be used to adjust geometric phase accumulation on a single mode.\\
\\

\subsection{Controlled joint parity}

The joint parity operator $e^{i\pi\left(\hat{a}^\dagger\hat{a}+\hat{b}^\dagger\hat{b}\right)}$ is the $ZZ_L$ operator for many bosonic codes, such as the Fock 01, dual-rail, and two-legged cat code. The controlled joint parity unitary can be written as
\begin{equation}    \mathbf{cJP}=\ket{g}\bra{g}\mathds{1}+\ket{f}\bra{f}e^{i\pi\left(\hat{a}^\dagger\hat{a}+\hat{b}^\dagger\hat{b}\right)},
\end{equation}
up to the rotation operator $e^{-i\frac{\pi}{2}\left(\hat{a}^\dagger\hat{a}+\hat{b}^\dagger\hat{b}\right)}$, we can write the ``symmeterized'' controlled joint parity unitary
\begin{equation}
     \textbf{cJP}_2 = \ket{g}\bra{g} e^{i\frac{\pi}{2}\left(\hat{a}^\dagger \hat{a} + \hat{b}^\dagger \hat{b}\right)} + \ket{f}\bra{f} e^{-i\frac{\pi}{2}\left(\hat{a}^\dagger \hat{a} + \hat{b}^\dagger \hat{b}\right)},   
\end{equation}
with the desired mode transformations
\begin{eqnarray}
\begin{pmatrix}
\hat{a}_g \\
\hat{b}_g \\
\end{pmatrix}
\rightarrow
\begin{pmatrix}
i\hat{a}_g \\
i\hat{b}_g \\
\end{pmatrix}
,\ \ \ \  
\begin{pmatrix}
\hat{a}_f \\
\hat{b}_f \\
\end{pmatrix}
\rightarrow
\begin{pmatrix}
-i\hat{a}_f \\
-i\hat{b}_f \\
\end{pmatrix}.
\end{eqnarray}
We can obtain these mode transformations with precession vectors $\vec{n}_g$ and $\vec{n}_f$ that are antipodal to one another, by setting $\Delta = 0$. This also means that the corresponding trajectories have equal precession rates given by $\Omega=\sqrt{g^2+\chi^2}$ and return to the poles at the same time. The final step is to set the magnitude of $g$ such thata solid angle $\phi$ is enclosed. For a precession vector with polar angle $\theta$, the general formula for the solid angle is $\phi=4\pi(1-\cos(\theta))$. Therefore, for these precession vectors $|\phi|=4\pi(1-\frac{\chi}{2\Omega})=\frac{\pi}{2}$, which is solved by $g=\frac{\sqrt{3}}{2}\chi$ in time $T=2\pi/\chi$. The trajectory for this operating point is that which is shown in Fig.~\ref{fig:JP and CSWAP}a.
\\
\\
\\
\\
\\

\subsection{Controlled joint 4-parity}
The joint 4-parity operator, $e^{i\frac{\pi}{2}\left(\hat{a}^\dagger\hat{a}+\hat{b}^\dagger\hat{b}\right)}$ is the $ZZ_L$ operator for bosonic codes with 2-fold rotational symmetry such as the binomial and 4-legged cat code. There exist different choices for symmetrized control joint 4-parity unitaries, each with their own operating point:
\begin{equation}
     \textbf{cJP}_4^{\text{slow}} = \ket{g}\bra{g} e^{i\frac{\pi}{4}\left(\hat{a}^\dagger \hat{a} + \hat{b}^\dagger \hat{b}\right)} + \ket{f}\bra{f} e^{-i\frac{\pi}{4}\left(\hat{a}^\dagger \hat{a} + \hat{b}^\dagger \hat{b}\right)},
\end{equation}
or
\begin{equation}
     \textbf{cJP}_4^{\text{fast}} = \ket{g}\bra{g} e^{i\frac{3\pi}{4}\left(\hat{a}^\dagger \hat{a} + \hat{b}^\dagger \hat{b}\right)} + \ket{f}\bra{f} e^{-i\frac{3\pi}{4}\left(\hat{a}^\dagger \hat{a} + \hat{b}^\dagger \hat{b}\right)}.
\end{equation}
One choice results in a faster unitary than the other. Both perform the controlled joint 4-parity unitary, up to a rotation $\hat{R}_\phi$. In the slow case, a solid angle of $\pi/4$ must be enclosed, which is achieved for $g=\frac{\sqrt{7}}{6}\chi$ in time $3\pi/\chi$. In the fast case, a solid angle of $3\pi/4$ must be enclosed, which is achieved with $g=\frac{\sqrt{15}}{2}$ in time $\pi/\chi$. The operator Bloch sphere trajectories are shown in \ref{fig:bonus_trajs}a and \ref{fig:bonus_trajs}b.
\subsection{Ancilla-controlled unitaries at smaller $g/\chi$ ratios}
The ancilla-controlled unitaries we have considered thus far all rely on the ability to tune the magnitude of the beamsplitter rate $g$ to a specific value. In experiment, there may be restrictions that prevent one from reaching the required $g/\chi$ ratio. For each ancilla-controlled unitary, one can find alternate sets of Hamiltonian parameters that use a smaller $g/\chi$ ratio to implement an equivalent ancilla-controlled unitary but with a longer total duration. For example, to perform $\textbf{cZZ}$ one could instead perform two $\textbf{cJP}_4^{\text{slow}}$ unitaries back-to-back, which requires total gate duration $6\pi/\chi$ but approximately halves the required $g/\chi$ ratio. 

The same logic can be applied to the $\textbf{cSWAP}$ unitary. The trajectory for $\hat{a}_g(t)$ may complete many orbits in the time it takes the $\ket{a}_f(t)$ trajectory to reach the south pole of the operator Bloch sphere. The geometric phase accumulated depends on the number of orbits completed. When $\Delta$ is fixed to $\chi/2$ this condition may be written as
\begin{equation}
    T=\frac{\pi}{g}=\frac{2\pi n}{\sqrt{g^2+\chi^2}}
\end{equation}
For $n=1,2,3...$
\subsection{Unconditional SWAP}
Unconditional swap (\textbf{uSWAP}) refers to implementing the mode transformations
\begin{eqnarray}
\begin{pmatrix}
\hat{a}_g \\
\hat{b}_g \\
\end{pmatrix}
\rightarrow
\begin{pmatrix}
\hat{b}_g \\
\hat{a}_g \\
\end{pmatrix}
,\ \ \ \  
\begin{pmatrix}
\hat{a}_f \\
\hat{b}_f \\
\end{pmatrix}
\rightarrow
\begin{pmatrix}
\hat{b}_f \\
\hat{a}_f \\
\end{pmatrix}.
\end{eqnarray}
In which the bosonic modes are swapped without regard to the state in the ancilla. In the operator Bloch sphere it is straightforward to realize the related mode transformation
\begin{eqnarray}
\label{eq:uSWAP}
\begin{pmatrix}
\hat{a}_g \\
\hat{b}_g \\
\end{pmatrix}
\rightarrow
\begin{pmatrix}
e^{i\phi}\hat{b}_g \\
e^{i\phi}\hat{a}_g \\
\end{pmatrix}
,\ \ \ \  
\begin{pmatrix}
\hat{a}_f \\
\hat{b}_f \\
\end{pmatrix}
\rightarrow
\begin{pmatrix}
e^{-i\phi}\hat{b}_f \\
e^{-i\phi}\hat{a}_f \\
\end{pmatrix}.
\end{eqnarray}
With appropriate delays before and after the unitary one can again realize a ``true'' unconditional SWAP. 

One way to realize the unitary described by Eq.~\ref{eq:uSWAP} is to set $g>\chi$ and $\Delta = 0$. After some time, $T_{eq}$, the trajectories $\hat{a}_g(t)$ and $\hat{a}_f(t)$ should reach the equator of the operator Bloch sphere simultaneously. At this time, one disables the beamsplitter interaction and applies a $\pi$-pulse on the ancilla to flip the $\ket{f}$ and $\ket{g}$ states. If the beamsplitter interaction is re-enabled for time $T_{eq}$, both trajectories will now travel towards and meet at the south pole. The complete trajectories are illustrated in \ref{fig:bonus_trajs}a with the shaded area in between the trajectories equal to $2\phi$. 

Performing an unconditional SWAP between bosonic modes can enable an ancilla coupled to one mode to interact with more than two different bosonic modes. As an example, if $\textbf{uSWAP}_{1i}$ and $\textbf{cZ}_1$ is available, one can measure $\hat{Z}_1\hat{Z}_2\hat{Z}_3\hat{Z}_4$ stabilizers directly on four bosonic modes by first preparing the ancilla in $\ket{+}_{gf}$, implementing the unitary sequence $\textbf{cZ}_1-\textbf{uSWAP}_{12}-\textbf{cZ}_1-\textbf{uSWAP}_{13}-\textbf{cZ}_1-\textbf{uSWAP}_{14}
-\textbf{cZ}_1$, and finally measuring the ancilla in the $\ket{\pm}_{gf}$ basis.
\section{Realization in other experimental platforms}
\label{app:other platforms}
The operator Bloch sphere is applicable to other hardware platforms with access to a beamsplitter interaction and an ancilla-controlled operation on a two-bosonic mode system.
In the main text, we have considered Hamiltonian terms can be easily engineered in cQED,
\begin{eqnarray}
\{ \hat{a}^{\dag}\hat{b} + \hat{a}\hat{b}^{\dag}, \hat{a}^{\dag}\hat{a}, \hat{\sigma}^z\hat{a}^{\dag}\hat{a} \},
\end{eqnarray}
where the dispersive interaction with an ancilla imparts a state-dependent frequency shift upon one of the bosonic modes.
This effectively produces an ancilla-dependent detuning between two bosonic modes, which we leverage in all proposed ancilla-controlled unitaries.

Alternatively, one can consider a gate set that includes a conditional beamsplitter interaction:
\begin{eqnarray}
\{ \hat{\sigma}^z(\hat{a}^{\dag}\hat{b} + \hat{a}\hat{b}^{\dag}), \hat{a}^{\dag}\hat{b} + \hat{a}\hat{b}^{\dag}, \hat{a}^{\dag}\hat{a} \}.
\end{eqnarray}
This gate set is realized in trapped-ion systems where phononic modes serve as bosons and hyperfine states play the role of the ancilla \cite{ortiz-gutierrez_continuous_2017, gan_hybrid_2020, katz_programmable_2022}.
By setting the amplitude, phase, and detuning of the unconditional beamsplitter drive one can engineer ancilla-state-dependent precession vectors and trajectories on the operator Bloch sphere.

\section{Constructing an arbitrary, excitation-preserving two-qubit gate}
\label{app:family of logical gates}
With parameterized eSWAP$(\theta)$ and $ZZ_L(\theta)$ gate, one can construct any two-qubit gate that conserves the number of excitations in the encoded subspace.
Excitation-preserving two-qubit gates can be parameterized as follows: 

\begin{center}
\begin{quantikz}
& \gate{Z(\theta_1)} & \gate[2]{ZZ(\theta_3)} & \gate[2]{\textsc{SWAP}(\theta_4)} & \qw \\
& \gate{Z(\theta_2)} & & & \qw
\end{quantikz}
\end{center}
With particular choices of $\theta_1, \theta_2, \theta_3$ and $\theta_4$ one can generate useful gate families:

\vspace{5mm}
CPHASE$(\theta)$=
\begin{adjustbox}{height=0.05\textwidth}
\begin{quantikz}
& \gate{Z(-\frac{\theta}{2})} & \gate[2]{ZZ(\theta)} & \qw \\
& \gate{Z(-\frac{\theta}{2})} & & \qw
\end{quantikz}
\end{adjustbox}

\vspace{5mm}
iSWAP$(\theta)$=
\begin{adjustbox}{height=0.05\textwidth}
\begin{quantikz}
& \gate[2]{ZZ(-\theta)} & \gate[2]{\textsc{SWAP}(2\theta)} & \qw \\
& & & \qw
\end{quantikz}
\end{adjustbox}

\vspace{5mm}
fSim$(\theta, \phi)$=
\begin{adjustbox}{height=0.05\textwidth}
\begin{quantikz}
& \gate{Z\left(-\frac{\phi}{2}\right)} & \gate[2]{ZZ\left(-\theta+\frac{\phi}{2}\right)} & \gate[2]{\textsc{SWAP}(2\theta)} & \qw \\
& \gate{Z\left(-\frac{\phi}{2}\right)} & & & \qw
\end{quantikz}
\end{adjustbox}
$\text{CZ} = \text{CPHASE}(\frac{\pi}{2})$ is locally equivalent to a CNOT gate, which is often used as the basic two-qubit gate in general quantum circuits.
On the other hand, parameterized iSWAP$(\theta)$ and fSim$(\theta_1, \theta_2)$ are useful for efficiently compiling near-term algorithms, which leverages the excitation-conserving nature of the gate to simulate particular quantum chemistry problems whose electronic structure involves number-conserving symmetry \cite{kivlichan_quantum_2018}. 

\section{Hamiltonian engineering for error-corrected/error-detected gates}
\subsection{The error closure condition}
\label{app:error_closure}

What are the formal requirements for hardware errors that occur during a gate to be detectable or correctable at the end of the gate? Previous work on error-correctable gates relied on error transparency\cite{ma_error-transparent_2020}, but in this work we recognize that error transparency is a stronger condition than that which is necessary for a gate to be error-correctable. Here, we describe a new, less stringent condition which still guarantees that a gate is error-correctable, which we call ``error closure''. For clarity, we set $\hbar=1$ in this section.

First, we reiterate the requirements for a Hamiltonian $\widehat{\mathcal{H}}_0$ to be transparent to a set of hardware errors $\{\hat{\epsilon}\}_{\text{hardware}}$ that may occur at any time during the gate:
\begin{align}
    [\widehat{\mathcal{H}}_0,\hat{\epsilon}] &= 0,\\
    \forall \hat{\epsilon} &\in \{\hat{\epsilon}\}_{\text{hardware}}.
\end{align}
If one is able to correct for errors from $\{\hat{\epsilon}\}_{\text{hardware}}$ affecting idle qubits, error transparency ensures that one can also correct these errors if they occur during the gate.

This condition generalizes to detectable errors as well as correctable errors. $\{\hat{\epsilon}\}_{\text{hardware}}$ is error-detectable if one can perform syndrome measurements that indicate whether an error in the set occurred, but one cannot know (or cannot implement) the correction unitary on the states. This can occur when different errors yield the same error syndrome, or when the unknown time of a jump error means that one cannot know what the appropriate correction unitary should be. In the main text, we define a gate to be error-detectable if one can detect any one jump error from $\{\hat{\epsilon}\}_{\text{hardware}}$ occurring during the gate, via syndrome measurements after the gate.

There exist counterexamples (such as the $\widehat{ZZ}_L$ measurement) that are error-correctable operations which are not error-transparent. The error closure formalism is used to evaluate the effects of jump errors from $\{\hat{\epsilon}\}_{\text{hardware}}$. The effects of no-jump back-action are not included in the error closure formalism and we consider them separately in our gate and measurement constructions later on, although their effect is usually small.

We first define a larger set of errors $\{\hat{\epsilon}\}_{\text{corr}}$ that contains all the errors one can correct for. We assume that one has independent error correction operations (syndrome measurement and recovery) for each error in $\{\hat{\epsilon}\}_{\text{hardware}}$. It follows that if $\hat{\epsilon}_i,\hat{\epsilon}_j \in \{\hat{\epsilon}\}_{\text{hardware}}$, one can correct superpositions of errors such as $c_i\hat{\epsilon}_i+c_j\hat{\epsilon}_j$ and products of errors such as $\hat{\epsilon}_i\hat{\epsilon}_j$ for $i\neq j$. The fact that $\{\hat{\epsilon}\}_{\text{corr}}$ encompasses a larger set of errors than $\{\hat{\epsilon}\}_{\text{hardware}}$ is what allows us to relax the error transparency condition and search for a weaker condition for error correction. 

We now define our error closure conditions for Hamiltonian $\widehat{\mathcal{H}}_0$ and error set $\{\hat{\epsilon}\}_{\text{hardware}}$. First we generate a new set of errors $\{\hat{\epsilon}\}_{\text{ext}}$ from every possible commutator between $\widehat{\mathcal{H}}_0$ and the elements of $\{\hat{\epsilon}\}_{\text{hardware}}$:
\begin{equation}
\label{eq:ext_generator}
    [\widehat{\mathcal{H}}_0,\hat{\epsilon}_{\text{hardware}}]=\hat{\epsilon}_{\text{ext}},
\end{equation}
as well as linear combinations of these commutators.
Error closure is satisfied if:
\begin{enumerate}
    \item $\hat{\epsilon}_{\text{ext}}\in \{\hat{\epsilon}\}_{\text{corr}}$,
    \item $[\widehat{\mathcal{H}}_0,\hat{\epsilon}_{\text{ext}}]\in\{\hat{\epsilon}\}_{\text{ext}}$.
\end{enumerate}
These conditions state that the errors generated by Eq.~\ref{eq:ext_generator} must form a closed set of correctable errors and ensure that hardware errors during the gate remain correctable errors after the gate. We now sketch the proof.

Jump evolution for a hardware error $\hat{\epsilon}$ occurring at time $T-t$ during the gate evolution can be written as
\begin{equation}
e^{-i\widehat{\mathcal{H}}_0 t} \hat{\epsilon}e^{-i\widehat{\mathcal{H}}_0 (T-t)}=e^{-i\widehat{\mathcal{H}}_0 t} \hat{\epsilon}e^{i\widehat{\mathcal{H}}_0 t}e^{-i\widehat{\mathcal{H}}_0 T}.
\end{equation}
The condition for hardware errors to be correctable is:
\begin{align}
e^{-i\widehat{\mathcal{H}}_0 t} \hat{\epsilon}e^{i\widehat{\mathcal{H}}_0 t} & \in \{\hat{\epsilon}\}_{\text{corr}}, \\
\forall \hat{\epsilon} & \in \{\hat{\epsilon}\}_{\text{hardware}},\\
\forall t &\in [0,T].
\end{align}
Our error closure conditions ensure this is satisfied. From the Baker–Campbell–Hausdorff (BCH) theorem, we may write: 
\begin{align}
& e^{-i\widehat{\mathcal{H}}_0 t} \hat{\epsilon}e^{i\widehat{\mathcal{H}}_0 t}= \\
&\hat{\epsilon} + i [\widehat{\mathcal{H}}_0,\hat{\epsilon}]t - [\widehat{\mathcal{H}}_0,[\widehat{\mathcal{H}}_0,\hat{\epsilon}]]\frac{t^2}{2!} - i [\widehat{\mathcal{H}}_0,[\widehat{\mathcal{H}}_0,[\widehat{\mathcal{H}}_0,\hat{\epsilon}]]]\frac{t^3}{3!}+...    
\end{align}

The error closure conditions ensure the nested commutation relations only produce errors that are in $\{\hat{\epsilon}\}_{\text{corr}}$ and hence the entire Taylor series is also in $\{\hat{\epsilon}\}_{\text{corr}}$ since it is a superposition of correctable errors. If $\widehat{\mathcal{H}}_0^{(1)}$ and $\widehat{\mathcal{H}}_0^{(2)}$ both satisfy error closure then so does $\widehat{\mathcal{H}}_0^{(1)}+\widehat{\mathcal{H}}_0^{(2)}$. Note that satisfying the error transparency condition automatically satisfies the error closure conditions. It also follows that if a correctable error occurs before the gate, it will still be correctable after the gate. We now give explicit examples to show how this framework can be used.
\subsection{Example A. - Photon loss in beamsplitter interactions}
Suppose one wishes to perform a SWAP operation between two bosonic modes by setting $\widehat{\mathcal{H}}_0=\frac{g}{2}(\hat{a}^\dagger\hat{b}+\hat{a}\hat{b}^\dagger)$ and evolving for time $T=\pi/g$. Let us choose a bosonic encoding that allows us to correct single photon loss after the SWAP via photon number parity measurements, such as the binomial or 4-legged cat code. The hardware errors we consider are $\{\hat{\epsilon}\}_{\text{hardware}}=\{\hat{a},\hat{b}\}$ and the errors one can correct on idle qubits are $\{\hat{\epsilon}\}_{\text{corr}}= \{\hat{a},\hat{b},\hat{a}\hat{b},c_0\hat{a}+c_1\hat{b}+c_2\hat{a}\hat{b}\}$. We will show how one can correct for photon loss errors at intermediate times, even when outside of the logical codespace. From Eq.~\ref{eq:ext_generator} one finds
\begin{align}
    [\hat{a}^\dagger\hat{b}+\hat{a}\hat{b}^\dagger,\hat{a}]&=-\hat{b}, \\
    [\hat{a}^\dagger\hat{b}+\hat{a}\hat{b}^\dagger,\hat{a}]&=-\hat{a},
\end{align}
but when evaluating the commutator with error $\hat{a}\hat{b}$ one finds

\begin{equation}   
    \left[\hat{a}^\dagger\hat{b}+\hat{a}\hat{b}^\dagger,\hat{a}\hat{b} \right] =-\hat{b}^2-\hat{a}^2
\end{equation}

which generates errors outside the set of correctable errors and so $\hat{a}\hat{b}$ is not in $\{\hat{\epsilon}\}_{\text{ext}}$. 
After evaluating all the commutators one finds $\{\hat{\epsilon}\}_{\text{ext}}=\{c_0\hat{a}+c_1\hat{b}\}\subset\{\hat{\epsilon}\}_{\text{corr}}$\\
\\
Thus $\widehat{\mathcal{H}}_0$ and $\{\hat{\epsilon}\}_{\text{hardware}}$ satisfy the error closure conditions for the SWAP operations. This means that one can correct for $\hat{a}$ and $\hat{b}$ errors during the gate without requiring error transparency and despite evolving outside the codespace.

\subsection{Example B. - Ancilla-controlled unitaries generated by $\widehat{\mathcal{H}}_{\chi\text{BS}}$}
The previously-mentioned gate and measurement constructions have favorable error detection/correction properties because the Hamiltonians used to generate the ancilla-controlled unitaries satisfy error closure; here we show why. We consider hardware errors $\{\hat{\epsilon}\}_{\text{hardware}}=\{\hat{a},\hat{b},\hat{\sigma}_z^{gf}\}$ where transmon ancilla decay is treated separately by using the $gf$ manifold.

We set $\widehat{\mathcal{H}}_0=\frac{g}{2}(\hat{a}^\dagger\hat{b}+\hat{a}\hat{b}^\dagger)+(\Delta+\frac{\chi}{2}\hat{\sigma}_z^{gf}) \hat{a}^\dagger \hat{a}$.
With an appropriate bosonic code one can detect photon loss via parity measurements and ancilla dephasing via repeated measurements or by using $\ket{f}$ as a flag state in the proposed gate construction, so
\begin{eqnarray}
\{\hat{\epsilon}\}_{\text{corr}}=\{\hat{a},\hat{b},\hat{\sigma}_z^{gf},\hat{a}\hat{\sigma}_z^{gf},\hat{b}\hat{\sigma}_z^{gf},\hat{a}\hat{b},\hat{a}\hat{b}\hat{\sigma}_z^{gf}\}.
\end{eqnarray}
Where we now omit the linear combinations from $\{\hat{\epsilon}\}_{\text{corr}}$ for clarity.
By writing out the commutation relations
\begin{align}
    \left[\widehat{\mathcal{H}}_0,\hat{a}\right]&=-\frac{g}{2}\hat{b}-(\Delta+\frac{\chi}{2}\hat{\sigma}_z^{gf})\hat{a},\\
    \left[\widehat{\mathcal{H}}_0,\hat{b}\right]&=-\frac{g}{2}\hat{a},\\
    \left[\widehat{\mathcal{H}}_0,\hat{\sigma}_z^{gf}\right]&=0.
\end{align}
We begin to find the elements in $\{\hat{\epsilon}\}_{\text{ext}}$. We calculate the next order of commutators, $[\widehat{\mathcal{H}}_0,[\widehat{\mathcal{H}}_0,\hat{\epsilon}]]$
to arrive at the closed error set
\begin{equation}
    \{\hat{\epsilon}\}_{\text{ext}}=\{\hat{a},\hat{b},\hat{\sigma}_z^{gf},\hat{a}\hat{\sigma}_z^{gf},\hat{b}\hat{\sigma}_z^{gf}\}\subset\{\hat{\epsilon}\}_{\text{corr}},
\end{equation}
and so the error closure conditions are satisfied for each of the ancilla-controlled unitaries $\hat{U}_c$. 

Hardware errors midway through $\hat{U}_c$ can lead to complicated errors. For example, photon loss at an unknown time results in the operator $(c_0\hat{a}+c_1\hat{b})e^{i\varphi(\ket{g}\bra{g}-\ket{f}\bra{f})}\hat{U}_c$ being applied to the system, where $c_0,c_1,\varphi$ all depend on the exact time of the photon loss. Despite not knowing this time, the error can still be corrected. 

No-jump backaction associated with photon loss does not form part of the error closure formalism but can be evaluated by considering whether the photon number populations depend on the ancilla states $\ket{g}$ and $\ket{f}$. For $\textbf{cZZ}_L$ unitaries, the photon number distributions of the bosonic modes are independent of the ancilla states. Visually, this means that trajectories on the operator Bloch sphere have the same lattitude at all points in time, and hence no-jump backaction due to photon loss or ancilla decay is absent. This is not true for \textbf{cSWAP}, although it is generally a small effect. 
\subsection{Suitable Hamiltonians for error closure}
One can engineer many different Hamiltonian terms on bosonic modes via processes such as four-wave mixing with a transmon in cQED. When exploring exclusively bosonic codes designed to correct discrete photon loss errors, we find that only the lowest order interactions are suitable as detailed in Table~\ref{tab:closure}. 

\setlength{\tabcolsep}{4pt} 
\renewcommand{\arraystretch}{1.5}

\begin{table}[h]
\begin{tabular}{ccc}
\toprule
$\widehat{\mathcal{H}}_0$ & $[\widehat{\mathcal{H}}_0,a]$ &$\widehat{\mathcal{H}}_0$ and $\{a,b\}$ satisfy error closure? \\ \hline
$a^\dagger a$ & $a$ & \ding{51} \\ 
$a + a^\dagger$ & 1 & \ding{51}\\ \hline
$a b^\dagger + a^\dagger b$ & $b$ & \ding{51}\\ 
$a^\dagger b^\dagger + ab$ & $b^\dagger$ & \ding{55}\\
${a^\dagger}^2 + a^2$ & $2a^\dagger$ & \ding{55}\\ \hline
$a^\dagger a (b + b^\dagger)$ & $a(b + b^\dagger)$ & \ding{55}\\ $(a + a^\dagger) b^\dagger b$ & $b^\dagger b$ & \ding{55}\\ \toprule
\end{tabular}
\caption{\textbf{Error closure for candidate Hamiltonians for bosonic codes designed to protect from single photon loss}. For bosonic codes designed to correct against single photon loss, we can evaluate the commutator $[\widehat{\mathcal{H}}_0,\hat{a}]$ to see if the Hamiltonian could satisfy error closure. We find only the lowest order Hamiltonians make good candidates for constructing gates. In the table, if $\widehat{\mathcal{H}}_0$ is a good candidate, so is $\widehat{\mathcal{H}}_0\otimes\hat{\sigma}_z^{gf}$
}
\label{tab:closure}
\end{table}

\setlength{\tabcolsep}{6pt} 
\renewcommand{\arraystretch}{1} 
\section{Error-corrected measurements}
\label{app:error-corrected measurement}
Here we examine the measurements that can be constructed from ancilla-controlled unitaries and the errors which can be corrected. 
\begin{figure}[t]
\centering
    \begin{tabular}{c c}
    \includegraphics[width=1\linewidth]{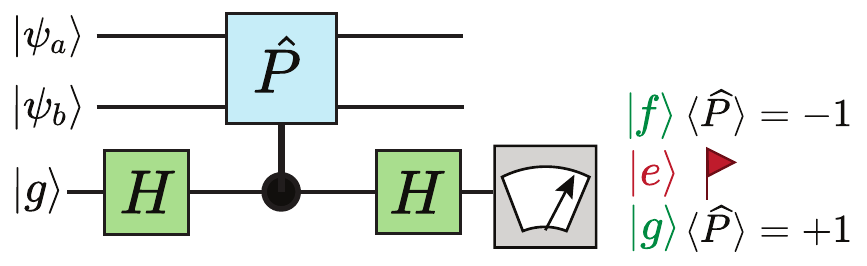}
    \end{tabular}
\caption{Circuit for performing QND measurements of binary-valued operator $\widehat{P}$ from ancilla-controlled unitaries. If $\chi$-matching is satisfied, we can error correct the dominant errors in the ancilla and bosonic modes.}
\label{fig:EC_msmts}
\end{figure}
The circuit used to perform this measurement is shown in Fig. \ref{fig:EC_msmts}. If $\hat{P}$ is a binary-valued operator with eigenvalues $\pm1$, this circuit performs a QND measurement of $\hat{P}$. 

The simpler structure of this circuit means that one can both detect \emph{and} correct errors from $\{\hat{a},\hat{b},\hat{\sigma}_z^{gf},\ket{e}\bra{f}\}$ that occur during the circuit. This measurement can be considered as the two-mode extension to the fault-tolerant parity measurement presented in \cite{rosenblum_fault-tolerant_2018}. 

To correct ancilla decay $\ket{e}\bra{f}$, one must now also engineer ``$\chi$-matching'' ($\chi=\chi_{ge}=\chi_{gf}$) as in \cite{rosenblum_fault-tolerant_2018,reinhold_error-corrected_2020}. If $\ket{e}$ is measured at the end of the circuit then decay likely happened during $\hat{U}_c$. By contextuality this means the ancilla started in $\ket{f}$ at the beginning of $\hat{U}_c$. $\chi$-matching ensures that one knows with certainty which unitary has been applied to the bosonic modes even though the precise time of the decay is unknown. This unitary is the one generated by $\widehat{\mathcal{H}}_f = \frac{g_{\mathrm{bs}}}{2}(\hat{a}^\dagger \hat{b} + \hat{a} \hat{b}^\dagger) + \left(\Delta + \frac{\chi}{2}\right) \hat{a}^\dagger \hat{a}=\widehat{\mathcal{H}}_e$ acting for the duration of $\hat{U}_c$. For measurements where $\widehat{P}=$SWAP or $ZZ_L$, this will either be identity, SWAP, or a unitary of the form $e^{i\phi\left( \hat{a}^\dagger \hat{a} +\hat{b}^\dagger \hat{b}\right)}$, all of which are straightforward to correct for in cQED. Afterwards one can retry the measurement. We now describe how to correct the remaining errors in $\{\hat{\epsilon}\}_{\text{hardware}}=\{\hat{a},\hat{b},\hat{\sigma}_z^{gf}\}$

We first consider errors from this set occurring during $\hat{U}_c$, the longest duration unitary in the measurement circuit. As a sequence of successive unitaries applied right to left, the gate sequence reads 
\begin{equation}
e^{-i\frac{\pi}{4}Y_{\text{gf}}}\hat{U}_c e^{i\frac{\pi}{4}Y_{\text{gf}}}.
\end{equation}
Hardware errors during $\hat{U}_c$ result in a sequence of unitaries equivalent to 
\begin{equation}
e^{-i\frac{\pi}{4}Y_{\text{gf}}}\hat{\epsilon}\hat{U}_c e^{i\frac{\pi}{4}Y_{\text{gf}}}
 = \hat{\epsilon}’ e^{-i\frac{\pi}{4}Y_{\text{gf}}}\hat{U}_c e^{i\frac{\pi}{4}Y_{\text{gf}}},
\end{equation}
where $\hat{\epsilon}'$ can be a superposition of errors from the set $\{\hat{a},\hat{b},\hat{a}\hat{\sigma}_x^{gf},\hat{b}\hat{\sigma}_x^{gf}\}$ which is also a set that satisfies the error closure condition for $\hat{\mathcal{H}}_{\chi\text{BS}}$.
(Note: commuting through $\pi/2$-pulses changes $\hat{\sigma}_z^{gf}$ for $\hat{\sigma}_x^{gf}$, meaning the measurement outcomes $\ket{g}$ and $\ket{f}$ can be flipped).
In words, this means ancilla dephasing during $\hat{U}_c$ does not affect the bosonic modes but results in a high chance (50\%) of observing $\ket{f}$ instead of $\ket{g}$ and vice versa.

Photon loss errors during $\hat{U}_c$ are equivalent to ancilla dephasing during the measurement accompanied by photon loss occurring after the measurement, which can be detected/tracked\cite{ofek_extending_2016} via fault-tolerant parity measurements \cite{rosenblum_fault-tolerant_2018} and then corrected.

For completeness, we also consider ancilla dephasing during the short ancilla rotations, which is equivalent to measuring/initializing the ancilla in the wrong basis. Again, the $\ket{g}$ and $\ket{f}$ measurement outcomes may be flipped but a backaction of the form $c_0\mathds{1} + c_1 \widehat{P}$ may also be imparted upon the bosonic modes for unknown $c_0$ and $c_1$. Once again, this error can be corrected by repeating the measurement until correctly projecting into an eigenstate of $\widehat{P}$. Importantly, this backaction conserves the $\widehat{P}$ eigenvalue which makes this error correctable. In general, $g-f$ measurement errors in the measurement construction are correctable by repeating the measurement and taking a majority vote on the outcomes. 

As an example, we describe how one might perform an error-corrected $ZZ_L$ measurement on two qubits encoded in the binomial code. First, one repeats $ZZ_L$ measurements until observing enough $\ket{g}$ or $\ket{f}$ outcomes to perform a majority vote and obtain a $ZZ_L$ eigenvalue. Then, one must check that no photon loss occurred during the measurements, which would invalidate the majority vote. This can be done by performing  parity measurements on individual modes.

If both modes are even parity, the measured $ZZ_L$ eigenvalue will be reliable (the $ZZ_L$ measurements only yield useful information if both bosonic modes are in even-photon-number states). If any modes have an odd number of photons, one must apply the error correction unitary for photon loss in the binomial code and repeat the entire measurement again until satisfying the photon loss checks. Although the sequence involves many measurements to majority vote on both the $ZZ_L$ measurements and parity measurements, this sequence corrects errors in $\{\epsilon\}_{\text{hardware}}$. 

\section{The error-detected gate construction}
\label{app:error detected gate construction}
Despite using the same $\hat{U}_c$ building blocks as the measurements, the gate construction can fundamentally only detect errors in $\{\epsilon\}_{\text{hardware}}$, not correct them. This is because of the way ancilla dephasing errors commute through the entire gate construction. Depending on whether they happen in the first or second $\hat{U}_c$, they have the same flag state $\ket{f}$ but impart different unitaries on the bosonic modes. Since we are designing a gate that only detects ancilla errors, we do not need to engineer $\chi$-matching. Whenever the ancilla is observed to be in $\ket{e}$ at the end of the gate, a decay error (which may have occurred at any point during the gate construction) is indicated. No-jump backaction associated with the $\ket{e}\bra{f}$ decay can be mitigated by over-rotating the ancilla by slightly more than $\pi/2$ in the first ancilla rotation.
The gate construction ideally performs the unitary
\begin{equation}
    \widehat{U}_{\text{gate}}(\theta)\otimes\ket{g}\bra{g}+\widehat{U}_{\text{gate}}(-\theta)\otimes\ket{f}\bra{f},
\end{equation}
where $\widehat{U}_{\text{gate}}(\theta)$ is the desired entangling gate on the bosonic modes. By sequentially applying the sequence of unitaries to a system with an ancilla starting in the ground state
\begin{equation}
e^{-i\frac{\pi}{4}\hat{Y}_{\text{gf}}}\widehat{U}_ce^{-i\frac{\theta}{2}\hat{X}_{\text{gf}}}\widehat{U}_ce^{i\frac{\pi}{4}\hat{Y}_{\text{gf}}}.
\end{equation}
We now investigate the effects of ancilla dephasing. Photon loss errors also dephase the ancilla, so analyzing this error resolves the issue of error propagation resulting from the hardware errors $\{\hat{a},\hat{b},\hat{\sigma}_z^{gf}\}$.
Ancilla dephasing $\hat{\sigma}_z^{gf}$ that happens during the first $\widehat{U}_c$ results in the overall gate unitary
\begin{align}
  &e^{-i\frac{\pi}{4}\hat{Y}_{\text{gf}}}\widehat{U}_ce^{-i\frac{\theta}{2}\hat{X}_{\text{gf}}}\widehat{U}_c\hat{\sigma}_z^{gf}e^{i\frac{\pi}{4}\hat{Y}_{\text{gf}}}\\&= e^{-i\frac{\pi}{4}\hat{Y}_{\text{gf}}}\widehat{U}_ce^{-i\frac{\theta}{2}\hat{X}_{\text{gf}}}\widehat{U}_ce^{i\frac{\pi}{4}\hat{Y}_{\text{gf}}}\hat{\sigma}_x^{gf}\\
&=\widehat{U}_{\mathrm{gate}}(\theta)\hat{\sigma}_x^{gf}\\
&=\hat{\sigma}_x^{gf}\widehat{U}_{\mathrm{gate}}(-\theta).
\end{align}
In other words, the ancilla ends in $\ket{f}$ and the incorrect unitary $\widehat{U}_{\mathrm{gate}}(-\theta)$ is applied to the bosonic modes.

Dephasing that occurs during the second $\hat{U}_c$ results in the overall gate unitary $\hat{\sigma}_z\widehat{U}_{\mathrm{gate}}(\theta)$. The ancilla ends in $\ket{f}$ and the correct gate unitary $\widehat{U}_{\mathrm{gate}}(\theta)$ is applied. From these two scenarios, if the ancilla is observed to be in $\ket{f}$ it is unknown whether the unitary $\widehat{U}_{\mathrm{gate}}(-\theta)$ or the unitary $\widehat{U}_{\mathrm{gate}}(\theta)$ was applied, this error is only detectable.

We also show that one can still detect ancilla dephasing even if it occurs during any of the three short ancilla rotations. This means that the error-detected gate fidelity should not be limited by ancilla decoherence during the rotations (i.e. transmon gate fidelities in cQED). If ancilla dephasing happens during the final $e^{-i\frac{\pi}{4}\hat{Y}_{\text{gf}}}$ rotation, the ancilla and bosonic modes are disentangled by this stage. One still performs $\widehat{U}_{\mathrm{gate}}(\theta)$ correctly but there is now a probability that the ancilla is detected in $\ket{f}$. Dephasing during the first $e^{i\frac{\pi}{4}\hat{Y}_{\text{gf}}}$ rotation is equivalent to starting the ancilla in a random $c_0\ket{g}+c_1\ket{f}$ superposition. Thus, when measuring the ancilla in $\ket{g}$, the correct gate unitary will have also been performed. Dephasing during the middle pulse $e^{-i\frac{\theta}{2}\hat{X}_{\text{gf}}}$ rotation results in applying the gate $\widehat{U}_{\text{gate}}({\theta_\text{random}})$, where $\theta_{\text{random}}$ depends on the exact time of the dephasing jump during the middle pulse. This error is also flagged because the ancilla will also end in the $\ket{f}$ state.

The gate construction is not robust to second-order ancilla dephasing. On roughly half of the occasions where two $\hat{\sigma}_z$ jumps occur during the gate, they will occur during different ancilla-controlled unitaries. The net result is that the incorrect unitary $\hat{U}(-\theta)$ is applied to the bosonic modes \emph{and} the ancilla is measured in state $\ket{g}$ at the end of the gate, so this error is not flagged. 

Photon loss errors also cause dephasing on the ancilla \cite{rosenblum_fault-tolerant_2018} and thus can only be detected at the end of the gate construction (e.g. via parity measurements).
\section{Using the gate construction in other contexts}
The construction in Fig. \ref{fig:exponential_gadget} can be used to generate many other useful error-detectable gates on bosonic modes. As a simple example, one can set $P_L=Z_L$ and engineer the \textbf{cZ} ancilla-controlled unitary. This is straightforward to achieve in cQED with the dispersive interaction and a wait time $T$. For bosonic codes with $n$-fold rotational symmetry, this is realized with $T=\pi/n\chi$. With this, one may implement $Z_L(\theta)$ gates on rotationally-symmetric codes as an alternative to SNAP gates\cite{reinhold_error-corrected_2020,heeres_cavity_2015}. When $n=1$, $P_L = \text{Parity}=e^{i\pi\hat{a}^\dagger \hat{a}}$ (which was used to implement the eSWAP$(\theta)$ gate in \cite{gao_entanglement_2019}). It follows that this realization has the same error-detection properties as the gate construction presented here.

The {$P_L(\theta)$ gate construction can also be used on GKP codewords. With conditional displacement Hamiltonians \cite{Campagne-Ibarcq2020} one can engineer \textbf{cZ}, \textbf{cZZ}, \textbf{cX}, \textbf{cXX} and thus implement $Z_L(\theta), ZZ_L(\theta), X_L(\theta), XX_L(\theta)$ logical rotations with the ability to detect ancilla errors. This construction resembles some of the circuits found in echoed conditional displacement control \cite{eickbusch_fast_2022}. In cQED, the controlled displacement is a highly switchable interaction that allows for the direct implementation of ancilla-controlled unitaries without the need for the operator Bloch sphere picture. 

\section{Simulation details}
\label{app:simulation}
A $ZZ_L(\theta)$ gate is simulated by numerically solving the Lindblad master equation under the static dispersive Hamiltonian and the controls needed to realize the constituent operations (specifically, beamsplitters and transmon pulses). 
\begin{equation}
    \begin{split}
        \widehat{\mathcal{H}}_{\chi \text{BS}} = \widehat{\mathcal{H}}_{\chi}+\widehat{\mathcal{H}}_{\text{BS}}(t)+\widehat{\mathcal{H}}_{\text{T}}(t).
    \end{split}
\end{equation}

The static Hamiltonian describes the dispersive coupling to the ancilla with a frequency shift for each ancilla basis state,

\begin{equation}
    \begin{split}
    \widehat{\mathcal{H}}_{\chi} / \hbar = -\hat{a}^\dagger \hat{a}\Big(\frac{\chi_{f}}{2} &\ket{g}\bra{g} \\
        + (\frac{\chi_{f}}{2}-\chi_{e}) &\ket{e}\bra{e} \\
        - \frac{\chi_{f}}{2}&\ket{f}\bra{f}\Big),
    \end{split}
\end{equation}
where we choose $\chi_{e}/2\pi= -0.5 \text{ MHz}$, $\chi_{f}/2\pi= -1 \text{ MHz}$. We have written this Hamiltonian in a frame where the dispersive interaction for $\ket{g}$ and $\ket{f}$ is symmetric.  
We assume the time-dependent beamsplitter and transmon drives to be piecewise-constant throughout the protocol and constant when realizing each unitary in the gate construction. This allows us to neglect the effects of any particular choice of pulse shape, since we wish to highlight how the overall protocol fidelity scales with various error rates. Furthermore, because the beamsplitter and transmon pulses are never simultaneous, we can define distinct time-independent Hamiltonians corresponding to beamsplitter operations and transmon operations: 
\begin{equation}
    \begin{split}
        \widehat{\mathcal{H}}_{\text{BS}}/\hbar &=\frac{g}{2}\left(\hat{a}^\dagger\hat{b} + \hat{a}\hat{b}^\dagger\right) + \Delta \hat{a}^\dagger\hat{a} \\
        \widehat{\mathcal{H}}_T/\hbar &= \epsilon_x \hat{\sigma}_x^{gf} + \epsilon_y \hat{\sigma}_y^{gf}, 
    \end{split}
\end{equation}

where $\epsilon_{x}$ and $\epsilon_{y}$ are the drive strengths of the two control quadratures coupled to the $g-f$ manifold Pauli operators $\hat{\sigma}_x^{gf} = |f \rangle \langle g| + |g \rangle \langle f|$ and $\hat{\sigma}_y^{gf} = i|f \rangle \langle g| - i|g \rangle \langle f|$. Because the $ZZ_L(\theta)$ gate uses $\textbf{cZZ}_L$, we take $\Delta=0$ throughout the gate sequence. During transmon operations, we neglect the dispersive coupling, as this can also be compensated for with appropriate pulse shaping.

We then use this Hamiltonian in the Lindblad master equation:

\begin{equation}
    \frac{d\rho}{dt} = -i\left[\widehat{\mathcal{H}}, \rho \right] + \Gamma_1^T \mathcal{D}[\hat{t}]\rho + \Gamma_\phi^T\mathcal{D}[\hat{t}^\dagger \hat{t}]\rho + \Gamma_1^C\mathcal{D}[\hat{a}]\rho,
\end{equation}
where $\hat{t} = |g\rangle \langle e| + \sqrt{2} |e \rangle \langle f|$ is the annihilation operator for the transmon mode and  $\mathcal{D[\hat{L}]\rho} = \hat{L} \rho \hat{L}^\dagger - \frac{1}{2}\hat{L}^\dagger \hat{L} \rho - \frac{1}{2} \rho \hat{L}^\dagger \hat{L} $ is the usual Lindblad dissipator.

The Lindblad equation can also be expressed in terms of a Liouvillian $\widehat{\mathcal{L}}$ as

\begin{equation}
    \frac{d\rho}{dt} = -i\widehat{\mathcal{L}}\rho,
\end{equation}
where $\widehat{\mathcal{L}}\rho = \left[\widehat{\mathcal{H}}, \rho \right] + i\Gamma_1^T \mathcal{D}[\hat{t}]\rho + i\Gamma_\phi^T\mathcal{D}[\hat{t}^\dagger \hat{t}]\rho + i\Gamma_1^C\mathcal{D}[\hat{a}]\rho$. Because the transmon and beamsplitter drives are either enabled or disabled for each of the five steps in a given protocol, we can express the final state density matrix after the whole sequence described in Fig.~\ref{fig:exponential_gadget} as
\begin{equation}
    \rho^f = \widehat{\mathcal{U}}_{T_3} \widehat{\mathcal{U}}_{B_2}  \widehat{\mathcal{U}}_{T_2}  \widehat{\mathcal{U}}_{B_1}  \widehat{\mathcal{U}}_{T_1} 
    \rho^i,
\end{equation}
where each $\widehat{\mathcal{U}}_j = e^{-i \widehat{\mathcal{L}}_j t}$ is the (generally \emph{non-unitary}) propagator under the time-independent Liouvillian corresponding to a transmon pulse $\widehat{\mathcal{U}}_{T}$ or beamsplitter pulse $\widehat{\mathcal{U}}_{B}$ in the presence of errors.

We then prepare the 36 cardinal states of the joint logical space $\rho^i_k$ as

\begin{equation}
    \begin{split}
        \{ \rho^i: & ~|\psi\rangle \otimes |\phi\rangle ~\forall ~ |\psi\rangle , |\phi\rangle \in \\
        \{ & |0_L\rangle, |1_L\rangle, \\
        & \frac{1}{\sqrt{2}}\left(|0_L\rangle + |1_L\rangle\right), \frac{1}{\sqrt{2}}\left(|0_L\rangle - |1_L\rangle\right), \\
        & \frac{1}{\sqrt{2}}\left(|0_L\rangle + i|1_L\rangle\right), \frac{1}{\sqrt{2}}\left(|0_L\rangle - i|1_L\rangle\right)\} \}
    \end{split}
\end{equation}
and simulate the evolution of each one under the sequence of Liouvillians for a given protocol, yielding a final density operator $\rho^f_k$. We choose these states because the average trace fidelity over these states yields the trace fidelity over entire joint logical subspace \cite{nielsen_simple_2002}. 

Next, we simulate measurement of the ancilla. In a small fraction of measurements, one will erroneously measure an ancilla in the $|e\rangle$ or $|f\rangle$ states as being in $|g\rangle$. Therefore, to simulate imperfect ancilla measurement we first calculate the system density matrices that would result from perfect ancilla measurements in its three basis states. The traces of these density matrices are the probabilities of observing the ancilla in a particular state, and therefore we take the gate failure probability to be $1-\mathrm{Tr}\left[\ket{g}\bra{g}\rho^f_k\right]$.

We then compute a weighted sum of these density matrices to find the mixed logical state resulting from an imperfect observation of $\ket{g}$, with weights given by the probability of misassigning a particular state. The overall model is expressed as
\begin{equation}
    \rho^{p}_k = \sum_{\psi\in g,e,f} \eta_{g\psi} |\psi\rangle \langle\psi| \rho^f_k |\psi \rangle \langle \psi|,
\end{equation}
where $\eta_{g\psi}$ is the probability of observing $|g\rangle$ when the state was $|\psi\rangle$. In cQED, most of the readout error comes from decay of the transmon during the readout pulse itself; therefore, we set $\eta_{gg} = 1-10^{-4}$ based on the distinguishability of ancilla pointer states in a typical integrated readout signal \cite{WallraffPRA2017readout}. Given typical readout pulse lengths, we compute the results for $\eta_{ge}=0.01$ and $0.05$. In all cases we assume that $\eta_{gf} = \eta_{ge}^2$ \cite{elder_high-fidelity_2020}. Tracing over the ancilla states yields the resulting mixed density matrix conditioned on an imperfect postselection measurement. 

We then simulate the detection of errors in the bosonic modes by using the appropriate syndrome measurements for the dual-rail and binomial codes. For the dual-rail code with basis states $\{\ket{01},\ket{10}\}$, the occurrence of decays can be detected by measuring joint photon-number parity on each pair of dual-rail modes and observing an odd outcome for both. We can therefore write

\begin{equation}
    \begin{split}
        \hat{M} & = |0_L \rangle \langle 0_L| \otimes |0_L \rangle \langle 0_L| \\
        & + |0_L \rangle \langle 0_L| \otimes |1_L \rangle \langle 1_L| \\
        & + |1_L \rangle \langle 1_L| \otimes |0_L \rangle \langle 0_L| \\
        & + |1_L \rangle \langle 1_L| \otimes |1_L \rangle \langle 1_L|,
    \end{split}
\end{equation}

where $\ket{0}_L = \ket{01}$ and $\ket{1}_L = \ket{10}$. For the binomial code, photon number parity measurements are used for syndrome measurements, and therefore the corresponding measurement operator is

\begin{equation}
    \hat{M} = \left(\frac{\mathds{1}+e^{i\pi(\hat{a}^\dagger \hat{a})}}{2}\right)\left(\frac{\mathds{1}+e^{i\pi(\hat{b}^\dagger \hat{b})}}{2}\right).
\end{equation}

After simulating idealized syndrome checks, the state of the modes conditioned on observing no error can be written as 

\begin{equation}
    \rho^{p,M}_k = \hat{M} \rho^{p}_k \hat{M}^\dagger.
\end{equation}

Finally, we renormalize to obtain $\tilde{\rho}^{p,M}_k$ and compute the overlap with the pure state obtained from applying the perfect gate unitary $\hat{U}$ to the initial state. We average over the 36 cardinal states to obtain 

\begin{equation}
    \bar{\varepsilon}_{ED} = \frac{1}{36}\sum_k \mathrm{Tr}\left( \tilde{\rho}_k^{p,M} \hat{U} \rho_k^i \hat{U}^\dagger\right).
\end{equation}

\section{Impact of device nonidealities}

\begin{figure*}[t]
    \centering
    \includegraphics[scale=0.5]{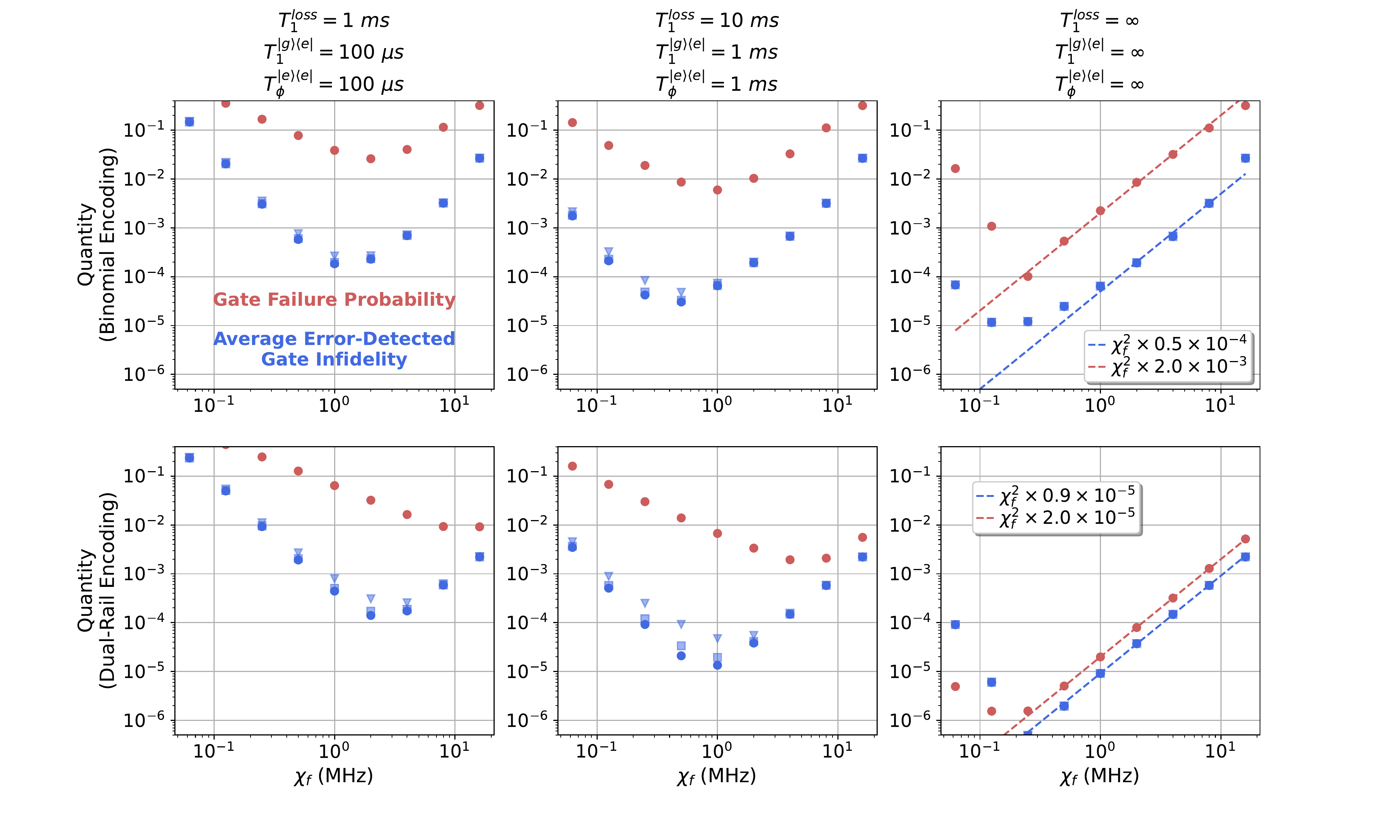}
    \caption{\textbf{The impact of undesired nonlinearities due to ancilla coupling.} Gate failure probability (red circles) and average error-detected gate infidelity (blue circles) for three sets of coherence parameters as functions of $\chi_f$. The resulting infidelities and error probabilities scale quadratically with $\chi_f$, as shown with fits to the simulation data (blue and red dashed lines). Reducing $\chi_f$ allows one to reduce the strengths of self-Kerr and higher-order corrections to the dispersive interaction, but increases the impact of decoherence and cross-Kerr, leading to the emergence of an optimal operating point. Having 99\% measurement error (blue squares) or 95\% measurement error has a small but sometimes noticeable impact on the resulting logical infidelity. }
    \label{fig:nonlinearity}
\end{figure*}

In cQED, the dispersive Hamiltonian is only an approximation, with known higher-order corrections. The next-highest-order terms are the self-Kerr and cross-Kerr between bosonic modes, as well as higher-order corrections to the dispersive interaction itself. Although these are typically a factor of 100-1000 weaker than $\chi$, they present a source of coherent error unaccounted for in the proposed gate design. Here we quantify their effects on gate performance for realistic experimental parameters. 

The higher-order corrections are modelled with the Hamiltonian

\begin{eqnarray}
\begin{split}
\hat{\mathcal{H}}_{NL}/\hbar &= -\frac{K_a}{2} \hat{a}^\dagger \hat{a}^\dagger \hat{a} \hat{a} - \frac{K_b}{2} \hat{b}^\dagger \hat{b}^\dagger \hat{b} \hat{b} \\
&+ \hat{a}^\dagger \hat{a}^\dagger \hat{a} \hat{a}\left(\chi_{f}^\prime \ket{f}\bra{f} + \chi_{e}^\prime \ket{e}\bra{e} \right) \\
&+ \chi_{ab}\hat{a}^\dagger \hat{a} \hat{b}^\dagger \hat{b},
\end{split}
\end{eqnarray}
where $K_a, K_b$ are the self-Kerrs of each bosonic mode, $\chi_{e}', \chi_{f}'$ are higher-order corrections to the dispersive interaction, and $\chi_{ab}$ is the cross-Kerr between the bosonic modes, which may result from both modes participating in the nonlinear element used to actuate the beamsplitter coupling. 

What is the expected error induced by these terms? We can write the rates associated with these corrections as $\tilde{K}\sim(K_a,K_b,\chi'_{f},\chi_{ab})$. As a rough approximation, these terms add a state-dependent detuning to the bosonic modes of order $\bar{n}\sim\tilde{K}$, where $\bar{n}$ is the average photon number in the modes. From our intuition for single qubit gates on a Bloch sphere, this detuning will cause the state to miss its target by a small distance proportional to $\bar{n}\sim\tilde{K}$, which includes the possibility of moving off of the surface of the sphere due to distortion of the logical state. Since fidelity is quadratic in the state overlap, our ancilla-controlled unitaries should only be quadratically sensitive to this detuning, with an infidelity expected to scale as $\sim(\bar{n}\tilde{K}\tau_{\text{gate}})^2\sim\left(\frac{\bar{n}\tilde{K}}{\chi_{f}}\right)^2$. 

When designing a device for experiment, we can control $\chi_{f}$ through the geometry of the device. Increasing this parameter causes $\tau_{\text{gate}}$ to decrease linearly but $\tilde{K}$ to increase \emph{quadratically}. As such, reducing $\chi_f$ will always reduce the gate error from $\hat{\mathcal{H}}_{NL}$. On the other hand, reducing $\chi_f$ will always increase the errors due to decoherence by increasing $\tau_{\text{gate}}$. These competing effects mean there is an optimal value of $\chi_f$ to engineer.

Additional simulations displayed in Fig.~\ref{fig:nonlinearity} highlight the competition between nonlinearity and decoherence. From typical values we measure in experiment \cite{reinhold_error-corrected_2020}, we assume an initial operating point where $\chi_{f}^\prime/2\pi = 2$ kHz, $\chi_{e}^\prime/2\pi = 1.125$ kHz and $K_a/2\pi = K_b/2\pi = 2$ kHz when $\chi_{f}/2\pi=-1$ MHz, and that these quantities scale quadratically if we were to vary $\chi_{f}$. We also assume $\chi_{ab}/2\pi=100$ Hz for all $\chi_{f}$, chosen to reflect values attainable in modern hardware \cite{StijnBS2022}. For the transmon pulse we again neglect the dispersive interaction and its higher-order corrections.

The rightmost column shows that in the absence of decoherence, the error associated with nonlinearity scales quadratically in $\chi_f$ as expected. The leftmost and center columns highlight that when decoherence is introduced, the benefits of a weaker coupling to the nonlinear ancilla are overridden by the increased incoherent error experienced by the slower gate. This leads to an optimum which may be found for different bosonic codes and decoherence rates. For the codes and parameters explored here, we find the optimal $\chi_{f}$ to be $\sim1\text{ MHz}$ and observe persistent error-detected gate infidelities of order $10^{-4}$ and below.

\section{Error scaling prefactors for different error channels and bosonic codes}
\label{app:sim_scalings}
The simulation results shown in Fig. \ref{fig:simulation} show that gate failure scales linearly whilst error-detected gate infidelity scales quadratically, with each hardware decoherence rate considered. The fits take the form $A_n\left(\frac{\tau_{\text{gate}}}{T_{\text{coh}}}\right)^n$. For a given error channel we have found $A_2$ is significantly smaller than $A_1$ (and often much smaller than 1), granting further protection against second-order hardware errors; here we explain this phenomenon. 

We begin with the case of photon loss. The probability of a single photon loss occurring determines $A_1$. For the binomial code, there are an average of two photons in each bosonic mode for a total of 4. Hence $A_1\approx4$ for the gate failure probability. Similarly, for the dual-rail code $A_1\approx2$. Double photon loss in the binomial code sets $A_2$ for the error-detected gate infidelity. The probability of double photon loss is $\frac{1}{2}\left(\frac{\bar{n}\tau_{\text{gate}}}{T_{\text{coh}}}\right)^2$, where the factor of $1/2$ comes from the fact that photon loss must occur sequentially in a given time window. Half of the time, double photon loss results in the detectable error $\hat{a}\hat{b}$. Overall, this means that $A_2=\frac{1}{4}\bar{n}^2\approx 1$

For the gate failure probability resulting from ancilla errors, $A_1\approx1$. A single ancilla error results in a failed gate. The values of $A_2$ for ancilla errors require more detailed analysis. For ancilla dephasing, if two $\sigma_z$ errors occur within the same ancilla-controlled unitary, they cancel each other out. Only if they occur in different ancilla-controlled unitaries do they cause a gate error and hence we pick up a factor of $1/2$. When this error happens, the applied gate is $ZZ_L(-\pi/2)$ which causes an error on half of the cardinal states, yielding another factor of $1/2$. Overall, this makes $A_2\approx 1/4$ for ancilla dephasing. 

Finally, $A_2$ for ancilla decay is the most involved to calculate. Double decay errors require decay to $\ket{e}$, then to $\ket{g}$. The $\ket{g}\bra{e}$ decay rate is half the $\ket{e}\bra{f}$ decay rate, yielding an initial factor of $1/2$. Decay to $\ket{e}$ must happen before decay to $\ket{g}$ in the same time window, giving the next factor of $1/2$. Most of the time, double decay to $\ket{g}$ will leave the system in a random ancilla state in the $gf$-manifold due to the ancilla rotations in the sequence, and hence double decays are detected as $\ket{f}$ at the end of the sequence half of the time. Finally, we assume that when double decay happens, the bosonic modes may be outside of the codespace, but still have some overlap with the target states. This quantity is difficult to calculate. We define it as $A_{\text{leak}}<1$. Overall, this means that $A_1 = A_{\text{leak}}/8<1/8$ for double ancilla decay errors. In general, these combinatoric factors help further suppress the effects of second-order ancilla errors.

\end{document}